\documentclass[12pt]{iopart}
\pdfoutput=1
\expandafter\let\csname equation*\endcsname\relax
\expandafter\let\csname endequation*\endcsname\relax

\usepackage{color}
\usepackage[breakwords]{truncate}
\usepackage[font={small}]{caption}
\usepackage{physics}
\usepackage{graphicx}
\usepackage{cite} 
\usepackage{subcaption}
\usepackage{etoolbox}
\usepackage[section]{placeins}
\usepackage[sectionbib]{chapterbib}
\usepackage{amsfonts}
\usepackage[bookmarks,bookmarksopen,bookmarksdepth=2]{hyperref}
\makeatletter
\newrobustcmd{\fixappendix}{%
  \patchcmd{\l@section}{1.5em}{7em}{}{}%
  \patchcmd{\l@subsection}{2.3em}{7em}{}{}%
}
\makeatother

\appto\appendix{
\addtocontents{toc}{\fixappendix}
\addtocontents{toc}{\protect\setcounter{tocdepth}{1}}}

\begin{document}
\newcommand{\ben}[1]{\textcolor{blue}{\textbf{#1}}}

\title{Survival probability of a run-and-tumble particle in the presence of a drift}
\author{Benjamin De Bruyne}
\address{LPTMS, CNRS, Univ.\ Paris-Sud, Universit\'e Paris-Saclay, 91405 Orsay, France}
\author{Satya N. Majumdar}
\address{LPTMS, CNRS, Univ.\ Paris-Sud, Universit\'e Paris-Saclay, 91405 Orsay, France}
\author{Gr{\'e}gory Schehr }
\address{Sorbonne Universit\'e, Laboratoire de Physique Th\'eorique et Hautes Energies, CNRS UMR 7589, 4 Place Jussieu, 75252 Paris Cedex 05, France}
\eads{\mailto{benjamin.debruyne@centraliens.net}, \mailto{satya.majumdar@universite-paris-saclay.fr},
\mailto{gregory.schehr@u-psud.fr}}
\begin{abstract}
We consider a one-dimensional run-and-tumble particle, or persistent random walk, in the presence of 
an absorbing boundary located at the origin. After each tumbling event, which occurs at a constant rate $\gamma$, the (new) velocity 
of the particle is drawn randomly from a distribution $W(v)$. We study the survival probability $S(x,t)$ of a particle starting from $x \geq 0$ up to time $t$
and obtain an explicit expression for its double Laplace transform (with respect to both $x$ and $t$) for an {\it arbitrary} velocity distribution $W(v)$, not necessarily symmetric. This result is obtained as a consequence of Spitzer's formula, which is well known in the theory of random walks and can be viewed as a generalization of  the Sparre Andersen theorem. We then apply this general result to the specific case of 
a two-state particle with velocity $\pm v_0$, the so-called persistent random walk (PRW), and in the presence of a constant drift $\mu$ and obtain an explicit expression for $S(x,t)$, for which we present more detailed results. Depending on the drift $\mu$, 
we find a rich variety of behaviours for $S(x,t)$, leading to three distinct cases: (i) {\it subcritical} drift $-v_0\!<\!\mu\!<\! v_0$, (ii) {\it supercritical} drift $\mu < -v_0$ and (iii) {\it critical} drift $\mu=-v_0$. In these three cases, we obtain exact analytical expressions for the survival probability $S(x,t)$ and establish connections with existing formulae in the mathematics literature. Finally, we discuss some applications of these results to record statistics and to the statistics of last-passage~times. 
\end{abstract}
\noindent{\it Keywords\/}: Run-and-tumble, Telegraphic process, Survival probability, First-passage time, Drifted process.

\submitto{\JSTAT}
\newpage
{\pagestyle{plain}
 \tableofcontents
\cleardoublepage}

\section{Introduction, model and summary of the main results}
\subsection{Introduction}
 Brownian motion (BM) is certainly the most popular stochastic process to model particles in interaction with their surrounding environment~\cite{einstein06}. In its simplest form, BM 
 is driven by an uncorrelated white noise $\xi(t)$ induced by thermal fluctuations. Thanks to its universality, BM has been shown to be at the heart of many complex systems ranging from colloidal solutions \cite{von1906kinetischen,krapivsky2010kinetic} to financial stock markets \cite{bachelier1900theorie,bouchaud2000theory}, and all the way to applications in astrophysics \cite{chandrasekhar1943stochastic}. On the theoretical side, BM has attracted a considerable amount of interest, in particular because of its numerous connections with other problems in theoretical physics and probability theory \cite{feller2008introduction,pitman2018guide} --- a case in point being the extreme statistics of 
 Brownian motion \cite{majumdar2020extreme}. Related to extreme value questions, the survival probability of a Brownian motion in the presence of an absorbing boundary has been extensively studied~\cite{bray2013persistence,redner2001guide,benichou2011intermittent,majumdar2007brownian,majumdar2010universal,aurzada2015persistence}. The first-passage time to an absorbing boundary plays a crucial role in various phenomena such as animals searching for food, financial stocks reaching a stop price or rivers overflowing their banks. Consider for instance a one-dimensional Brownian motion with diffusion constant $D$, starting from $x>0$ and in the presence of an absorbing boundary located at the origin. It is well known that the survival probability $S(x,t)$ of this Brownian particle up to time $t$ is given by \cite{redner2001guide,bray2013persistence}
\begin{align}
  S(x,t) = \text{erf}\left(\frac{x}{\sqrt{4\,D\,t}}\right)\,,
\end{align}
where $\text{erf}(z) = \frac{2}{\sqrt{\pi}}\int_0^{z} \e^{-u^2}\,du$ is the error function. 
Starting from $x>0$, the BM inevitably crosses the origin as the survival probability decays to 0 as $S(x,t) \propto t^{-1/2}$ at late times. Biasing the motion away from the absorbing boundary, e.g., by turning on a positive drift, increases the survival probability and has been the subject of recent works \cite{majumdar2012record,majumdar2002exact,mounaix2018asymptotics,de2020optimization,de2020optimization2}.

 While the survival probability of the Brownian motion has been known since a long time, general results are few and far between in the case when the particles are driven by correlated noise~\cite{hanggi1995colored}. An example of much current interest concerns the so-called active particles that naturally emerge in the context of living matter such as \emph{E. coli} bacteria \cite{berg2008coli} and fish schools or bird flocks~\cite{marchetti2013hydrodynamics}. A tremendous amount of numerical and experimental work has been devoted to them \cite{marchetti2013hydrodynamics,cates2015motility,bechinger2016active,tailleur2008statistical,
 berg2008coli,vicsek1995novel}. The ability of active particles to move autonomously renders them inherently different from the Brownian particles which are usually driven by collisions with the molecules in the surrounding medium. One model of active particle, currently of much interest, is the run-and-tumble particle (RTP), also known as the telegraphic process \cite{kac1974stochastic} or the persistent random walk \cite{weiss2002some, masoliver2017continuous},  which are driven by exponentially correlated noise. Interacting active particles  are known to exhibit a plethora of collective phenomena. Interestingly, active particles also display quite rich behaviors, already at the level of a {\it single} particle or of {\it noninteracting} RTP's. These include non-trivial density profiles \cite{bijnens2020pushing,martens2012probability,basu2019long,basu2020exact,dhar2019run,singh2020run,santra2020run}, dynamical phase transitions \cite{doussal2020velocity,gradenigo2019first} or anomalous transport properties \cite{dor2019ramifications,demaerel2019asymptotic,banerjee2020current,doussal2020velocity}.

 Of particular relevance are the first-passage properties of RTP models, which have been widely studied both in mathematics \cite{orsingher1990probability,orsingher1995motions,lopez2014asymmetric,cinque2020exact,cinque2020negative} as well as in physics \cite{foong1992first,masoliver1992solutions,angelani2014first,artuso2014sparre,angelani2015run,evans2018run,PhysRevA.36.1435,malakar2018steady,le2019noncrossing,mori2020universal,moriE2020universal}. Remarkably, it was recently realized \cite{mori2020universal,moriE2020universal} that for a wide class of {\it symmetric} RTP models, the survival probability $S(x=0,t)$ for a particle starting from the origin, i.e., exactly where the absorbing boundary is located, exhibits a universal behavior for all
 time $t$, reminiscent of the celebrated Sparre Andersen theorem, well known in the literature on one-dimensional discrete time random walks \cite{feller2008introduction,andersen1954fluctuations}. Such universal behavior was then shown to hold for a certain class of discrete time persistent random walk models~\cite{lacroix2020universal}. In fact, most of the results for the survival probability of RTP models concern the case of {\it symmetric} RTP models, where the velocity distribution $W(v)$ of the particle is symmetric, i.e. $W(v) = W(-v)$. In contrast, much less is known about the survival probability for {\it asymmetric} velocity distributions. There has been however a few studies devoted to first-passage times of asymmetric RTP's both in physics \cite{dor2019ramifications,doussal2020velocity} (e.g. for the mean first-passage time) as well as in the mathematics literature \cite{stadje2004telegraph,lopez2014asymmetric,cinque2020exact}. In particular, in \cite{lopez2014asymmetric}, $S(x,t)$ was obtained for the special simpler case of the two-state RTP in the presence of a constant drift [see (\ref{eq:eom}) below]. Even in this special case, the physical implications of the obtained formulae in \cite{lopez2014asymmetric} were not discussed in detail. 
 
 In this paper, we obtain an explicit formula for the double Laplace transform of the survival probability $S(x,t)$ -- with respect to both $x$ and $t$ --  
 for an RTP in one-dimension with an arbitrary velocity distribution $W(v)$, which can be either symmetric or asymmetric [see (\ref{intro_S})]. Our formula thus generalizes, to the case of RTP, the well-known Spitzer's formula valid for one-dimensional discrete time random walks with arbitrary jump distribution. We then apply this general formula to various examples of $W(v)$. This includes in particular the simpler case of the two-state RTP in the presence of a constant drift $\mu$, where our method recovers the previously known result \cite{lopez2014asymmetric}. In addition, we discuss the physical implications of the behaviour of $S(x,t)$ as a function of both the starting position $x$ as well as the time $t$, unveiling very rich behaviours depending on the strength of the drift $\mu$.

 
\subsection{The generalized run-and-tumble model}
 
  We consider an RTP in one-dimension starting from the initial position $x \geq 0$. The initial point is considered as a tumble. The particle chooses
a velocity $v$ from a distribution $W(v)$ (which can be asymmetric) and runs ballistically with this chosen velocity $v$ during a random run time $\tau$
drawn from an exponential distribution $p(\tau) = \gamma \,\e^{-\gamma \tau}$ where $\gamma^{-1}$ is the persistence time, i.e. the typical life-time of a run between
two consecutive tumblings.  At the end of the
run, the particle tumbles instantaneously and chooses a new velocity $v$ drawn again from the same distribution $W(v)$. It then runs again during an 
exponentially distributed random time $\tau$ drawn from the same $p(\tau)$. This run-and-tumble process continues till the fixed time $t$. 

\begin{figure}[t]
        \centering
        \includegraphics[width=0.4\textwidth]{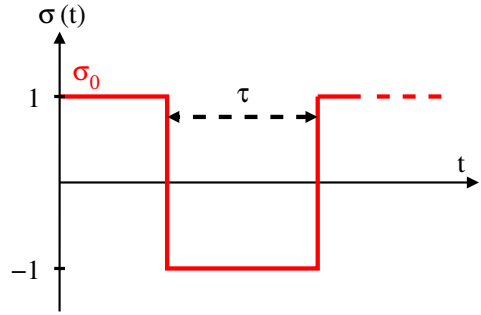}
    \caption{Telegraphic noise $\sigma(t)$ with $\sigma_0\equiv\sigma(0)=1$ whose evolution is given in (\ref{eq:telegraphic}). The time between two consecutive switches $\tau$ is drawn from an exponential distribution $p(\tau)=\tilde \gamma \, \rme^{-\tilde \gamma \tau}$.}
            \label{fig:telegraphic}
\end{figure}

Another well studied model in the literature is the so-called persistent random walk (PRW) \cite{kac1974stochastic,weiss2002some, masoliver2017continuous} where the position $x(t)$ of a particle, starting at $x(0)=x$,
evolves stochastically as 
 \begin{align}
    \dot{x}(t) = \mu + v_0\, \sigma(t)\, ,
    \label{eq:eom}
\end{align}
where $\mu$ is a drift and $\sigma(t)$ is a telegraphic noise that switches between the values $1$ and $-1$ according to a Poisson process with rate $\tilde \gamma$ (see figure \ref{fig:telegraphic}). The initial value of the telegraphic signal is denoted by $\sigma_0\equiv \sigma(0)$. During an infinitesimal time interval $dt$, the signal changes sign with probability $\tilde \gamma\, dt$ and remains constant with the complementary probability $1-\tilde \gamma\, dt$
\begin{align}
  \sigma(t+dt) = \left\{\begin{array}{rl}\sigma(t)\, ,  & \text{prob.~ }=1-\tilde \gamma\, dt\, , \\
  -\sigma(t)\, , &\text{prob.~ } =\tilde \gamma\, dt\, . \end{array}\right. \label{eq:telegraphic}
\end{align}
The time $\tau$ between two consecutive switches is thus distributed according to an exponential distribution $p(\tau)=\tilde \gamma\,\rme^{-\tilde \gamma\,\tau}$. The autocorrelation function of the telegraphic noise can be easily computed (see e.g. \ref{app:Poisson}) and one obtains
\begin{align}
   \langle \sigma(t_1)\sigma(t_2) \rangle = \rme^{-2\,\tilde \gamma\, (t_2-t_1)}\,. \label{eq:autocorell}
 \end{align}
The autocorelation function (\ref{eq:autocorell}) is said to be \emph{colored} because it has a finite correlation time $\tilde \gamma^{-1}$ which is called the \emph{persistence} time. This will be reflected in the motion of the run-and-tumble particle $x(t)$ (\ref{eq:eom}) which will in turn exhibit memory effects. This persistence, also called \emph{activity}, renders the process non-Markovian and hence does not fall into the universality class of the Brownian motion, which makes this process challenging to study. Nevertheless, it is possible to recover the Brownian diffusive regime by taking the scaling limit
\begin{align}
\tilde \gamma\rightarrow \infty\,, \quad v_0\rightarrow \infty\,,\quad \frac{v_0^2}{2\,\tilde \gamma}\equiv D\, ,\label{eq:brownianlimit}
\end{align}such that the effective diffusion coefficient $D$ is finite. In this limit, the persistence time $\tilde \gamma^{-1}$ tends to zero and the run-and-tumble particle behaves like a Brownian motion. Indeed, in this limit, the driving noise in the equation of motion (\ref{eq:eom}) becomes
\begin{align}
   \langle  v_0\,\sigma(t_1)\, v_0\,\sigma(t_2)\rangle = v_0^2\, \rme^{-2\tilde \gamma (t_2-t_1)} \rightarrow 2D\, \delta(t_2-t_1)\, ,
\end{align}
which is the well-known uncorrelated white noise.

It is easy to see that this PRW model is a special case of the more general RTP model defined earlier. Indeed it corresponds to choosing
a velocity distribution and a tumbling rate
\begin{align}
  W(v) = \frac{1}{2}\,\delta(v-\mu-v_0) + \frac{1}{2}\,\delta(v-\mu+v_0)\;, \quad {\rm and}  \quad \gamma = 2 \,\tilde \gamma \;. \label{eq:Wintro}
\end{align}
Note that in the general model, $\gamma$ denotes the rate of tumbling, i.e., the process renews with rate $\gamma$. Thus, after each tumbling, the velocity may either flip sign or retain the same sign with equal probability. Therefore, the rate at which the velocity changes sign is $\gamma/2$. This explains the relation $\tilde \gamma = \gamma/2$ in (\ref{eq:Wintro}).


In the present work, we are interested in computing the survival probability $S(x,t)$ of a general RTP with velocity distribution $W(v)$. Here $S(x,t)$ is the probability that
the particle, starting at $x \geq 0$, does not cross the origin up to time $t$. Since the paper is long, it is useful to provide a summary of our main results. After deriving a general
result for $S(x,t)$ with arbitrary $W(v)$, we will focus on the special case of the PRW in~(\ref{eq:Wintro}). For this case, we are able to derive detailed results for $S(x,t)$. Furthermore, 
to illustrate the usefulness of our general formula valid for arbitrary $W(v)$, we show how explicit results for $S(x,t)$ can be derived in another example of $W(v)$.

\begin{figure}[t]
    \centering
    \begin{subfigure}[t]{0.325\textwidth}
        \centering
        \includegraphics[width=\textwidth]{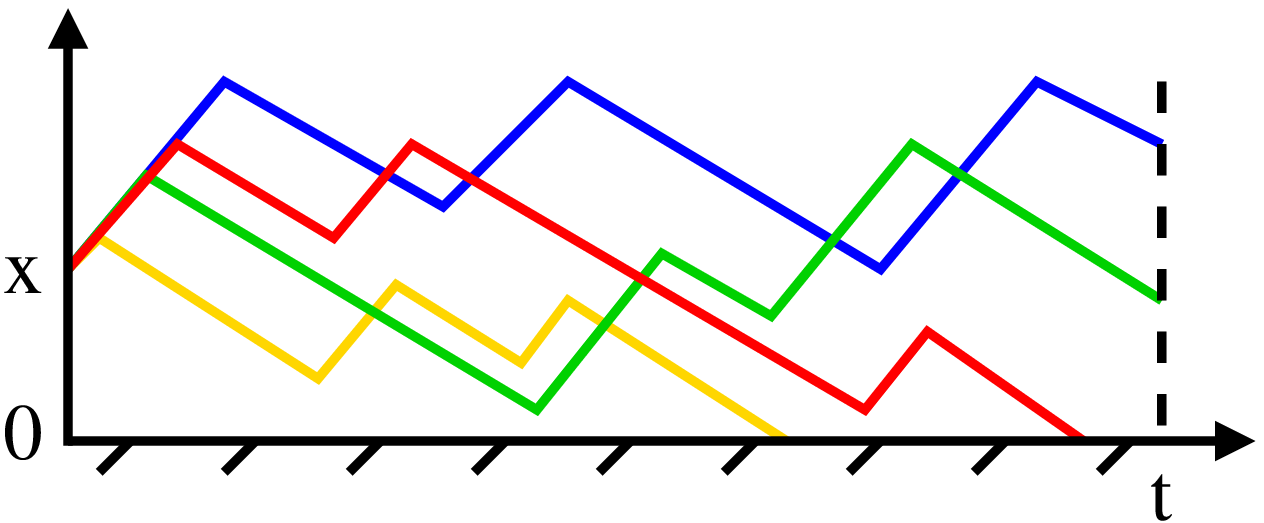}
        \caption{\scriptsize{Subcritical drift $-v_0<\mu<v_0$}}
        \label{fig:survivalw}
    \end{subfigure}%
    ~ 
    \begin{subfigure}[t]{0.32\textwidth}
        \centering
        \includegraphics[width=\textwidth]{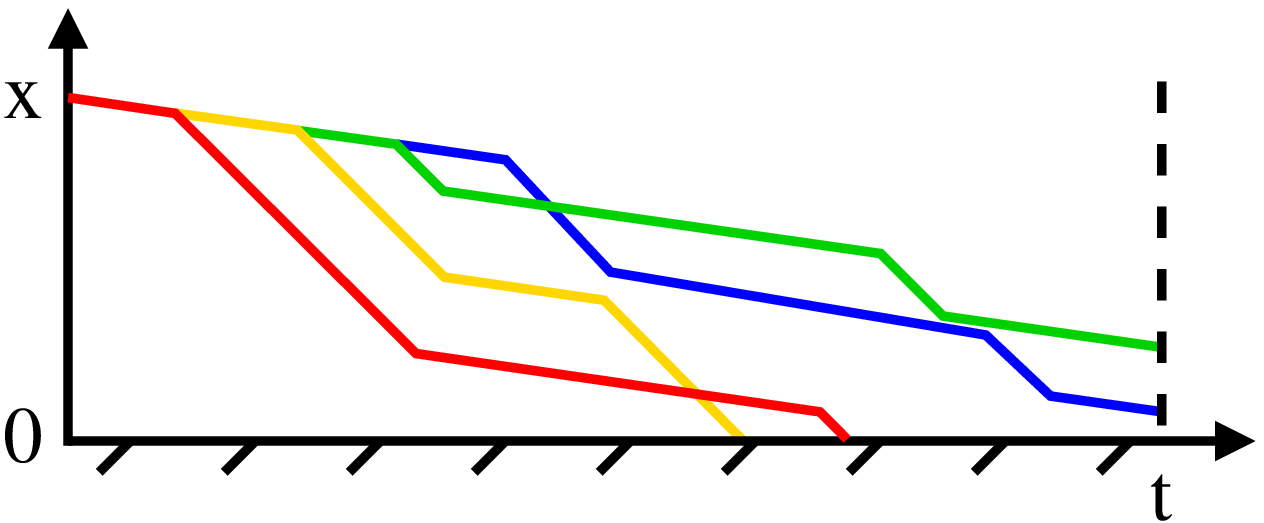}
        \caption{\scriptsize{Supercritical drift $\mu<-v_0$}}
        \label{fig:survivals}
    \end{subfigure}
    \begin{subfigure}[t]{0.32\textwidth}
        \centering
        \includegraphics[width=\textwidth]{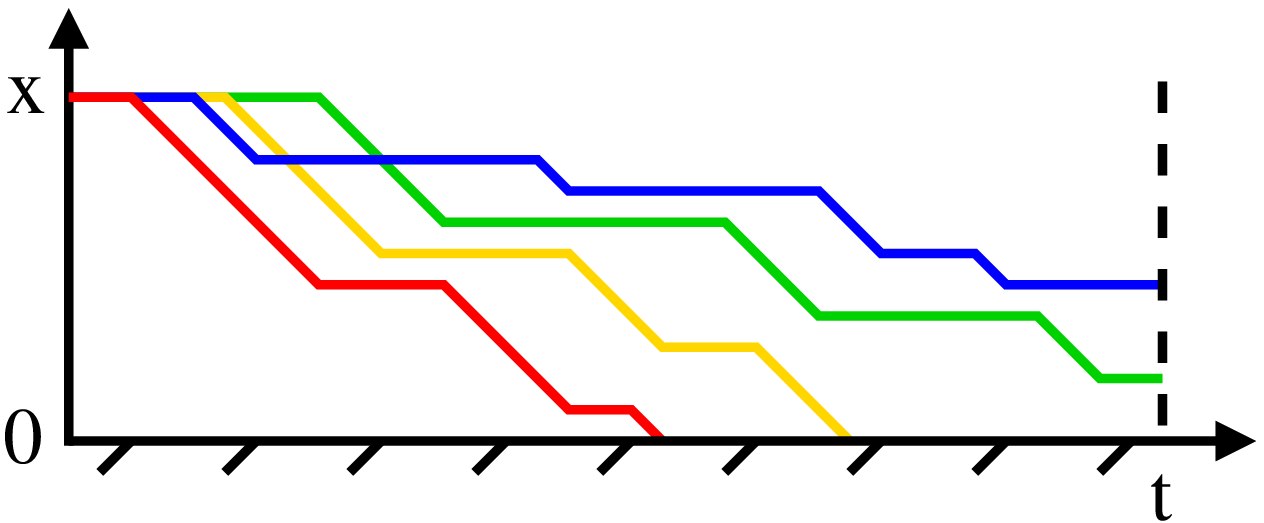}
        \caption{\scriptsize{Critical drift $\mu=-v_0$}}
        \label{fig:survivalc}
    \end{subfigure}
    \caption{Typical trajectories $x(t)$ of an RTP, starting at $x(0)=x$, with velocity $v_0$ in the presence of a drift $\mu$ [see (\ref{eq:eom})] and an absorbing boundary at the origin. The velocity of the particle $\dot x(t)$ given by the equation of motion (\ref{eq:eom}) can take two values: $\mu+v_0$ and $\mu-v_0$. Depending on the strength of the drift, the sign of these two values can either be positive and negative (a), both negative (b) or zero and negative (c). In the three panels, the blue and the green trajectories have not crossed the origin at time $t$ and will contribute to the survival probability $S_{\sigma_0}(x,t)$ while the red and the yellow ones will not (with $\sigma_0 = \sigma(t=0)$).}
    \label{fig:survival}
\end{figure}

\subsection{Summary of the main results}
\label{sec:summaryres}

For an RTP with arbitrary $W(v)$, we 
obtain a closed-form expression for the double Laplace transform of the survival probability $S(x,t)$
\begin{eqnarray} \label{intro_S}
 \hspace*{-2.6cm}\int_0^\infty dx\, \int_0^\infty dt \,S(x,t)\,\rme^{-u\,x - s \, t} && \nonumber \\
&&\hspace*{-2cm} = \frac{\gamma+s}{\gamma \, u\,s}\exp(-\frac{\rmi}{2\pi}\int_{\rmi\, \mathbb{R}} dz \ln(\frac{z+u}{z}) \frac{\int_{-\infty}^\infty   dv\, \frac{v\,W(v)}{(\gamma+s+z\,v)^2}}{\frac{1}{\gamma}-\int_{-\infty}^\infty   dv\, \frac{W(v)}{(\gamma+s+z\,v)}})-\frac{1}{\gamma u}\,, \label{eq:gformula_intro}
\end{eqnarray}
where $\rmi \mathbb{R}$ denotes the imaginary axis in the complex $z$ plane. Even though this formula may look a bit formal, one of the goals of this paper is to show that in various special cases of $W(v)$, it is possible to extract $S(x,t)$ explicitly from (\ref{intro_S}), an example being the two-state PRW model defined in (\ref{eq:eom}). As shown in \ref{app:spitzerpm}, we also obtain formulae similar to (\ref{eq:gformula_intro}) for the survival probabilities for a generic RTP conditioned to start with a negative or positive velocity [see (\ref{doubleLT_Sdown}) an (\ref{doubleLT_Sup})]. 

\begin{figure}[ht]
    \centering
    \includegraphics[width=0.8\textwidth]{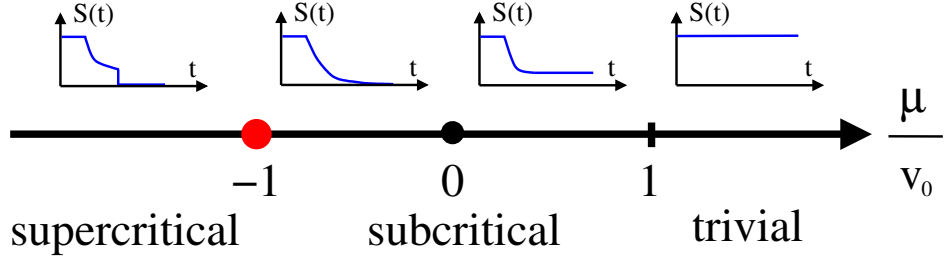} 
    \caption{Phase diagram illustrating the schematic decay profiles of the survival probability $S(t) \equiv S(x\!=\!1,t)$ at a fixed starting point, say $x\!=\!1$, as a function of $t$, for different values of the ratio $\mu/v_0$. When $-v_0\!<\!\mu\!<\!v_0$, the drift is subcritical and the motion of the particle consists of up and down runs (see figure \ref{fig:survivalw}). When $\mu\!<\!-v_0$, we say that the drift is supercritical and the particle always moves towards the origin (see figure \ref{fig:survivals}). At the critical point $\mu\!=\!-v_0$, the drift exactly balances the internal speed of the particle leading to an effective wait-and-run motion (see figure \ref{fig:survivalc}). When $\mu\!>\!v_0$, the particle always moves away from the origin and the survival probability is trivially $1$ at all times. }
    \label{fig:phase_diagram}
\end{figure}

Let us present our results for the specific PRW [see (\ref{eq:eom})], where the velocity $\dot x(t)$ can take only two values $\mu+v_0$ and $\mu-v_0$. In this case, it is also
natural to consider the survival probabilities $S_{\pm}(x,t)$, which are the survival probabilities up to time for a PRW starting from $x$ with an initial velocity given respectively by $\dot x(0) = \mu \pm v_0$, i.e.,
\begin{subequations} 
\begin{eqnarray}
S_{+}(x,t) &=& {\rm Pr.}\left[x(\tau) \geq 0 \,, \forall \tau \in [0,t] \, \Big | \, x(t=0) = x, \, \dot x(0) = \mu + v_0 \right] \;, \label{eq:def_Sp} \\
S_{-}(x,t) &=& {\rm Pr.}\left[x(\tau) \geq 0 \,, \forall \tau \in [0,t] \, \Big | \, x(t=0) = x, \,\dot x(0) = \mu - v_0\right] \label{eq:def_Sm}.
\end{eqnarray}
\end{subequations}
We find that these probabilities display a rich behavior, as function of $x$ and $t$, depending on the two parameters $\mu$ and $v_0$ (see figure \ref{fig:phase_diagram}): 
\begin{itemize}
  \item $\mu\geq v_0$ (trivial): $\dot x(t)$ is always positive which means that the particle always moves away from the origin. The survival probability in this case is trivially $1$.
  \item $-v_0\!<\!\mu\!<\!v_0$ (subcritical): $\dot x(t)$ can be positive or negative which means that the particle alternates between up runs, away from the origin, and down runs, towards the origin (see figure \ref{fig:survivalw}). Due to the down runs, it is now possible that the particle crosses the origin and its survival probability decays exponentially with time. More precisely, it decays to $0$ if $-v_0\!<\!\mu\!<\!0$ or to a finite positive value if $0\!<\!\mu\!<\!v_0$.
  \item $\mu\!<\!-v_0$ (supercritical): $\dot x(t)$ is always negative which means that the particle always moves towards the origin (see figure \ref{fig:survivals}) and the survival probability decays to $0$ in a finite time.
  \item $\mu\!=\!-v_0$ (critical): the two possible values for $\dot x(t)$ are $0$ and $-2\,v_0$ which means that the particle alternatively waits and runs towards the origin (see figure \ref{fig:survivalc}). In this case, the survival probability decays anomalously to $0$. The precise form of this anomalous decay will be discussed in section \ref{sec:survc}.
\end{itemize}   
This leads us to divide the presentation of our results into three parts: (i) subcritical drift $-v_0\!<\!\mu\!<\!v_0$, (ii) supercritical drift $\mu\!<\!-v_0$  and (iii) critical drift $\mu\!=\!-v_0$. These three parts are summarized in the phase diagram in figure \ref{fig:phase_diagram}.  

\begin{figure}[t]
    \centering
    \begin{subfigure}[t]{0.5\textwidth}
        \centering
        \includegraphics[width=\textwidth]{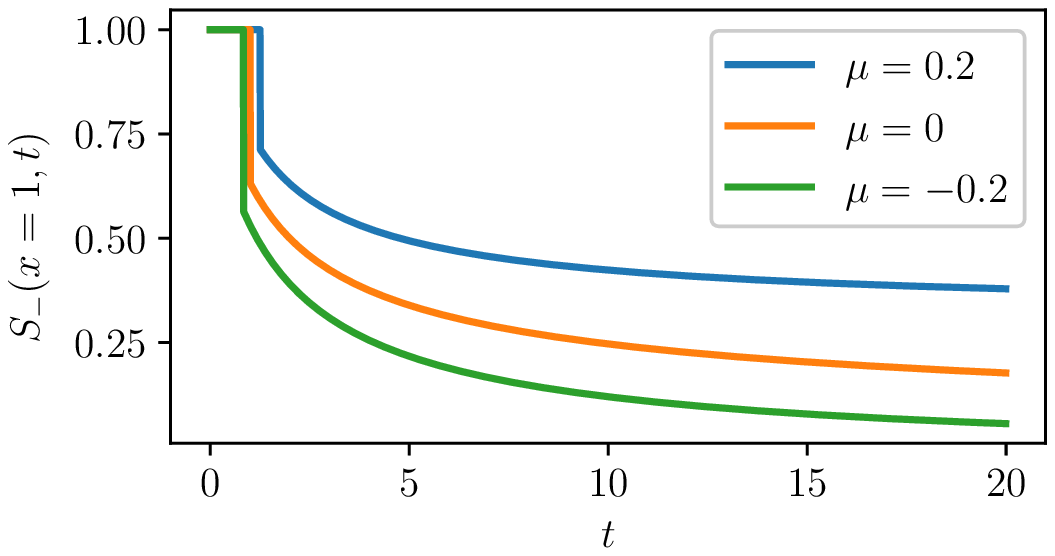}
        \caption{$\sigma_0=-1$}
        \label{fig:survivalsub1}
    \end{subfigure}%
    ~ 
    \begin{subfigure}[t]{0.5\textwidth}
        \centering
        \includegraphics[width=\textwidth]{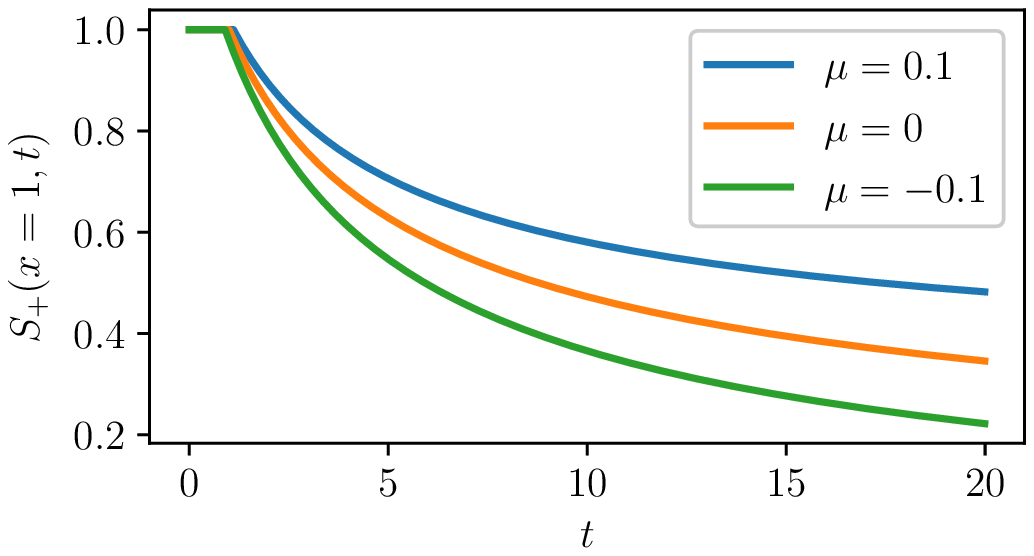}
        \caption{$\sigma_0=+1$}
        \label{fig:survivalsub2}
    \end{subfigure}
    \caption{Survival probability $S_{\sigma_0}(x\!=\!1,t)$, with $\sigma_0 = \pm 1$, in the presence of a subcritical drift $-1\!<\!\mu\!<\!1$ for a particle starting in the state $\sigma_0$.  $S_{\sigma_0}(x\!=\!1,t)$ only starts to decrease after a time $t_m = 1/(1-\mu)$ (\ref{eq:tm}) which is the minimum time it takes for the particle to reach the origin. $S_-(1,t)$ has a sudden drop at $t=t_m$ which is due to the trajectory that goes straight to the origin without tumbling (this happens with a finite probability weight $\rme^{-t_m}$). This trajectory is not present for $S_+(x,t)$. In the long-time limit, both $S_+(x,t)$ and $S_-(x,t)$ tend towards a finite probability when $0\!<\!\mu\!<\!1 $ (\ref{eq:longtpos}) or decay exponentially when $-1\!<\!\mu\!\leq\!0$~(\ref{eq:longtneg}).}
    \label{fig:survivalsub}
\end{figure}

Before presenting our results, let us set $v_0\!=\!1$ and $\tilde \gamma\!=\!1$ for the remaining of this section. This simply amounts to rescale all the times by $\tilde \gamma$ and all the positions by $v_0/\tilde \gamma$. It is always possible to reintroduce the units by performing the replacements
 \begin{align}
   x &\rightarrow \frac{\tilde \gamma\, x}{v_0}\,, \quad
   t \rightarrow \tilde \gamma\, t\,, \quad
   \mu\rightarrow \frac{\mu}{v_0}\,. 
   \label{eq:units}
 \end{align}
 In these dimensionless units, the velocity $\dot x(t)$ of the run-and-tumble particle evolving according to the equation of motion (\ref{eq:eom}) can take the two values $\mu+1$ and $\mu-1$. The drift is subcritical when $-1\!<\!\mu\!<\!1$ (figure \ref{fig:survivalw}), supercritical when  $\mu\!<\!-1$ (figure \ref{fig:survivals}) and critical when $\mu\!=\!-1$ (figure \ref{fig:survivalc}). Our main results for the two-state RTP in the presence of a drift can be summarized as follows:

\paragraph{Subcritical drift ($-1\!<\!\mu\!<\!1$).} In the case of a subcritical drift (see figure \ref{fig:survivalw} and \ref{fig:phase_diagram}), we find that the survival probability is 
 \begin{subequations}
\begin{align}
    S_+(x,t) &= \Bigg\{\begin{array}{lr}
    1\,, & t < t_m\,, \\
         1- \int_{t_m}^{t}\, dt' \, \frac{\rme^{-t'}}{g(t',x)}\left(x \,I_0\left[h(t',x)\right] + (1+\mu)\,\sqrt{\frac{f(t',x)}{g(t',x)}}\,I_1\left[h(t',x)\right]\right)\,, & t\geq t_m\,,
        \end{array} \label{eq:Spsub}\\
        S_-(x,t) &= \Bigg\{\begin{array}{lr}
    1\,, & t < t_m\,, \\
         1-\rme^{-t_m} - \int_{t_m}^{t}\, dt' \,\,\rme^{-t'} \frac{x}{h(t',x)} \, I_1\left[h(t',x)\right]\,, & t\geq t_m\,, 
        \end{array} \label{eq:Spmt}
         \end{align}\label{eq:S11}
\end{subequations}
 where  $I_0(z)$ and $I_1(z)$ are the modified Bessel functions and
 \begin{subequations}
 \begin{align}
 t_m&=\frac{x}{1-\mu}\,, \label{eq:tm}\\
 f(t,x)&= t\,(1-\mu)-x\,,\label{eq:f}\\
 g(t,x)&=t\,(1+\mu)+x\,,\label{eq:g}\\
 h(t,x)&=\sqrt{f(t,x)\, g(t,x)}\,.\label{eq:h}
 \end{align}
 \label{eq:tmfgh}
 \end{subequations}
 These results in (\ref{eq:Spsub}) and (\ref{eq:Spmt}) are in agreement with the ones derived in \cite{lopez2014asymmetric} by a different method using coupled Fokker-Planck equations which works only for this particular class of two-state models, where the magnitude of the velocity is a constant $v_0$. However, our method is more general and applies to RTP models with arbitrary velocity distribution $W(v)$. The survival probability is illustrated in~figure~\ref{fig:survivalsub}. The first-passage time distribution $F_{\pm}(x,t)=-\partial_t S_{\pm}(x,t)$  has a simple analytical expression that enables us to compute the mean first-passage time to the origin which is infinite when $0\!<\!\mu\!<\!1$ and finite when $-1\!<\!\mu\!<\!0 $ (\ref{eq:Tavgwn}). Finally, we remark that for $\mu=0$ (the unbiased case), our result in (\ref{eq:Spsub}) and (\ref{eq:Spmt}) reduces to the well-known result for unbiased PRW \cite{{orsingher1990probability,orsingher1995motions,foong1992first,masoliver1992solutions,angelani2014first,evans2018run,PhysRevA.36.1435,malakar2018steady,le2019noncrossing,mori2020universal,moriE2020universal}}.

\begin{figure}[t]
    \centering
    \begin{subfigure}[t]{0.5\textwidth}
        \centering
        \includegraphics[width=\textwidth]{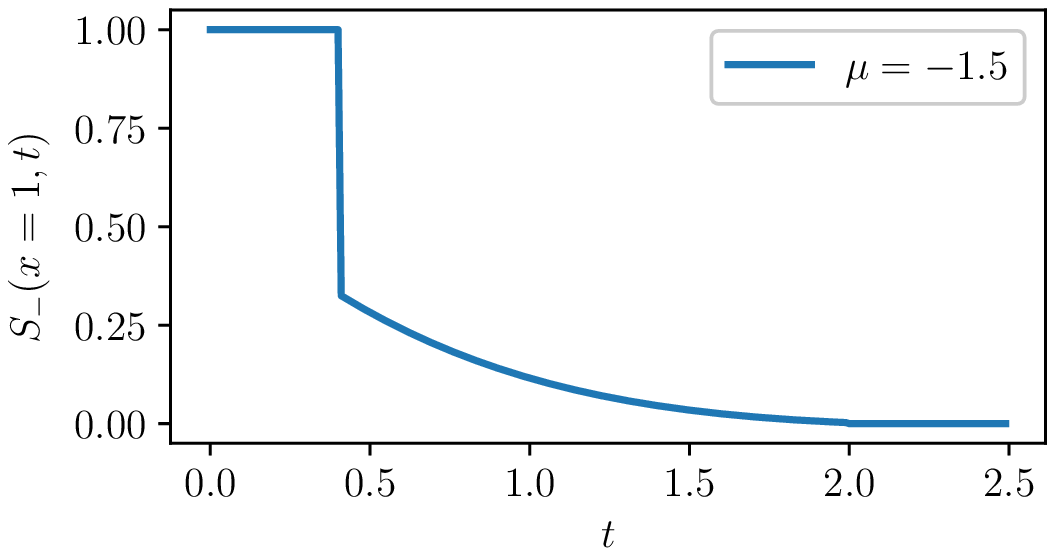}
        \caption{$\sigma_0=-1$}
        \label{fig:survivalsup1}
    \end{subfigure}%
    ~ 
    \begin{subfigure}[t]{0.5\textwidth}
        \centering
        \includegraphics[width=\textwidth]{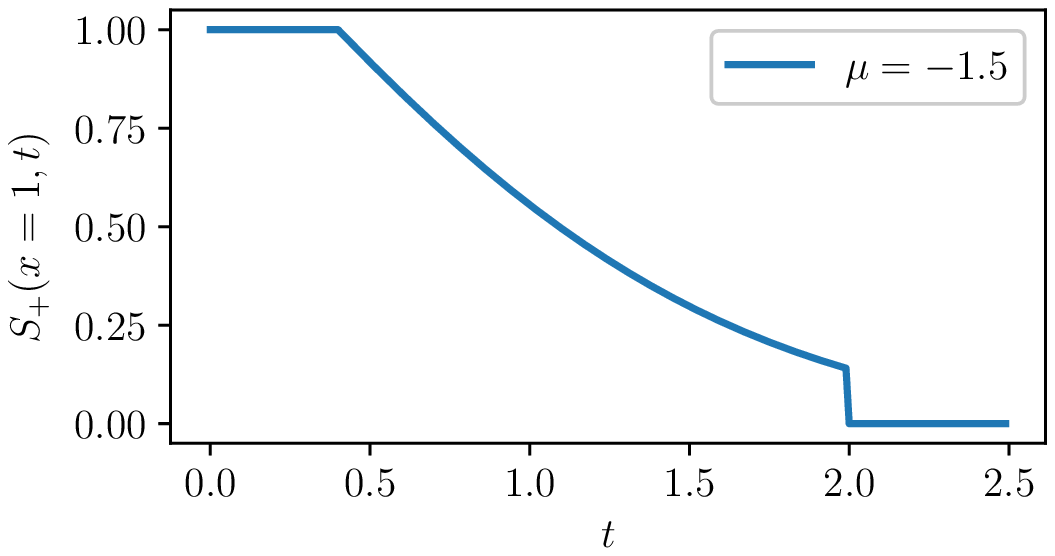}
        \caption{$\sigma_0=+1$}
        \label{fig:survivalsup2}
    \end{subfigure}
    \caption{Survival probability $S_{\sigma_0}(x\!=\!1,t)$ in the presence of a supercritical drift $\mu \!=\! -3/2\!<\!-1$ for a particle starting in the state $\sigma_0$.  The particle certainly reaches the origin after the minimum time $t_m = 1/(1-\mu)=0.4$ (\ref{eq:tm}) and before the maximum time $t_M = -1/(1+\mu) = 2$ (\ref{eq:tM}). $S_-(x\!=\!1,t)$ has a sudden drop at $t=t_m$ and $S_+(x\!=\!1,t)$ has a sudden drop at $t=t_M = 2$. These drops are due to the trajectories that go straight to the origin without tumbling: these events happen respectively with probabilities $\rme^{-t_m}$ and~$\rme^{-t_M}$.}
    \label{fig:survivalsup}
\end{figure}

\paragraph{Supercritical drift ($\mu<-1$).} In the case of a supercritical drift (see figure \ref{fig:survivals} and \ref{fig:phase_diagram}), the drift outweighs the internal velocity of the particle $v_0\!=\!1$, making it effectively always move in the direction of the drift but with two different speeds $\mu+1$ and $\mu-1$. The particle will surely not survive and reach the origin before a maximum time $t_M$. Indeed, even if the internal velocity of the particle remains opposite to the drift during the whole motion, it will eventually reach the origin by a time $t_M=-x/(1+\mu)$. The final result for the survival probability in the presence of a supercritical drift is 
 \begin{subequations}
\begin{align}
     S_+(x,t)  &= \left\{\begin{array}{ll}1\,, & t<t_m\, , \\
      1-\int_{t_m}^{t} dt'\, \rme^{-t'}\left( \frac{1-\mu}{2}\,I_0\left[h(t',x)\right] -\frac{1+\mu}{2}\sqrt{\frac{f(t',x)}{g(t',x)}}\,I_1\left[h(t',x)\right]\right)\,, & t_m\leq t<t_M\, ,\\
       0\,, & t\geq t_M\, ,\end{array}\right.\label{eq:Sstrongpi} \\[1em]
      S_-(x,t)  &= \left\{\begin{array}{ll}1\,, &\;\hspace*{2.3cm} t<t_m\, , \\
       1-\rme^{-t_m}-\int_{t_m}^{t} dt'\, \rme^{-t'}\Big( \frac{1-\mu}{2}\sqrt{\frac{g(t',x)}{f(t',x)}}\,I_1\left[h(t',x)\right] \\
       \hspace*{1.6cm}-\frac{1+\mu}{2}\,I_0\left[h(t',x)\right]\Big)\,, & \;\hspace*{2.3cm}t_m\leq t<t_M\, ,\\
      0\, ,& \;\hspace*{2.3cm}t\geq t_M\, ,\end{array} \right.
\end{align}   
\label{eq:Sstrongintro}
 \end{subequations}
 where $t_m$, $f(t,x)$, $g(t,x)$ and $h(t,x)$ are given in (\ref{eq:tmfgh}) and 
 \begin{align}
 t_M=-\frac{x}{1+\mu}\, \label{eq:tM}
 \end{align} 
 is the maximum time to reach the origin.  The survival probability is illustrated in figure~\ref{fig:survivalsup}. The first-passage time distribution $F_{\sigma_0}(x,t)=-\partial_t S_{\sigma_0}(x,t)$ has a simple expression that enables us to compute the mean first-passage time to the origin, which is always finite in this case (\ref{eq:Tavgsn}).

 \paragraph{Critical drift ($\mu\!=\!-1$). } 
  \begin{figure}[ht]
    \centering
    \begin{subfigure}[t]{0.5\textwidth}
        \centering
        \includegraphics[width=\textwidth]{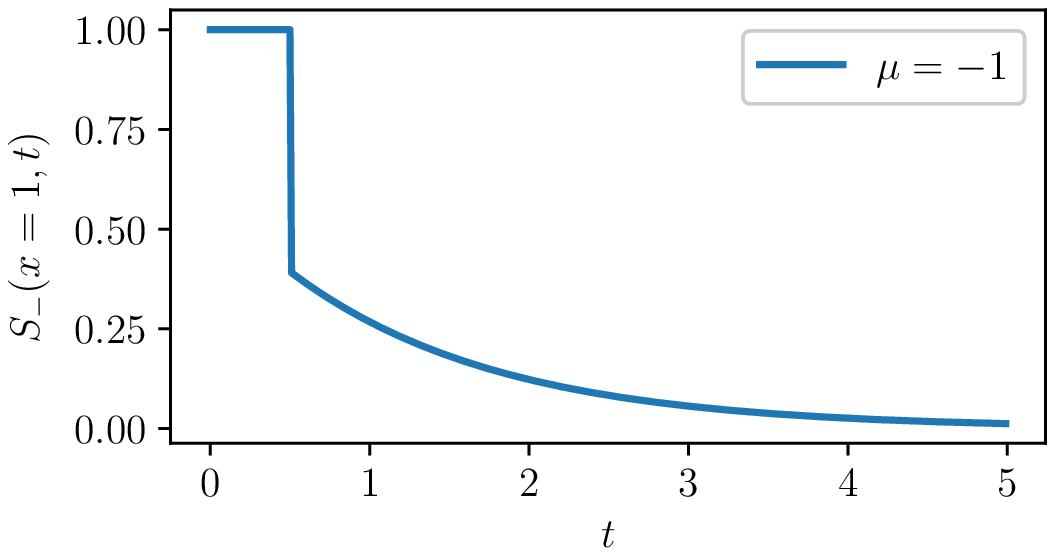}
        \caption{$\sigma_0=-1$}
        \label{fig:survivalcrit1}
    \end{subfigure}%
    ~ 
    \begin{subfigure}[t]{0.5\textwidth}
        \centering
        \includegraphics[width=\textwidth]{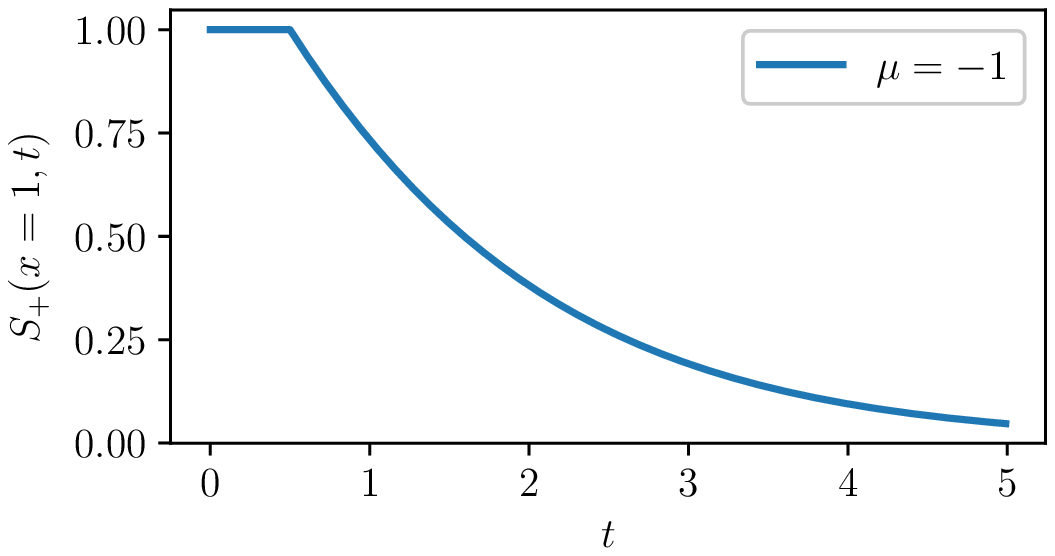}
        \caption{$\sigma_0=+1$}
        \label{fig:survivalcrit2}
    \end{subfigure}
    \caption{Survival probability $S_{\sigma_0}(x\!=\!1,t)$ in the presence of a critical drift $\mu\!=\!-1$ for a particle starting in the state $\sigma_0$.  $S_{\sigma_0}(x\!=\!1,t)$ only starts to decrease after a time $t=t_m = 1/(1-\mu)=0.5$ (\ref{eq:tm}) which is the minimum time it takes for the particle to reach the origin. $S_-(x\!=\!1,t)$ has a sudden drop at $t=t_m=0.5$ which is caused by the 
trajectory that goes straight to the origin without tumbling: this event happens with probability $\rme^{-t_m}$. This trajectory is not present for $S_+(x,t)$. In the long-time limit, $S_+(x\!=\!1,t)$ and $S_-(x\!=\!1,t)$ decay anomalously to $0$.}
    \label{fig:survivalcrit}
\end{figure}
In this case, for the survival probabilities $S_{\pm}(x,t)$, we obtain the exact results
\begin{subequations}
\begin{align}
    S_+(x,t)&=\left\{\begin{array}{ll} 1\,, & t < x/2\,, \\
     1- \int_{x/2}^{t} dt' \, \rme^{-t'} \,I_0\left[\sqrt{x\,(2t'-x)}\right] \,,& t\geq x/2,\end{array}\right. \\[1em]
     S_-(x,t)&= \left\{\begin{array}{ll} 1\,, & t < x/2\,, \\
     1-\rme^{-x/2} - \int_{x/2}^{t} dt' \,\rme^{-t'}\, \frac{\sqrt{x}}{\sqrt{2t'-x}} \, I_1\left[\sqrt{x\,(2t'-x)}\right]\,,& t\geq x/2 \,.
     \end{array}\right. 
\end{align}
\end{subequations}
These functions are plotted in figure \ref{fig:survivalcrit} and their asymptotic behaviors are discussed in detail in section \ref{sec:survc}. Finally, in the same section, 
we also discuss the cross-over between the critical and off-critical behaviors as time progresses by setting $\mu$ close to the critical value $\mu = -1$.

The rest of the paper is organized as follows. In section \ref{sec:SA}, we outline a derivation of the survival probability of a particle with an arbitrary velocity distribution and apply it on the PRW model in the presence of a drift. In section \ref{sec:subcr} we study the survival probability for the case of a subcritical drift $-1\!<\!\mu\!<\!1$ and discuss its long-time limit along with the mean first-passage time to the origin. In section, \ref{sec:sSs}, we study the survival probability for the case of a supercritical drift $\mu\!<\!-1$. We discuss the mean first-passage time to the origin as well as an alternative derivation for the survival probability based on the propagator of the particle. In section \ref{sec:survc}, we study the survival probability in the presence of a critical drift $\mu\!=\!-1$ and derive some scaling functions close to criticality. In section \ref{sec:other}, we show how our method can be applied to another asymmetric velocity distribution. In section \ref{sec:appli}, we discuss applications of our results to emptying times and record statistics. Finally, we conclude in section \ref{sec:summary}. Some technical calculations are relegated in~\ref{app:Poisson}~to~I. 

\section{Survival probability for a particle with an arbitrary velocity distribution}
\label{sec:SA}

We start with a generalized RTP model, with an arbitrary velocity distribution $W(v)$ and tumbling rate $\gamma$. Following \cite{mori2020universal,moriE2020universal}, we will first map the RTP motion to a discrete time random walk on the line
with a jump distribution that depends on the velocity distribution $W(v)$. Under this mapping, the survival probability (in the Laplace space
with respect to time $t$) of the RTP gets related to the survival probability of a one-dimensional discrete-time random walk with a 
specified jump distribution. The latter can then be computed by adapting Spitzer's formula \cite{spitzer1957wiener}, valid for one-dimensional random
walk with arbitrary jump distribution. This will lead to the result mentioned in (\ref{intro_S}). We then use this general formula to derive explicitly
$S(x,t)$ for a specific $W(v)$ corresponding to the two-state RTP mentioned earlier (\ref{eq:Wintro}).

\subsection{Mapping of the run-and-tumble process to an effective discrete-time random walk} \label{sec:mapping}

\begin{figure}[ht]
\centering
\includegraphics[width = 0.5\linewidth]{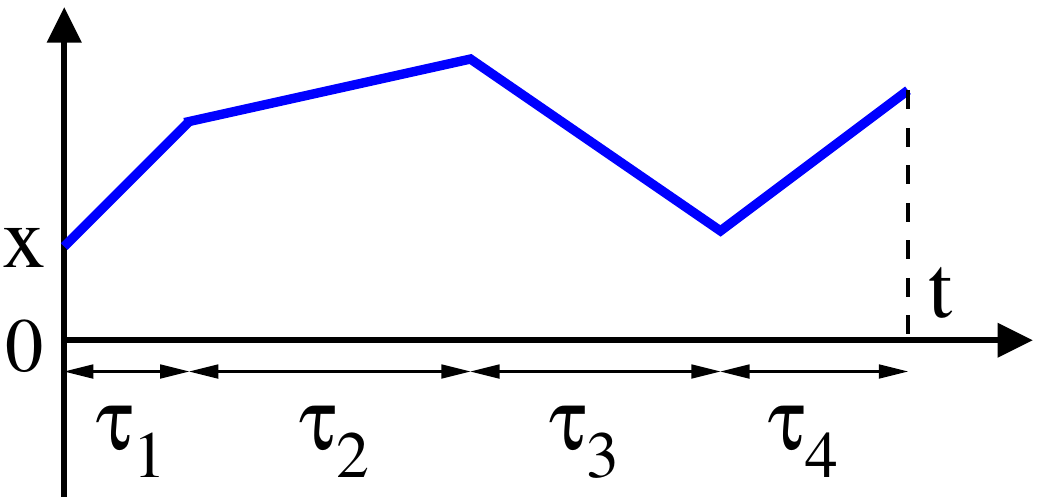}
\caption{Trajectory of a generalized run-and-tumble model starting at $x \geq 0$ with $n=4$ tumblings (by convention, the first tumbling happens at $t=0$). The time intervals 
 $\tau_1, \tau_2$ and $\tau_3$ are called the run-times -- and they are identically distributed exponential random variables of parameter $\gamma$. The last run $\tau_4$ is not finished and thus behaves differently from the first run times (see the discussion in the text).}\label{fig_trajectory}
\centering
\end{figure}

In this model, since the total time is fixed, the actual number of tumblings $n$ in the trajectory of
the process is a random variable and fluctuates from one trajectory to another. Consider a particular
trajectory of total duration $t$ with $n$ runs, $n \geq 1$. Of these runs, the first $n-1$ are complete while
the last one is not since the process has been stopped exactly at the observation time $t$. For each of the first
$n-1$ runs, the run time $\tau_i$ is chosen independently from the distribution $p(\tau_i) = \gamma\,\e^{-\gamma \tau_i}$, $i=1,2, \ldots, (n-1)$.
In contrast, the probability weight associated to the last run $\tau_n$ is  $\int_{\tau_n}^\infty d\tau\,p(\tau) = \e^{-\gamma\,\tau_n}$ (see figure \ref{fig_trajectory}). 
During each run, the velocity $v_i$ is chosen independently from $W(v)$. Consequently, the run lengths $\ell_i = v_i \tau_i$ are
also random variables. For the brevity of notations, we will denote the collection of $n$ run lengths together by the vector 
$\vec{\ell}=\{\ell_1, \ell_2, \cdots, \ell_n \}$. We first note that the survival probability $S(x,t)$ of the continuous time process $x(\tau)$ starting at $x$ is
identical to the probability of the event that the positions at the end of each tumbling up to time $t$ is nonnegative. This is easy to see because if two consecutive tumbling positions are both positive, the particle could not have gone to the negative side in between. Similarly, if the position of the RTP during a run between two successive tumblings stays positive, then obviously the two end-positions of the run are also nonnegative. Hence we just have to compute the probability of the event that these tumbling positions $\{x+\ell_1, x+\ell_1+ \ell_2, \ldots, x+\ell_1+\ell_2+\ldots + \ell_n \}$ are simultaneously nonnegative. Hence,    
\begin{equation} \label{Survival_theta}
S(x,t) = \sum_{n} \int d\vec{\ell}\,P(\vec{\ell}, n |t)\, \theta(x+\ell_1) \theta(x+\ell_1+\ell_2) \cdots \theta(x+\ell_1 + \ell_2 + \cdots + \ell_n)  \;,
\end{equation}
where $\theta(z)$ is the Heaviside theta function and $P(\vec{\ell}, n |t)$ denotes the joint distribution of run lengths 
$\vec{\ell}$ and the number of tumblings $n$, given a fixed time $t$. The product of
the theta functions ensures that the trajectory stays nonnegative up to time $t$. This joint distribution $P(\vec {\ell}, n |t)$ can be written down explicitly 
as \cite{mori2020universal,moriE2020universal}
\begin{eqnarray} \label{eq:jPDF}
P(\vec {\ell}, n |t) &=& \left[\prod_{i=1}^{n-1} \int_0^\infty d \tau_i\, \int_{-\infty}^\infty d v_i \, W(v_i) \gamma \, \e^{-\gamma \tau_i} \, \delta(\ell_i - v_i \tau_i)  \right] \times \nonumber \\
&\times& \left[\int_0^\infty d \tau_n\, \int_{-\infty}^\infty d v_n  \, W(v_n)\, \e^{-\gamma \tau_n}\right] \delta\left(t-\sum_{i=1}^n\tau_i \right) \;.
\end{eqnarray}
This formula is easy to understand. The product in the first line corresponds to the probability weight of the first $(n-1)$ complete
and independent runs. On the second line, the first factor corresponds to the probability weight of the last (incomplete) run. Finally,
the global delta function ensures that the total time $t$ is fixed. Note the inequivalence between the probability weight of any of 
the $(n-1)$ complete runs and the one of the last run since they differ by a factor $\gamma$. In order to make them equivalent, it is 
convenient to divide by a global factor $\gamma$ and rewrite (\ref{eq:jPDF}) as
\begin{equation}\label{eq:jPDF2}
P(\vec {\ell}, n |t) =\frac{1}{\gamma}\, \left[\prod_{i=1}^{n} \int_0^\infty d \tau_i\, \int_{-\infty}^\infty d v_i  \, W(v_i) \gamma \, \e^{-\gamma \tau_i} \delta(\ell_i - v_i \tau_i)  \right] \delta\left(t-\sum_{i=1}^n\tau_i \right) 
\end{equation}
The presence of the global delta function in (\ref{eq:jPDF2}) suggests that the joint distribution factorizes in the Laplace space with respect to time $t$. Thus 
 taking the Laplace transform of (\ref{eq:jPDF2}) we get
 \begin{eqnarray}
\hspace*{-2.8cm}\int_0^\infty P(\vec {\ell}, n |t) e^{-s\,t}\,dt &=& \frac{1}{\gamma} \left( \frac{\gamma}{\gamma+s}\right)^n\prod_{i=1}^n \int_0^\infty d \tau_i\, \int_{-\infty}^\infty d v_i  \, (\gamma+s)\,\e^{-(\gamma+s)\tau_i}\, W(v_i) \delta(\ell_i - v_i\tau_i)  \label{eq:jPDF3} \nonumber \\
\hspace*{-2.8cm}&=& \frac{1}{\gamma} \left( \frac{\gamma}{\gamma+s}\right)^n\prod_{i=1}^n \phi_s(\ell_i) \;, \label{eq:jPDF4}
 \end{eqnarray}
where in the first line we have multiplied and divided by a factor $(\gamma+s)$ and defined 
\begin{eqnarray}\label{eq:stepdist0}
\phi_s(\ell) = \int_0^\infty d\tau \, \int_{-\infty}^\infty d v\, (\gamma+s)\,\e^{-(\gamma+s)\tau}\, W(v) \delta(\ell - v\,\tau)  \;.
\end{eqnarray}
The reason behind this manipulation is that one sees immediately that $\phi_s(\ell)$ in (\ref{eq:stepdist0}) is nonnegative for all $\ell$ and normalized to unity, i.e. $\int_{-\infty}^\infty d\ell \, \phi_s(\ell) = 1$ using $\int_{-\infty}^\infty dv W(v)=1$ as well as $\int_0^\infty (\gamma + s)\e^{-(\gamma+s)\tau}\, d\tau =1$. Hence one can interpret $\phi_s(\ell)$ as a PDF of the random variable $\ell$ which is parametrized by $s$. Indeed the double integral in $\phi_s(\ell)$ can be reduced to a single integral by performing the integral over $v$. This gives
\begin{align}
    \phi_s(\ell)&=(\gamma+s)\int_0^\infty \frac{d\tau}{\tau} \,\rme^{-(\gamma+s)\,\tau}\,W\left(\frac{\ell}{\tau}\right)\, .
\label{eq:stepdist}
\end{align}

Finally taking the Laplace transform (\ref{Survival_theta})
with respect to $t$, defining $\tilde S(x,s) = \int_0^\infty S(x,t) \, \e^{-s t}\, dt $ and using (\ref{eq:jPDF4}) one obtains
\begin{equation}
\tilde S(x,s) = \frac{1}{\gamma}
\sum_{n=1}^\infty\left(\frac{\gamma}{\gamma+s}\right)^n\int d\vec{\ell} \, \theta(x+\ell_1) \theta(x+\ell_1+\ell_2) \cdots \theta(x+\ell_1+\ell_2+\cdots + \ell_n)  \,\prod_{i=1}^n\, \phi_s(\ell_i)\, .\label{eq:multint} 
\end{equation}
We now recognize the integral 
\begin{equation} \label{def_qn}
q_n(x) = \int d\vec{\ell} \, \theta(x+\ell_1) \theta(x+\ell_1+\ell_2) \cdots \theta(x+\ell_1+\ell_2+\cdots + \ell_n)  \,\prod_{i=1}^n\, \phi_s(\ell_i)
\end{equation}
as the survival probability up to step $n$ of a discrete time random walk starting at $x$ and with jumps $\{\ell_1, \ell_2, \cdots, \ell_n\}$, each drawn 
independently at each step from the PDF $\phi_s(\ell)$. Therefore the expression in (\ref{eq:multint}) reduces to
\begin{align}
  \tilde  S(x,s) = \frac{1}{\gamma}\sum_{n=1}^\infty \left(\frac{\gamma}{\gamma+s}\right)^n\,q_n(x)\, . \label{eq:sumSs}
\end{align}
Defining the 
generating function of the survival probability of the effective random walk
\begin{align}
\bar q(x,r) =  \sum_{n=0}^{\infty}\, r^n \,q_n(x) \,,\label{eq:genfun}
\end{align}
we can then express the Laplace transform of the survival probability of the original RTP problem in terms of the generating function of the
$1d$ random walk problem with jump distribution $\phi_s(\ell)$
\begin{align}
 \tilde  S(x,s) = \frac{1}{\gamma}\bar q\left(x,\,\frac{\gamma}{\gamma+s}\right) -\frac{1}{\gamma}\, ,
    \label{eq:Sqn}
\end{align}
 where the term $-1/\gamma$ appears because the sum in (\ref{eq:sumSs}) only starts at $n=1$ and we used $q_0(x) = 1$. This relation (\ref{eq:Sqn})
 is very general and holds for arbitrary velocity distribution $W(v)$, including asymmetric ones. The dependence on $W(v)$ enters only through the
 jump distribution $\phi_s(\ell)$ of the random walk problem [as in (\ref{eq:stepdist})].

 \subsection{Survival probability of the effective discrete-time random walk using Spitzer's formula}
 Spitzer established a general formula to obtain the distribution of the minimum of a discrete-time random walk after $n$ steps \cite{spitzer1957wiener}. As the survival probability of the discrete-time random walk $q_n(x)$ after $n$ steps starting from $x$  can be seen as the probability that its minimum remains above the origin during $n$ steps, his formula can be used to obtain $q_n(x)$. The formula states that the Laplace transform with respect to $x$ of the generating function $\bar q(x,r)$ in (\ref{eq:genfun}) is given by (see \ref{app:spitzer} for details)
 \begin{subequations}
   \begin{align}
  \int_0^\infty dx \,\bar q(x,r)\,\rme^{-u\,x}  = \frac{1}{u(1-r)}\exp(\frac{\Phi_s(0,r)-\Phi_s(u,r)}{2\pi})
   \label{eq:spitzer}\, , 
\end{align}
where the function $\Phi_s(u,r)$ reads
\begin{align}
  \Phi_s(u,r) = \int_{-\infty}^\infty \frac{dk}{u+\rmi k} \ln(1-r\,\hat\phi_s(k))\,, \label{eq:Xi}
\end{align}
and $\hat\phi_s(k)$ is the Fourier transform of the jump distribution $\phi_s(\ell)$
\begin{align}
  \hat\phi_s(k) &= \int_{-\infty}^\infty d\ell\, \phi_s(\ell) \,\rme^{-\rmi \,k\,\ell}\, . \label{FourierPhi_s}
  \end{align}
 \end{subequations}
 Inserting the expression of the jump distribution $\phi_s(\ell)$ (\ref{eq:stepdist}), we find that its Fourier transform is given by
  \begin{align}
 \hat\phi_s(k) &=\int_{-\infty}^\infty  d\ell\, \rme^{-\rmi \,k\,\ell} (\gamma+s)\int_0^\infty \frac{d\tau}{\tau} \,\rme^{-(\gamma+s)\,\tau}\,W\left(\frac{\ell}{\tau}\right)\,,\\
 & \\
  &= (\gamma+s)\int_{-\infty}^\infty   dv\, \frac{W(v)}{(\gamma+s+\rmi\,k\,v)}\,.\label{eq:fourierTFO}
\end{align}
We can slightly simplify the integral in the expression of $\Phi_s(u,r)$ (\ref{eq:Xi}) upon integrating it by parts. Evaluating the difference $ \Phi_s(u,r)-\Phi_s(0,r)$ (the boundary terms do vanish) we get
\begin{align}
  \Phi_s(u,r)-\Phi_s(0,r) &= -\rmi\, \int_{-\infty}^\infty   dk \ln(\frac{\rmi k+u}{\rmi k}) \frac{ \partial_k\hat\phi_s(k)}{\frac{1}{r}-\hat\phi_s(k)}\,. \label{eq:xid}
\end{align}
Inserting the Fourier transform $\hat\phi_s(k)$ (\ref{eq:fourierTFO}) in this function (\ref{eq:xid}) and changing variable $z=\rmi\, k$, we find
\begin{align}
  \Phi_s(u,r) - \Phi_s(0,r)  &= \rmi \int_{\rmi\, \mathbb{R}} dz \ln(\frac{z+u}{z}) \frac{\int_{-\infty}^\infty   dv\, \frac{v\,W(v)}{(\gamma+s+z\,v)^2}}{\frac{1}{r(\gamma+s)}-\int_{-\infty}^\infty   dv\, \frac{W(v)}{(\gamma+s+z\,v)}}\,,\label{eq:xiua}
\end{align}
where the integration domain $\rmi\, \mathbb{R}$ is now the imaginary axis.
Finally plugging this function (\ref{eq:xiua}) in Spitzer's formula (\ref{eq:spitzer}), we find
   \begin{align}
  \int_0^\infty dx \,\bar q(x,r)\,\rme^{-u\,x}  = \frac{1}{u(1-r)}\exp(-\frac{\rmi}{2\pi}\int_{\rmi\, \mathbb{R}} dz \ln(\frac{z+u}{z}) \frac{\int_{-\infty}^\infty   dv\, \frac{v\,W(v)}{(\gamma+s+z\,v)^2}}{\frac{1}{r(\gamma+s)}-\int_{-\infty}^\infty   dv\, \frac{W(v)}{(\gamma+s+z\,v)}})
   \label{eq:spitzerf}\, .
\end{align}

\subsection{Survival probability of the run-and-tumble particle using the mapping}
Using Spitzer's formula (\ref{eq:spitzerf}) and the relation between the survival probability of the run-and-tumble process and the discrete-time random walk (\ref{eq:Sqn}), we obtain

\begin{eqnarray} 
&&\int_0^\infty dx \, \int_0^\infty dt \, S(x,t) \e^{-st - x u } = \nonumber \\
&=&  \frac{\gamma+s}{\gamma \, u\,s}\exp(-\frac{\rmi}{2\pi}\int_{\rmi\, \mathbb{R}} dz \ln(\frac{z+u}{z}) \frac{\int_{-\infty}^\infty   dv\, \frac{v\,W(v)}{(\gamma+s+z\,v)^2}}{\frac{1}{\gamma}-\int_{-\infty}^\infty   dv\, \frac{W(v)}{(\gamma+s+z\,v)}})-\frac{1}{\gamma u}\,, \label{eq:gformula}
\end{eqnarray}
where $\rmi \mathbb{R}$ denotes the imaginary axis in the complex $z$-plane. 

The formula (\ref{eq:gformula}) is very general and can be used to compute the survival probability of a particle with an arbitrary velocity distribution $W(v)$. In particular, when $W(v)$ has a positive bias, the particle will have a finite non-zero survival probability which can be easily computed from (\ref{eq:gformula}) by extracting the $1/s$ factor in its small-$s$ limit and yields
\begin{align}
  \int_0^\infty dx\, \lim_{t\rightarrow\infty}   S(x,t)\,\rme^{-u\,x} = \frac{1}{u}\exp(-\frac{\rmi}{2\pi}\int_{\rmi\, \mathbb{R}} dz \ln(\frac{z+u}{z}) \frac{\int_{-\infty}^\infty   dv\, \frac{v\,W(v)}{(\gamma+z\,v)^2}}{\frac{1}{\gamma}-\int_{-\infty}^\infty   dv\, \frac{W(v)}{(\gamma+z\,v)}})-\frac{1}{u}\,,
\end{align}
which in principle can be analyzed for various velocity distributions with a positive bias. 

It is also interesting to compute the survival probability $S_{\uparrow}(x,t)$ (respectively $S_{\downarrow}(x,t)$) for an RTP with an arbitrary velocity distribution $W(v)$, starting from $x \geq 0$ with a {\it positive} (respectively negative) initial velocity. One expects indeed that they behave rather differently: for instance, $S_{\uparrow}(0,t) > 0$ while $S_{\downarrow}(0,t) = 0$ for all time $t>0$ since the particle starting at $x=0$ with a {\it negative} velocity gets immediately absorbed. As we show in \ref{app:spitzerpm}, 
explicit expressions for the double Laplace transforms of $S_{\uparrow}(x,t)$ and $S_{\downarrow}(x,t)$ can be obtained for arbitrary velocity distributions $W(v)$. They are given respectively in (\ref{doubleLT_Sdown}) and (\ref{doubleLT_Sup}).

\subsection{Survival probability of a two-state persistent random walk}

Let us now apply this general result (\ref{eq:gformula}) to the case describing the two-state PRW, corresponding to the velocity distribution given in (\ref{eq:Wintro}). In this case, one finds that the formula in (\ref{eq:gformula}) reads
\begin{align}
 \hspace*{-0.5cm} \int_0^\infty dx\, \int_0^\infty dt \,  S(x,t)\,\rme^{-u\,x - s\, t} = \frac{\gamma+s}{\gamma u\,s}\exp(-\frac{\rmi}{2\pi}\int_{\rmi\, \mathbb{R}} dz \ln(\frac{z+u}{z}) \sum_{j=1}^4 \frac{(-1)^j}{z-z_j})-\frac{1}{\gamma \, u}\,. \label{eq:gformm}
\end{align}
 with
 \begin{subequations}
 \begin{align}
 z_1 &= \frac{ (\gamma+s)}{v_0-\mu}\,,\\
  z_2  &=  \frac{\mu 
    (\gamma+2s)-\sqrt{4\,s^2 v_0^2+4 \, \gamma \, s v_0^2+\mu^2 \gamma^2}}{2 \left(v_0^2-\mu
   ^2\right)}\,, \\
    z_3&= -\frac{ (\gamma+s)}{\mu +v_0}\,,\\
  z_4 &=  \frac{\mu 
    (\gamma+2s)+\sqrt{4\,s^2 v_0^2+4 \, \gamma \, s v_0^2+\mu^2 \gamma^2}}{2 \left(v_0^2-\mu
   ^2\right)}\, .
 \end{align}
 \label{eq:poles}
 \end{subequations}
 The integral in the general formula for the survival probability (\ref{eq:gformm}) can now be easily done by using the residue theorem and closing the integration contour from the right. We find
\begin{eqnarray}
\hspace*{-2cm}\int_0^\infty dx \int_0^\infty dt \,  \tilde S(x,s) \, \e^{-u\,x - s \, t}=\frac{\gamma+s}{\gamma \, u\,s}\exp\left(\sum_{j=1}^4 \theta(z_j) \, (-1)^{j+1} \ln{\frac{z_j+u}{z_j}} \right) - \frac{1}{\gamma \, u} \,. \label{eq:gform2}
\end{eqnarray}
The result (\ref{eq:gform2}) is quite different depending on the strength of the drift as it will affect the signs of the poles $z_j$ (\ref{eq:poles}). This leads us to treat separately 
three different cases: (i) subcritical drift $-v_0\!<\!\mu\!<\!v_0$, (ii) supercritical drift $\mu\!<\!-v_0$  and (iii) critical drift $\mu\!=\!-v_0$. 
 \section{Survival probability in the presence of a subcritical drift $-v_0<\mu<v_0$}
 \label{sec:subcr}
In this section we consider a subcritical drift (see figure \ref{fig:survivalw} and \ref{fig:phase_diagram}). In this case, we note that the poles $z_1$ and $z_4$ are positive while $z_2$ and $z_3$ are not (\ref{eq:poles}). The survival probability (\ref{eq:gform2}) simplifies to
  \begin{align}
   \int_0^\infty dx \int_0^\infty \, dt  \, S(x,s)\,\rme^{-u\,x - s \, t}  = \frac{\gamma+s}{\gamma\, u\,s}\,\,\frac{u+ z_1}{u+ z_4}\,\,\frac{ z_4}{ z_1}-\frac{1}{\gamma \, u}\,.
 \end{align}
 The Laplace transform with respect to $x$ is easily inverted, yielding the result
    \begin{align}
 \tilde S(x,s) = \int_0^\infty dt\, S(x,t)\, \e^{-s\,t}= \frac{\gamma+s}{\gamma \, s}\,\,\frac{z_1+(z_4-z_1)\,\rme^{-z_4\, x}}{z_1}-\frac{1}{\gamma}\,.
   \end{align}
Upon injecting the expressions for $z_1$ and $z_4$ (\ref{eq:poles}) which we write in terms of $\tilde \gamma = \gamma/2$, we find
   \begin{align}
 \tilde  S(x,s) = \frac{1}{2\,s}\left(2-\tilde \gamma \frac{\rme^{-x \,(\eta +\lambda )}}{(\mu +v_0) (\lambda+v_0\,\eta/\mu)}-\rme^{-x\, (\eta +\lambda )}\right)\, ,\label{eq:lapsumw}
\end{align}
where
\begin{subequations}
\begin{align}
\lambda &= \frac{\sqrt{s\,v_0^2\, (2 \tilde\gamma+s)+\mu^2\tilde \gamma^2}}{v_0^2-\mu^2}\,\label{eq:lambda} ,\\
\eta &= \frac{\mu\, (\tilde \gamma+s)}{v_0^2-\mu^2}\,.\label{eq:eta}
\end{align}
  \label{eq:lambdaeta}
\end{subequations}
 For clarity purposes, we denote $\lambda$ and $\eta$ as single variables but it is important to keep in mind that they depend explicitly on the Laplace variable $s$ and the drift $\mu$. For later purposes, we also provide an explicit expression for the generating function $\bar q(x,r)$ defined in (\ref{eq:genfun}). Using the relation (\ref{eq:Sqn}) and the result in (\ref{eq:lapsumw}) we get, e.g. for $x=0$
 \begin{align}\label{qbar0}
 \bar q\left(x=0, \frac{\gamma}{\gamma+s}\right) = \frac{\mu(\gamma+2s)+\sqrt{4s(\gamma+s)v_0^2 + \gamma^2 \mu^2}}{2s(\mu+v_0)} \;.
 \end{align}

For this two-step PRW, it is also interesting to study the survival probabilities $S_{\pm}(x,t)$ conditioned on the initial velocity $\dot x(0) = \mu \pm v_0$, defined respectively in (\ref{eq:def_Sp}) and (\ref{eq:def_Sm}). Since in this subcritical case one has $\mu + v_0 >0$ and $\mu - v_0 <0$, $S_+(x,t)$ coincides with the survival probability $S_\uparrow(x,t)$ (which is the survival probability for a particle starting with a positive velocity, see (\ref{eq:def_Sup})) while $S_-(x,t)$ coincides with the survival probability $S_\downarrow(x,t)$ (which is the survival probability for a particle starting with a negative velocity, see (\ref{eq:def_Sdown})). Specifying our general results (\ref{doubleLT_Sdown}) and (\ref{doubleLT_Sup}) to the two-state PRW corresponding to the velocity distribution in (\ref{eq:Wintro}), we find
\begin{subequations}
\begin{align}
    \tilde S_+(x,s)&= \int_0^\infty dt \, \e^{-s\,t} S_+(x,t) = \int_0^\infty dt \, \e^{-s\,t} S_\uparrow (x,t) = 
     \frac{1}{s}\left(1-\tilde \gamma \frac{\rme^{-x \,(\eta +\lambda )}}{(\mu +v_0) (\lambda+v_0\,\eta/\mu)}\right)\, ,\label{eq:sps}\\
   \tilde S_-(x,s)&=  \int_0^\infty dt \, \e^{-s\,t} S_-(x,t) = \int_0^\infty dt \, \e^{-s\,t} S_\downarrow (x,t) = \frac{1}{s}\left(1-\rme^{-x\, (\eta +\lambda )}\right)\, ,\label{eq:spm}
\end{align}
\label{eq:Spm}
\end{subequations}
where $\lambda$ and $\eta$ are given in (\ref{eq:lambdaeta}) and we stress again that they depend explicitly on the Laplace variable $s$. These results coincide with the one obtained in \cite{lopez2014asymmetric} by a completely different method. By inverting these Laplace transforms (see \ref{app:lapinv}), we find the survival probabilities presented in the introduction in (\ref{eq:Spsub}) and (\ref{eq:Spmt}).

In the remainder of this section, we will set $v_0 = \tilde \gamma =1$ for simplicity of notations. 
 
\subsection{Long-time limit of the survival probability}
\label{sec:lgtlimsurvival}
We now discuss the long-time limit of the survival probability (\ref{eq:Spmt}) separately for the case of a positive subcritical drift $0\!<\!\mu\!<\!1$ and the case of a negative subcritical drift $-1\!<\!\mu \!<\!0$.
\paragraph{Positive subcritical drift ($\,0\!<\!\mu\!<\!1$).}
 In the case of a positive subcritical drift, the particle might eventually survive due to the fact that the drift pushes it away from the origin. The large time behavior of $S_{\pm}(x,t)$ can be conveniently obtained
 by analyzing the small $s$ behavior of the Laplace transforms $\tilde S_\pm(x,s)$ in (\ref{eq:sps}) and (\ref{eq:spm}). Using that $\eta \to \mu/(1-\mu^2)$ as well as $\lambda \to |\mu|/(1-\mu^2)$ as $s \to 0$ [from (\ref{eq:eta}) and (\ref{eq:lambda}], setting $v_0 = \tilde\gamma =1$), the expressions in (\ref{eq:sps}) and (\ref{eq:spm}) yield straightforwardly
\begin{subequations}
\begin{align}
   \lim_{t\rightarrow \infty} S_+(x,t)=&\, 1 - 
    \frac{1-\mu}{1+\mu}\,\rme^{-x\frac{2\,\mu}{1-\mu^2}}\, ,\\
    \lim_{t\rightarrow \infty} S_-(x,t)=&\, 1 -\rme^{-x\frac{2\,\mu}{1-\mu^2}}\, .
\end{align}
 \label{eq:longtpos}
\end{subequations}
\paragraph{Negative subcritical drift ($-1\!<\!\mu \!<\! 0$).} In the case of negative subcritical drift, the particle will surely die and $S_{\pm}(x,t)$ both decay to $0$ at large time. To obtain the large time behavior of $S_\pm(x,t)$, it is convenient to analyze its derivative with respect to $t$, i.e., the first-passage time density $F_{\pm}(x,t)  = - \partial_t S_{\pm}(x,t)$, which has a simpler expression [see (\ref{eq:Spsub}) and (\ref{eq:Spmt})], and integrate it back to obtain $S_{\pm}(x,t)$. Using the asymptotic expansion of the Bessel function $I_{0,1}(x)\sim \rme^{x}\,/\sqrt{2\pi x}$ for $x\rightarrow \infty$ we find
\begin{subequations}
\begin{align}
S_+(x,t) &\sim \frac{x+\sqrt{1-\mu^2}}{\left[(1-\mu^2)^{1/4}-(1-\mu^2)^{3/4}\right]}\frac{1}{1+\mu}\frac{1}{\sqrt{2\pi}}\,\rme^{-\frac{x\,\mu}{\sqrt{1-\mu^2}}}\,\frac{\rme^{-t\left(1-\sqrt{1-\mu^2}\right)}}{t^{3/2}}\,, \label{eq:longtposs}\\
    S_-(x,t) &\sim \frac{x}{\left[(1-\mu^2)^{3/4}-(1-\mu^2)^{5/4}\right]} \frac{1}{\sqrt{2\pi}}\,\rme^{-\frac{x\,\mu}{\sqrt{1-\mu^2}}}\frac{\rme^{-t\left(1-\sqrt{1-\mu^2}\right)}}{t^{3/2}}\, .
\end{align}\label{eq:longtneg}
\end{subequations}
Note that, contrary to $S_-(x,t)$, the survival probability $S_+(x,t)$ does not vanish when the particle starts right at the origin $x\!=\!0$. This is a signature of the ``activity'' of the PRW and it is reminiscent of the fact that an initial positive velocity gives the particle a chance to survive. As pointed out in \cite{le2019noncrossing}, one needs to extrapolate $x$ to $x=-\xi_\text{Milne}\equiv -\sqrt{1-\mu^2}$ for the survival probability to vanish. The notation $\xi_\text{Milne}$ is borrowed from neutron scattering and we refer to \cite{le2019noncrossing} for further explanations. Another interesting fact is that the ratio of $S_+(x,t)$ and $S_-(x,t)$ in (\ref{eq:longtneg}) tends, when $t \to \infty$, to a non-trivial function of $x$ and $\mu$ which shows that the influence of the initial condition persists up to arbitrary large times. Finally, upon reintroducing the units (\ref{eq:units}) and taking the Brownian limit (\ref{eq:brownianlimit}) in the survival probability (\ref{eq:longtpos}) and (\ref{eq:longtneg}), we recover the well-known Brownian results \cite{majumdar2002exact,redner2001guide}
\begin{subequations}
\begin{align}
S_{\pm}(x,t)&\sim 1-\rme^{-\frac{\mu x}{D}}\,, & 0<\mu<1\, ,\\
S_{\pm}(x,t)&\sim \sqrt{\frac{4Dt}{\pi}}\, \frac{x}{(\mu t)^2}\,\rme^{-\frac{\mu\,x}{2D}} \rme^{-\frac{\mu^2 t}{4D}}\,, &-1<\mu<0 \;,
\end{align}
\end{subequations}
as expected.

\subsection{Mean first-passage time to the origin}
\label{sec:mfpt}
A key information to extract from the survival probability is the mean first-passage time to reach the origin, which is infinite in the case of zero drift. A convenient way to compute this integral is to consider the Laplace transform of the first-passage time distribution $F_{\pm}(x,t) = - \partial_t S_\pm(x,t)$ and expand it for small $s$. We get
\begin{align}
  \tilde F_{\pm}(x,s) &= \int_0^\infty\, dt\, F_{\pm}(x,t)\, \rme^{-s\, t}\, , \\
  &\sim \int_0^\infty\, dt\, F_{\pm}(x,t)\, + s\, \,\langle T(x) \rangle_{\pm} \,,\\
  &\sim \left(1 - \lim_{t\rightarrow\infty}S_{\pm}(x,t) \right) + s\, \,\langle T(x) \rangle_{\pm} \,. \label{eq:Fseries}
\end{align}
We observe that the constant term in the expansion (\ref{eq:Fseries}) is the probability that the particle eventually reaches the origin and the coefficient of the linear term gives the mean first-passage time directly. On the other hand, we can compute the exact expression of the Laplace transform $\tilde F_{\pm}(x,s)$ based on the one of the survival probability $\tilde S_{\pm}(x,s)$. In Laplace domain, the relationship between $F_{\pm}(x,t)$ and $S_{\pm}(x,t)$ becomes, using integration by parts, 
\begin{align}
\tilde F_{\pm}(x,s)&= -\int_0^\infty \,dt\, \partial_t S_{\pm}(x,t) \rme^{-s\, t}\, , \\
&=  1- s\,\tilde S_{\pm}(x,s)\,, \label{eq:FSs}
\end{align}
where we used that $S_{\pm}(x,t\!=\!0)\!=\!1$. Making use of the relation between the first-passage time and the survival probability (\ref{eq:FSs}) in the Laplace transforms of the survival probabilities (\ref{eq:Spm}), we find
\begin{subequations}
\begin{align}
\tilde F_+(x,s)&= \frac{1}{(\mu+1)\,(\lambda+\eta/\mu)}\rme^{-x (\eta +\lambda )}\,,\label{eq:Fpsweak}\\
\tilde F_-(x,s)&= \rme^{-x (\eta +\lambda )}\,\label{eq:Fmsweak}.
\end{align}
\label{eq:Fpm}
\end{subequations}
 We can now expand the first-passage distributions (\ref{eq:Fmsweak}) around $s\!=\!0$ to find the average time $\langle T(x)\rangle_{\pm}$ using the observation made in (\ref{eq:Fseries}). We find quite different behaviors depending on the sign of the drift.
 
\paragraph{Negative subcritical drift ($-1\!<\!\mu\!<\!0$).} In the case of a negative subcritical drift, we find that the particle will certainly reach the origin with average times
\begin{subequations}
\begin{align}
  \langle T(x) \rangle_+ &= \frac{x}{|\mu|}+\frac{1}{|\mu|}\,\label{eq:Tpwn} ,\\
  \langle T(x) \rangle_- &= \frac{x}{|\mu|}\,.\label{eq:Tmwn}
\end{align} 
  \label{eq:Tavgwn}
\end{subequations}
The first term in the average times (\ref{eq:Tavgwn}) originates from the mean ballistic motion. The second term in (\ref{eq:Tpwn}) is a correction accounting for the positive initial velocity of the particle. Apart from this correction, the mean first-passage time is the same as the one for a drifted Brownian motion. Note that the average time $\langle T(x) \rangle_+$ does not vanish when the particle starts exactly at the origin. One needs to extrapolate to $x\!=\!-1$ for the average time to vanish. This is another appearance of the ''Milne extrapolation length'' \cite{le2019noncrossing} that arises from the persistent feature of the motion. 

\paragraph{Positive subcritical drift ($\,0\!<\!\mu\! <\! 1$).} 
In the case of a positive subcritical drift, the constant term in the expansion of the Laplace transform of the first-passage distribution (\ref{eq:Fseries}) is less than $1$ which means that there is a non-zero probability that the particle never reaches the origin. This is of course due to the fact that the drift blows the particle away from the origin. It is still possible to define a first-passage time by conditioning the trajectories to eventually reach the origin. This is done by dividing the linear term by the constant term in the expansion of the Laplace transform of the first-passage distribution (\ref{eq:Fseries}) and yields 
\begin{subequations}
\begin{align}
  \langle T(x) \rangle_{c,+} &= \frac{x\,(1+\mu^2)}{\mu\,(1-\mu^2)} + \frac{1}{\mu}\, ,\\
  \langle T(x) \rangle_{c,-} &= \frac{x\,(1+\mu ^2)
   }{\mu  \,(1-\mu
   ^2)}\,,
\end{align} 
  \label{eq:Tavgwp}
\end{subequations}
where the subscript $c$ refers to the conditioned average over the trajectories that eventually reach the origin. In this case, the conditional mean first-passage time differs from the one of the Brownian motion $\langle T \rangle = x/\mu$ due to the $\mu^2$ terms which means that the duality found in \cite{krapivsky2018first} does not extend to the run-and-tumble process. Nevertheless, the Brownian conditional mean first-passage time is recovered upon reintroducing the units (\ref{eq:units}) and taking the Brownian limit (\ref{eq:brownianlimit}).

\section{Survival probability in the presence of a supercritical drift $\mu<-v_0$}
\label{sec:sSs}
In this section, we consider a subcritical drift (see figure \ref{fig:survivalw} and \ref{fig:phase_diagram}) and we start with the formula for the double Laplace transform of the survival probability given in (\ref{eq:gform2}). In this case, the poles $z_1$, $z_2$, $z_3$ and $z_4$ are all positive (\ref{eq:poles}) and the survival probability~(\ref{eq:gform2}) simplifies to
  \begin{align}
   \int_0^\infty dx \,\tilde S(x,s)\,\rme^{-u\,x}  = \frac{\gamma+s}{\gamma\,u\,s}\,\,\frac{(u+z_1)(u+z_3) }{(u+ z_2)(u+ z_4)}\,\,\frac{ z_2\,z_4}{ z_1\,z_3}-\frac{1}{\gamma\, u}\,,
 \end{align}
 which, after an inverse Laplace transform from $u$ to $x$, reads
    \begin{equation}
 \tilde S(x,s) = \frac{\gamma+s}{\gamma\,s}\,\, \frac{z_4 (z_2-z_1) (z_2-z_3)\rme^{- z_2
   x}+z_2 (z_1-z_4) (z_4-z_3)\rme^{- z_4 x}+z_1
   z_3 (z_2-z_4)}{z_1 z_3  (z_2-z_4)}-\frac{1}{\gamma}
   \end{equation}
where $z_1$, $z_2$, $z_3$ and $z_4$ are given in (\ref{eq:poles}). Rewriting it in terms of $\tilde \gamma = \gamma/2$ [see (\ref{eq:Wintro})], we find     
   \begin{align}
 \tilde  S(x,s) = \frac{1}{2\,s}\left(2-\frac{\tilde \gamma\,\e^{-\eta\,x}}{s}\left(\cosh(\lambda\,x)-\frac{\mu\,\tilde\gamma}{\lambda (\mu^2-v_0^2)}\,\sinh(\lambda\,x)\right)\right)\,,
\end{align}
where $\lambda$ and $\eta$ are given in (\ref{eq:lambdaeta}). The inverse Laplace transform with respect to $s$ can be performed explicitly (see \ref{app:lapinv}) to obtain $S(x,t) = (S_{+}(x,t)+S_-(x,t))/2$ with $S_{\pm}(x,)$ given -- setting $v_0 = \tilde \gamma = 1$ -- in (\ref{eq:Sstrongintro}). In fact, in this supercritical case, $S(x,t)$ can be computed in an alternative way, which as we show below, also allows us to compute 
easily the conditioned survival probability $S_{\pm}(x,t)$ \footnote{Note also that in this case $S_{+}(x,t)$ and $S_-(x,t)$ do not coincide anymore with $S_\uparrow(x,t)$ and $S_\downarrow(x,t)$, as they do in the subcritical case since in this case the initial velocity is always negative.}. For simplicity, we set $v_0 = \tilde \gamma = 1$ for the rest of 
this section. 

As mentioned in the introduction, the supercritical drift forces the particle to always move in the same direction (see figure \ref{fig:survivals}). Therefore, once the particle reaches the origin for the first time, it will also be the last time that it reaches it. This means that the probability that the particle reaches the origin for the first time at time $t$ is equal to the probability that the particle is located at the origin at time $t$ (see also \cite{lopez2014asymmetric})
\begin{align}
   F_{\sigma_0}(x,t)\, dt = P(y=0,t|x,\sigma(0)=\sigma_0)\, dy \,, \label{eq:FPrel}
\end{align}
where $F_{\sigma_0}(x,t) = - \partial_t S_{\sigma_0}(x,t)$ is the first-passage time distribution while $P(y,t|x,\sigma_0)$ is the probability that the particle is located at $y$ at time $t$ given that it started at $x$ in the state $\sigma(0) = \sigma_0 = \pm 1$. When the particle is located at the origin at time $t$, it can be in either states $\sigma(t)=1$ or $\sigma(t)=-1$, so we expand the right hand side of (\ref{eq:FPrel}) over these two cases, namely
\begin{align}
  F_{\sigma_0}(x,t)\, dt =  P(y=0,t,+|x, \sigma_0)\, dy + P(y=0,t,-|x,\sigma_0)\, dy\, ,\label{eq:FPexprel}
\end{align}
where $P(y,t,\sigma|x, \sigma_0)$ is the probability that the particle is located at $y$ at time $t$ in the state $\sigma(t) = \sigma = \pm 1$ given that it started at $x$ in the state $\sigma_0$. This propagator is well known \cite{lopez2014asymmetric} and is also reproduced in \ref{app:kernel} for the sake of completeness. Using that $dy =(\mu+\sigma(t))\, dt$ in (\ref{eq:FPexprel}), we find
%
\begin{align}
  F_{\sigma_0}(x,t) = |\mu+1|\, P(y=0,t,+|x,\sigma_0) + |\mu-1|\, P(y=0,t,-|x,\sigma_0)\, .
\end{align}
Inserting the expressions for the propagator (\ref{eq:kernel}), and integrating over $t$ we obtain the expression for $S_{\pm}(x,t)$ given in~(\ref{eq:Sstrongintro}).

From these expressions for the survival probabilities $S_{\pm}(x,t)$ (\ref{eq:Sstrongintro}), one can compute  
the mean first-passage time to reach the origin $\langle T(x) \rangle_{\sigma_0}$ for a particle starting at $x$ in the state $\sigma_0$ in the presence of a supercritical drift $\mu<-1$. It can be computed in the same way as we did for the case of a subcritical drift in section \ref{sec:mfpt}. We find (we recall that $\mu < -1$)
\begin{subequations}
\begin{align}
  \langle T(x) \rangle_+ &= \frac{x}{|\mu|} +\frac{(1-\mu)\,(1- \rme^{\,\frac{2 \,\mu
   x}{\mu ^2-1}})}{2\,\mu ^2}\,, \\
  \langle T(x) \rangle_- &= \frac{x}{|\mu|} +\frac{(1+\mu)\,(1- \rme^{\,\frac{2 \,\mu
   x}{\mu ^2-1}})}{2 \,\mu ^2}\, .
\end{align} 
  \label{eq:Tavgsn}
\end{subequations}
The first term in both expressions in (\ref{eq:Tavgsn}) comes from the mean ballistic motion and the second term is a correction that decays exponentially as the intensity of the supercritical drift is increased.

\section{Survival probability in the presence of a critical drift $\mu=-v_0$}
\label{sec:survc}

In this part, we set $v_0=1$ for convenience, and study the particular case of a critical drift $\mu=-1$ (see figure \ref{fig:survivalc} and \ref{fig:phase_diagram}) which is the transition point between a subcritical drift $-1\!<\!\mu\!<\!1$ and a supercritical drift $\mu\!<\!-1$. Such drift exactly balances the internal velocity of the particle when it is in the $\sigma(t)\!=\!+1$ state, making it effectively motionless. In the other state $\sigma(t)\!=\!-1$, the particle moves towards the origin with an effective speed that is twice its internal speed. The process is therefore equivalent to a \emph{wait-and-run} model, where the particle waits for a random time after which it runs towards the origin for another random time and starts this cycle over again [see figure (\ref{fig:survivalc})]. Setting $\mu\!=\!-1$ in the expressions for the survival probability in the presence of a subcritical drift (\ref{eq:S11}) or in the presence of a supercritical drift (\ref{eq:Sstrongintro}), we obtain in both cases the same result
\begin{subequations}
\begin{align}
    S_+(x,t)&=\left\{\begin{array}{ll} 1\,, & t < x/2\,, \\
     1- \int_{x/2}^{t} dt' \, \rme^{-t'} \,I_0\left[\sqrt{x\,(2t'-x)}\right] \,,& t\geq x/2,\end{array}\right. \label{eq:ltzsp}\\[1em]
     S_-(x,t)&= \left\{\begin{array}{ll} 1\,, & t < x/2\,, \\
     1-\rme^{-x/2} - \int_{x/2}^{t} dt' \,\rme^{-t'}\, \frac{\sqrt{x}}{\sqrt{2t'-x}} \, I_1\left[\sqrt{x\,(2t'-x)}\right]\,,& t\geq x/2 \,.
     \end{array}\right. \label{eq:ltzsm}
\end{align}
\label{eq:ltSws}
\end{subequations}
It is then natural to ask how the survival probabilities $S_{\pm}(x,t)$ decay at large times, for fixed $x$. We show in \ref{app:ltsurvivalz} that, indeed, this late time behavior has a unusual form given by 
\begin{subequations}
\begin{align}
S_+(x,t) &\sim \frac{x^{-1/4}}{2^{3/4}\,\sqrt{\pi}}\,\frac{1}{t^{1/4}}\, \rme^{-t+\sqrt{2\, x\, t}}\,, \label{eq:longtzp}\\
    S_-(x,t) &\sim \frac{x^{1/4}}{2^{5/4}\,\sqrt{\pi}}\,\frac{1}{t^{3/4}}\, \rme^{-t+\sqrt{2\, x\, t}}\, .
    \label{eq:longtm}\end{align}
    \label{eq:longtz}
\end{subequations}
These results are anomalous in two ways: (i) the decay is not purely exponential but rather is an exponential modulated by a time-dependent amplitude 
that grows anomalously as $t^{-\alpha}\, \e^{\sqrt{2x\,t}}$, (ii) the exponent $\alpha$ is different for $S_+(x,t)$ and $S_-(x,t)$. In the former case, $\alpha = 1/4$ while in the latter case $\alpha = 3/4$. 

The above behavior holds at late times $t$ and at fixed $x$. One can also ask how the survival probability behaves for $t$ large but fixed and $x$ small ($x = O(1/t)$) or $x$ large ($x = O(t)$). In the first case, when $x= O (1/t)$, by analyzing (\ref{eq:ltzsp}), we find that there is a scaling behavior 
\begin{align} \label{S+smallx}
S_+(x,t) \sim \e^{-t} I_0(\sqrt{2 x\,t}) \;.
\end{align}  
 This scaling form holds for $t$ large, $x \to 0$ but with the product $x\,t$ fixed. In the limit where $t \gg 1/x$ -- or equivalently $z = x\,t \to \infty$ -- using the asymptotic behavior of the Bessel function $I_0(z) \sim e^{z}/\sqrt{2 \pi z}$, our scaling result (\ref{S+smallx}) indeed reduces to (\ref{eq:longtzp}). 
 
  \begin{figure}[ht]
    \centering
    \includegraphics{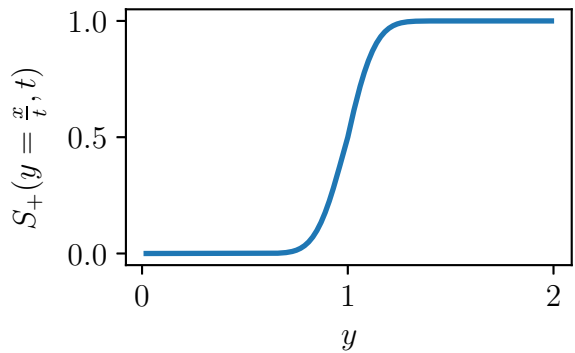}
    \caption{Plot of the exact survival probability $S_+(x,t)$ (\ref{eq:ltzsp}) as a function of the scaled initial distance $y=x/t$ evaluated at large time, e.g. here for $t=50$. As $t$ grows, the fluctuations are washed out and the survival probability tends towards a step function corresponding to the ballistic behavior.}
    \label{fig:Sldev}
\end{figure} 
We now consider the opposite limit when $t$ is large and $x=O(t)$. This means that in figure \ref{fig:Sldev}, we are zooming in near the ``shoulder'' of the curve when $y=x/t = 1$. It turns out that the typical width of the shoulder regime scales as $1/\sqrt{t}$ for large $t$. Analysing (\ref{fig:Sldev}) close to this point, we find the following behavior of $S_+(x,t)$ 
\begin{align}
 S_+\left(y=\frac{x}{t},t\right) &\sim\left\{\begin{array}{lll} \frac{\sqrt{\varphi(y)}}{\sqrt{2\pi\, y}\,(\varphi(y)-1)}\frac{1}{\sqrt{t}} \rme^{-t\,\Psi(y)}\,,  & y< 1 \text{ and } &|y-1|=O(1)\,,\\[1em]
\frac{1}{2}\, \text{erfc}\left(\frac{\sqrt{t} (1-y)}{\sqrt{2}}\right)\,,& &  |y-1|=O(1/\sqrt{t})\,,\\[1em]
   1-\frac{\sqrt{\varphi(y)}}{\sqrt{2\pi\, y}\,(\varphi(y)-1)}\frac{1}{\sqrt{t}} \rme^{-t\,\Psi(y)}\,,  & y> 1 \text{ and }& |y-1|=O(1)\,,
  \end{array}\right. 
  \label{eq:zerodrifttrans}
\end{align}
where $\varphi(y)=\sqrt{(2-y)/y}$. Here the second line provides the late time behavior of the typical survival probability, i.e., when $(y-1) = O(1/\sqrt{t})$. When $|y-1| \gg 1/O(\sqrt{t})$, the survival probability is described by a large deviation form as in the first and third line of (\ref{eq:zerodrifttrans}). The large deviation function $\Psi(y)$ is given by 
\begin{align}
\Psi(y)=1-\sqrt{y\,(2-y)}\,.  \label{eq:largedev}
\end{align}

In fact, the central regime $|y-1| = O(1/\sqrt{t})$ in (\ref{eq:zerodrifttrans}) matches smoothly with the regimes $|y-1| = O(1)$ both on the left and right. Consider first the left regime, where $1-y \gg O(1/\sqrt{t})$. In this case, we can use the asymptotic behaviour $\text{erfc}(z)\sim \rme^{-z^2}/\sqrt{2\pi} z$ as $z \to +\infty$. Plugging this behaviour in the second line of (\ref{eq:zerodrifttrans}), we get   
\begin{align}
  \frac{1}{2}\, \text{erfc}\left(\frac{\sqrt{t} (1-y)}{\sqrt{2}}\right)\sim \frac{1}{\sqrt{2\pi}\,(1-y)}\frac{1}{\sqrt{t}}\rme^{-t\,\frac{(1-y)^2}{2}}\,,\quad (1-y) \gg O(1/\sqrt{t}) \, .\label{eq:limr2}
\end{align}
In contrast, setting $(1-y) \ll O(1)$ in the first line of (\ref{eq:zerodrifttrans}), and using that $\Psi(y) \sim (1-y)^2/2$ in this limit [from (\ref{eq:largedev})], we get
\begin{align}
  \frac{\sqrt{\varphi(y)}}{\sqrt{2\pi\, y}\,(\varphi(y)-1)}\frac{1}{\sqrt{t}} \rme^{-t\,\Psi(y)} \sim \frac{1}{\sqrt{2\pi}\,(1-y)}\frac{1}{\sqrt{t}}\rme^{-t\,\frac{(1-y)^2}{2}}\,, \quad y\rightarrow 1\,,\label{eq:limr1}
\end{align}
where we used that $\varphi(y)\sim 1-y$ when $y\rightarrow 1$. The limiting behaviours in (\ref{eq:limr1}) and (\ref{eq:limr2}) agree and therefore the first and the second regimes in the scaling (\ref{eq:zerodrifttrans}) match. Similarly, when $(y-1) \gg O(1/\sqrt{t})$, using the asymptotic behavior ${\rm erfc}(z) \sim 2 \,-\, \rme^{-z^2}/\sqrt{2\pi} |z|$ as $z \to -\infty$, we can
easily verify that the second and the third lines in (\ref{eq:zerodrifttrans}) also match smoothly with each other.

 \vspace*{0.5cm}
 \noindent{\it Scaling behavior near the critical point.} So far, we discussed the late time behavior of the survival probability exactly at the critical point $\mu = -1$, as well as
 in the off-critical phases where $\mu \neq -1$. Thus it is interesting to ask what happens, as a function of time, if one stays in the off-critical phase by setting $\mu$ close
 to the critical value $\mu = -1$. In this case, one would expect that the system, at relatively early times, behaves as if it is at the critical point and beyond a cross-over
 time scale the system ``realizes'' that it is off-critical. This cross-over from early time critical behaviour to late time off-critical behaviour for $\mu$ close to $\mu=-1$ can
 be described by cross)-over scaling functions as we show below. This can be done, by setting the value of $\mu$ either on the subcritical side ($\mu>-1$) or in the supercritical 
 side ($\mu < -1$). 

We start with the sub-critical side and evaluate the survival probability in the scaling limit $t \to \infty$, $x \to 0$ and $\mu \to -1$ while keeping the scaling
variables  
\begin{align}
  z &= t\, x\,, \\
  u &= \sqrt{|1+\mu|}\,t \,,
\end{align}
fixed. To do so, we first rewrite, from (\ref{eq:Spsub}), the survival probability as $S_+(x,t)=\int_t^\infty dt'\,F_+(x,t')$ 
\begin{align}
S_+(x,t)=\int_{t}^{\infty}\, dt' \, \frac{\rme^{-t'}}{g(t',x)}\left(x \,I_0\left[h(t',x)\right] + (1+\mu)\,\sqrt{\frac{f(t',x)}{g(t',x)}}\,I_1\left[h(t',x)\right]\right)\,,
\end{align}
which is valid for $t\!>\!t_m$. Then, we change variable $t'=t+v$ and note that in the scaling limit considered here, we have
\begin{align}
  \frac{x}{g(t+v,x)} &\sim \frac{z}{u^2+z}\,,\\
  \frac{1+\mu}{g(t+v,x)}\,\,\sqrt{\frac{f(t+v,x)}{g(t+v,x)}} &\sim \frac{\sqrt{2}\,u^2}{(u^2+z)^{3/2}}\,,
\end{align}
so that $ S_+(x,t)$ can be written in terms of a scaling function $\mathcal{S}_{\rm sub}(z, u)$ as
\begin{align}
  S_+(x,t) &\sim \rme^{-t}\, \mathcal{S}_{\rm sub}(z=t\, x, u=\sqrt{|1+\mu|}\,t)\,,
\end{align}
with
\begin{align}
   \mathcal{S}_{\rm sub}(z, u) = \frac{z}{u^2+z}I_0(\sqrt{2(u^2+z)}) + \frac{\sqrt{2}u^2}{(u^2+z)^{3/2}} I_1(\sqrt{2(u^2+z)})\,.
\end{align}
We can then check that the scaling function contains the different regimes that we have already found. For instance, when $u\rightarrow 0$ and $z\rightarrow 0$, we find $S_+(x,t)\sim\rme^{-t}$ which is indeed the no-tumbling probability. Another check is to take the limit $u\rightarrow 0$ and $z\rightarrow \infty$, in which case we recover the long time limit of the survival probability in the presence of a critical drift (\ref{eq:longtzp}). Finally, we can take the limit $u\rightarrow \infty$ and $z\rightarrow 0$ so that we recover the long time limit of the survival probability in the presence of a subcritical drift (\ref{eq:longtposs}) when $x\rightarrow 0$ and $\mu\rightarrow -1$. Physically, the scaling variable $u$ gives us the cross-over time 
\begin{align}
  t_c = \frac{1}{\sqrt{|1+\mu|}}\,. \label{eq:tc}
\end{align}
When the drift is close to criticality $\mu=-1+\epsilon$, with $\epsilon>0$, the particle behaves like if the drift was critical $\mu=-1$ until a time $t_c$ after which it behaves like if the drift was subcritical $\mu >  -1$. 

An analogous scaling function can be found upon evaluating the survival probability $S_+(x,t)$ in the presence of a supercritical drift (\ref{eq:Sstrongpi}). Performing similar steps, we find a scaling function $\mathcal{S}_{\rm sup}(z,u)$ such that 
\begin{align}
  S_+(x,t) \sim \e^{-t}\, \mathcal{S}_{\rm sup}(z=t\,x,u=\sqrt{|1+\mu|}\,t)\,,
\end{align}
with 
\begin{align}
 \mathcal{S}_{\rm sup}(z,u)= I_0(\sqrt{2(u^2+z)})\,,\quad u^2 < z\,,
\end{align}
where the constraint $u^2<z$ comes from the fact that we must have $t<t_M$ in the presence of a supercritical drift (\ref{eq:Sstrongintro}). Analogously, when the drift $\mu=-1-\epsilon$, with $\epsilon>0$,  
is close to the critical value $\mu=-1$ (in the limit $\epsilon \to 0$), the particle behaves like if the drift was critical until the time $t_c$ (\ref{eq:tc}) after which it behaves like if the drift was supercritical $\mu < -1$.

\section{Another example of the velocity distribution $W(v)$} \label{sec:other}
To illustrate the generality of the formula obtained for the survival probability (\ref{eq:gformula}), we apply it to another two-state velocity distribution with unequal weights
\begin{align}
  W(v)=\left(\frac{1}{2}+c\right)\,\delta(v-\mu-1) +\left(\frac{1}{2}-c\right)\,\delta(v-\mu+1)\,,
  \label{eq:wvg}
\end{align}
 with $-\frac{1}{2}<c<\frac{1}{2}$. When $c=0$, we recover the velocity distribution (\ref{eq:Wintro}) discussed in the main part of this work. Using the general formula for the survival probability (\ref{eq:gformula}), we find a similar pole structure as for the survival probability (\ref{eq:gformm}). For the sake of conciseness, we will restrict ourselves to a subcritical drift $-1\!<\!\mu\!<\!1$ in this section. In this case, only two poles are positive. Performing similar steps to the ones done in section (\ref{sec:subcr}), we find that the Laplace transform of the survival probability is given by
%
\begin{align}
 \tilde S(x,s)= \frac{1}{2s}\left(2 - \left(\frac{\frac{4 c (s+1)}{\mu -1}+1}{(\mu +1) \left(-\frac{2 c}{1-\mu ^2}+\frac{\eta }{\mu }+\lambda_c
   \right)}+1\right)\exp(-x
   \left(\lambda_c +\eta +\frac{2\,c}{1-\mu^2}\right))\right)\,,\label{eq:Ssc}
\end{align}
where $\eta$ is given in (\ref{eq:eta}) and
\begin{align}
  \lambda_c = \frac{\sqrt{4 c^2+4 c \mu  (s+1)+\mu ^2+s (s+2)}}{1-\mu^2}\,.
\end{align}
 As we did in section \ref{sec:subcr}, we can use the identities (\ref{eq:qnp}) and obtain the survival probabilities conditioned over the initial state
\begin{subequations}
\begin{align}
 \tilde S_+(x,s)&= \frac{1}{s}\left(1 - \frac{\frac{4 c (s+1)}{\mu -1}+1}{(\mu +1) \left(-\frac{2 c}{1-\mu ^2}+\frac{\eta }{\mu }+\lambda_c
   \right)}\exp(-x
   \left(\lambda_c +\eta +\frac{2\,c}{1-\mu^2}\right))\right)\,,\\
   \tilde S_-(x,s)&= \frac{1}{s}\left(1 -\exp(-x
   \left(\lambda_c +\eta +\frac{2\,c}{1-\mu^2}\right))\right)\,.
\end{align}
\end{subequations}
Inverting these Laplace transforms, we find
\begin{subequations}
\begin{align}
  S_-(x,t)= \left\{\begin{array}{ll}1\,, &t<t_m\, ,\\
1- \rme^{-t_m\, (2\, c\, \mu +1)-2\, c \,x}&\\ \quad-\sqrt{1-4\,c^2}\int_0^t dt'\,\rme^{-t' \,(2 \,c\, \mu +1)-2 \,c\, x}\frac{ x
  }{h(t',x)} I_1\left[\sqrt{1-4\, c^2}\, h(t',x)\right]\,,& t\geq t_m\,,
  \end{array}\right.\label{eq:smcd}
\end{align}
and
\begin{align}
  S_+(x,t)= \left\{\begin{array}{ll}1\,, &t<t_m\, ,\\
  1-(1-2\,c)\int_0^t \,dt'\,\frac{\rme^{-t\, (2\, c\, \mu +1)-2\, c\, x}}{g(t',x)} &\\
  \quad \times \bigg(x \,I_0\left[\sqrt{1-4\,c^2}\,h(t',x)\right]  &\\
 \qquad\quad +\,  \frac{1+\mu}{\sqrt{1-4\,c^2}}\,\sqrt{\frac{f(t',x)}{g(t',x)}}\,I_1\left[\sqrt{1-4c^2}\,h(t',x)\right]\bigg)\,,& t\geq t_m\,,
  \end{array}\right.
\end{align}
\end{subequations}
where $t_m$, $f(t,x)$, $g(t,x)$ and $h(t,x)$ are given in (\ref{eq:tmfgh}).
When $c\!=\!0$, we recover the survival probabilities (\ref{eq:Spmt}) obtained previously.
The finite weight of the trajectory that crosses the origin at $t\!=\!t_m$ and creates a drop in the survival probability (\ref{eq:smcd}), is now given by
\begin{align}
  \rme^{-t_m (2 c \mu +1)-2 c x} = \sum_{n=0}^\infty \left(\frac{1}{2}-c\right)^n \frac{(2\,t_m)^n}{n!}e^{-2\,t_m}\,.
\end{align}
We see that it is a sum over all possible numbers of tumbling events $n$. For $n$ tumbling events, it is a product of the probability that the particle has tumbled $n$ times during a time $t_m$, given by the Poisson distribution (\ref{eq:poissondist}) (with $\tilde \gamma=2$), and the probability that the velocity $v=\mu-1$ was chosen at every tumbling events.

We find that the condition for the particle to have a finite survival probability is now given by
\begin{align}
  2\,c+\mu >0\,,
\end{align}
which yields the following finite survival probabilities
\begin{subequations}
\begin{align}
 \lim_{t\rightarrow\infty} S_-(x,t)&=1-\rme^{-2\,x\frac{ (2 c+\mu )}{1-\mu
   ^2}}\,,\\
  \lim_{t\rightarrow\infty}  S_+(x,t)&=1-\frac{1-4\,c-\mu}{1+\mu}\rme^{-2\,x\frac{ (2 c+\mu )}{1-\mu
   ^2}}\,.
\end{align}
\end{subequations}
We see that the uneven weight $c$ in the velocity distribution (\ref{eq:wvg}) creates an additional bias in the run-and-tumble motion.
\section{Applications}
\label{sec:appli}

\subsection{Maximum of a persistent random walk with a drift in a given time interval}

From a general point of view of extreme value statistics of correlated variables, it is interesting to study the maximum
of a stochastic process in a given time interval \cite{majumdar2020extreme}. For this PRW model with a drift, the distribution of the maximum $M(t)$
was recently studied by Cinque and Orsingher \cite{cinque2020exact}. 
One would expect that this survival probability studied here
is closely related to the cumulative distribution of the maximum. This is actually very general and is true for any stochastic process. 
To see this, consider a process $y(t)$ starting at $y(0) = y_0$. Let $M(t) = \max_{0 \leq \tau \leq t} \{ y(\tau)\}$ denote the maximum of the process $y(\tau)$. Let us consider the cumulative distribution of this maximum
\begin{equation} \label{CDF_max}
{\rm Pr.} \left[M(t) \leq M | y(0) = y_0 \right] = {\rm Pr.}\left[y(\tau) \leq M, \quad \forall \tau \in [0,t] \; | y(0) = y_0\right] \;.
\end{equation} 
We now consider a new process $x(t) = M - y(t)$. In terms of the $x$-process, the cumulative distribution in (\ref{CDF_max}) can be expressed as
\begin{equation}\label{CDF_2}
{\rm Pr.} \left[M(t) \leq M | y(0) = y_0 \right] = {\rm Pr.}\left[x(\tau) \geq 0 , \quad \forall \tau \in [0,t] \; | x(0) = M-y_0\right] \;.
\end{equation}  
Thus the cumulative distribution of the maximum of the process $y(\tau)$ is identical to the survival probability $S(x,t)$ of the process $x(\tau) = M-y(\tau)$, starting
at $x = M-y_0$. Therefore, we would expect that, for the PRW with a drift, the survival probability studied here must coincide with the distribution of the maximum
studied in \cite{cinque2020exact}. However, at first sight, the expressions provided in \cite{cinque2020exact} seem rather different from ours. 
We show, however, that they are indeed identical. This requires some nontrivial intermediate steps that are given in \ref{app:equiv}. We believe
that the mathematical tricks used here to prove the equivalence of the two approaches might be useful in other related problems.

\subsection{Last-passage time and emptying time}

\begin{figure}[ht]
  \begin{center}
    \includegraphics[width=0.4\textwidth]{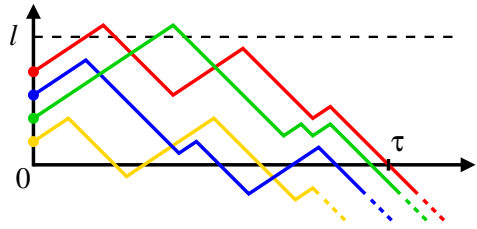}
    \caption{Typical trajectories of $N=4$ independent particles initially uniformly distributed in a box $[\,0,\,\ell\,]$. The emptying time $\tau$ is the time it takes for all the particles to permanently leave the box. A negative drift $\mu<0$ makes this notion well defined as it ensures that the process is transient and that the particles will eventually never return to the box.}
    \label{fig:emptying_box}
  \end{center}
\end{figure}

Let us consider $N$ independent run-and-tumble particles initially uniformly distributed in a box, defined by the region $[\,0,\,\ell\,]$. We study the emptying time $\tau$ of this box in the presence of a drift $\mu$ (see figure \ref{fig:emptying_box}). The {\it emptying time} is the time it takes for all the particles to permanently leave the box. The presence of a drift makes this notion well defined as it ensures that the process is transient and that the particles will eventually never return to the box.  We take the number of particles $N$ and the size of the box $\ell$ to be large while the density of particles $\rho\equiv N/\ell$ is fixed. Without loss of generality, we will choose a negative drift $\mu < 0$. In this case the particles can exit the box several times from either sides $0$ or $\ell$ but the drift will force the last exit to be made at the origin (see figure \ref{fig:emptying_box}).

 A key quantity to study this emptying time is the no-return probability $\Pi_{\sigma_0}(x_i,t)$ for a single particle, which is defined as the probability that a particle never returns to the origin after a time $t$ given that it started at $x$ in the state $\sigma_0$.  The cumulative distribution for the emptying time ${\rm Pr.}(\tau < t\,|\, \{x_i\}_N|\sigma_0)$ given the initial positions $\{x_i\}_N$ of the particles in the box $[\,0,\,\ell\,]$, can be related to $\Pi_{\sigma_0}(x_i,t)$ by stating that for the box to be empty at a time $t$, all the particles must never return to the origin again after a time $t$:
\begin{align}
  {\rm Pr.}(\tau \leq t \,| \,\{x_i\}_N,\sigma_0) = \prod_i^N \Pi_{\sigma_0}(x_i,t) \, .
  \label{eq:probxis}
\end{align}
The subscript $\sigma_0$ refers to the initial state which, for the sake of simplicity, is considered to be the same for all particles. Integrating over uniformly distributed initial positions $\{x_i\}_N$ in the box $[\,0,\ell\,]$ and taking the limit $N\rightarrow \infty $ and $\ell\rightarrow \infty$ with $\rho\equiv N/\ell$ fixed, the cumulative distribution of the emptying time is given by
\begin{align}
  {\rm Pr.}(\tau \leq t|\sigma_0) = \lim_{\ell\rightarrow \,\infty }\prod_i^{N=\rho\, \ell} \frac{1}{\ell}\int_0^\ell dx_i\, \Pi_{\sigma_0}(x_i,t)\, .
  \label{eq:Pempty}
\end{align}
 As a side note, the average taken over the initial positions in (\ref{eq:Pempty}) is referred to as an \emph{annealed} average in the context of disordered systems. It is also possible to study the \emph{quenched} average which is obtained by averaging the logarithm of the cumulative distribution of the emptying time (\ref{eq:probxis}) over the initial positions. We will restrict ourselves to the annealed average but we expect similar results for the quenched average.

\begin{figure}
  \begin{center}
    \includegraphics[width=0.4\textwidth]{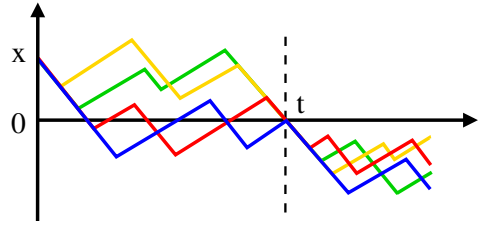}
    \caption{Typical trajectories that reach the origin for the last time at time $t$ given that the particle started at $x$ in the state $\sigma_0\!=\!-1$. These trajectories are those that start from $x$ in the state $\sigma_0\!=\!-1$ and end at the origin at time $t$ followed by all the trajectories that start at the origin at time $t$ and never reach the origin again. The motion can therefore be decomposed into a free particle that is initially at $x$ in the state $\sigma_0=-1$ and propagates to the origin in a time $t$ followed by a particle that starts from the origin and is constrained to stay on $]-\infty,\,0\,[$ forever. }
    \label{fig:last_passage}
  \end{center}
\end{figure}

 Similarly to the survival probability, the no-return probability $\Pi_{\sigma_0}(x,t)$ can be expressed in terms of its probability density function, the last-passage time distribution
\begin{align}
  \Pi_{\sigma_0}(x,t) = \int_0^t dt'\, L_{\sigma_0}(x,t')\, .
\end{align}
The last-passage distribution $L_{\sigma_0}(x,t)$ is an interesting observable per se since $L_{\sigma_0}(x,t)\, dt$ is the probability that the particle reaches the origin for the last time in the time interval $[\,t,t+dt]$ given that it started at $x$ in the state $\sigma_0$. The last-passage time distribution $L_{\sigma_0}(x,t)$ can be obtained from our previous results on the survival probability by observing that for a particle to reach the origin for the last time at time $t$, it must first be located at the origin at time $t$ and then never reach the origin again by surviving in the interval $]-\infty,0\,[$ (see figure \ref{fig:last_passage}).
Taking into account that the particle can either be in the state $\sigma(t)\!=\!+1$ or $\sigma(t)\!=\!-1$ when it is located at the origin at time $t$, it reads \cite{singh2019generalised}
\begin{align}
  L_{\sigma_0}(x,t)\, dt = P(y=0,t,+|x,\sigma_0)\,dy\, S^*_+ \; +  \; P(y=0,t,-|x,\sigma_0)\,dy\, S^*_-\,,
  \label{eq:lastpass}
\end{align}
 where $P(y,t,\sigma|x,\sigma_0)\, dy$ is the probability that the particle is located at $y$ at time $t$ in the state $\sigma$ given that it started at $x$ in the state $\sigma_0$ and $S^*_{\sigma_0}$ is the probability that a particle survives forever in the region $]-\infty,\,0\,[$ given that it started at the origin in the state $\sigma_0$. Note that it is possible to obtain the distribution of the last-passage time during a finite observation time window and generalize the results obtained in \cite{singh2019generalised} in the presence of a drift (see \ref{app:lastfin}).
 
 The propagator $P(y,t,\sigma|x,\,\sigma_0)$ is computed in \ref{app:kernel} and $S^*_{\sigma_0}$ is related to our results on the survival probability derived in the previous sections. One last step to perform in the relation (\ref{eq:lastpass}) is to match the time volume element $dt$ in the left hand side with the space volume element $dy$ in the right hand side. To do so, we proceed as in section \ref{sec:sSs} and use the fact that $dy/dt=\mu+\sigma(t)$, where $\sigma(t)$ is the state of the particle at time $t$ to obtain
 \begin{align}
  L_{\sigma_0}(x,t) = |\mu+1|\,P(y=0,t,+|x,\sigma_0)\, S^*_+ +  |\mu-1|\,P(y=0,t,-|x,\sigma_0)\, S^*_-\, .
  \label{eq:lastpassf}
\end{align}
We will now rely on the relation (\ref{eq:lastpassf}) and on our previous results to derive the last-passage time distribution for the case of a subcritical and supercritical negative drift.

\begin{figure}[t]
    \centering
    \includegraphics{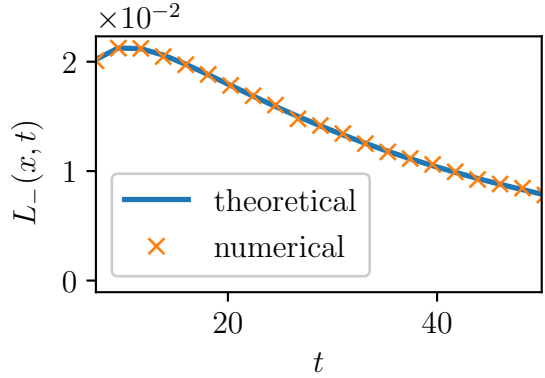}
    \caption{The probability distribution $L_-(x,t)$ of the last-passage time to the origin, starting at $x=5$ and $\sigma_0=-1$, plotted as a function of $t$ for a subcritical drift $\mu=-0.2$. The numerical results (crosses) agree perfectly with the analytical results given in (\ref{eq:lastpw}). Note that the Dirac delta function in (\ref{eq:lastpw}) is not shown to fit the data within the limited window size.}
    \label{fig:lptsol}
\end{figure}

\paragraph{Subcritical negative drift ($-1\!<\!\mu\!<\!0$).} For the case of a subcritical negative drift (see figure \ref{fig:survivalw} and \ref{fig:phase_diagram}), we use our results on the long-time limit of the survival probability found in section \ref{sec:lgtlimsurvival}. As we are interested in the survival probability in the region $(-\infty,\,0\,[$, we evaluate the eventual survival probability (\ref{eq:longtpos}) for $x=0$ and substitute 
 $\mu\to -\mu$ to get
 \begin{subequations}
 \begin{align}
   S_-^*&= 1-\frac{1+\mu}{1-\mu}\,, \\
   S_+^*&= 0\, .
 \end{align} 
 \label{eq:sstar}
 \end{subequations}
  Using the relation (\ref{eq:lastpassf}) and the propagator derived in \ref{app:kernel}, we find
 \begin{subequations}
 \begin{align}
  L_{+}(x,t) &= \left\{\begin{array}{ll} 0\,, & t<t_m\, ,\\ |\mu|\,\rme^{-t}\, I_0(h(t,x))\,, & t\geq t_m\, ,
\end{array} \right.\\
   L_{-}(x,t) &=  \left\{\begin{array}{ll} 0\, , & t<t_m ,\\
   2\,|\mu|\,\rme^{-t}\left(\delta(t-x + |\mu| t)+\frac{\sqrt{g(t,x)}}{\sqrt{f(t,x)}}\,I_1(h(t,x))/2\right)\, , & t\geq t_m\,  ,\label{eq:lastpw}\\
   \end{array}\right.  
 \end{align}
 \label{eq:lastw}
 \end{subequations}
 where $f(t,x)$, $g(t,x)$, $h(t,x)$ and $t_m$ are given in (\ref{eq:tmfgh}). Upon reintroducing the units (\ref{eq:units}) and taking the Brownian limit, we recover the well-known result for the Brownian motion (see e.g. \cite{comtet2020last})
\begin{align}
     L_{\sigma_0}(x,t) &\sim \frac{|\mu|}{\sqrt{4\pi D t}}\,\rme^{-\frac{1}{4Dt}\,\left(x+\mu\, t\right)^2}\, .
\end{align}
A numerical check shows that our results (\ref{eq:lastw}) are in excellent agreement with simulations (see figure \ref{fig:lptsol}).
It is interesting to compute the mean last-passage time. This is done by taking the average of the last-passage distribution (\ref{eq:lastw}). As we saw in section \ref{sec:mfpt}, it is easier to extract this information from a series expansion of the Laplace transform $\tilde L_{\sigma_0}(x,s)$ for small $s$, which is given by
\begin{subequations}
\begin{align}
\tilde L_+(x,s) &= \frac{|\mu|}{(1+\mu^2)\,\lambda} \, \rme^{-(\eta +\lambda)\, x }\, , \\
  \tilde L_-(x,s) &= -\frac{\eta+\mu\,\lambda}{(1-\mu)\,\lambda}\, \rme^{-(\eta +\lambda)\, x }\, , \label{eq:plsa}
\end{align}
\label{eq:Lpms}
\end{subequations}
where $\eta$ and $\lambda$ are given in (\ref{eq:lambdaeta}). To obtain (\ref{eq:Lpms}), we inserted the Laplace transform of the propagator, given in \ref{app:kernels} into the Laplace transform of the last-passage time distribution given in (\ref{eq:lastpassf}). As we did in section \ref{sec:mfpt}, we expand the Laplace transform (\ref{eq:Lpms}) close to $s=0$ and we find that the average last-passage time $\langle \hat T(x) \rangle_{\sigma_0}$ to the origin, given that the particle started at $x$ in the state $\sigma_0$, is given by
\begin{subequations}
\begin{align}
  \langle \hat T(x) \rangle_+ &= \frac{1}{\mu^2} - \frac{x}{\mu} + \frac{1}{\mu}\, ,\label{eq:tlastavg+} \\
  \langle \hat T(x) \rangle_- &= \frac{1}{\mu^2} - \frac{x}{\mu}\, .
\end{align}
\end{subequations}
 Upon reintroducing the units (\ref{eq:units}), we notice that beside the correction $1/\mu$ in $\langle \hat T(x) \rangle_{+}$ (\ref{eq:tlastavg+}), the mean last-passage time is the same as the one of a drifted Brownian motion (see e.g. \cite{comtet2020last})
 \begin{align}
   \langle \hat T(x)\rangle  &= \frac{2D}{\mu^2}-\frac{x}{\mu}  \, .
\end{align} Therefore, as it was the case for the mean first-passage time in section \ref{sec:mfpt}, the mean last-passage time is not enough to distinguish a run-and-tumble particle from a Brownian motion. The signature of persistence of the run-and-tumble particle will only be observed in higher order cumulants.

The no-return probability is then obtained as the cumulative density function of the last-passage distributions (\ref{eq:lastw}):
\begin{subequations}
   \begin{align}
   \Pi_{+}(x,t) &=  \left\{\begin{array}{ll} 0\,, & t<t_m\, ,\\  |\mu|\int_0^t dt'\,\rme^{-t'}\, I_0(h(t',x))\,, & t\geq t_m\, ,
\end{array} \right. \\
   \Pi_{-}(x,t) &=\left\{\begin{array}{ll} 0\, , & t<t_m ,\\
  2\,|\mu|\,\rme^{-\frac{x}{1-\mu}} + |\mu| \int_0^t dt'\,\rme^{-t'}\frac{\sqrt{g(t',x)}}{\sqrt{f(t',x)}}\,I_1(h(t',x))\, , & t\geq t_m\,  .\\
   \end{array}\right.
\end{align}
\label{eq:piw}
\end{subequations}
Because the emptying time of the box (\ref{eq:Pempty}) will be essentially governed by particles starting far away from the origin, we consider the limit $x\gg 1$ of the no-return probability (\ref{eq:piw}). The last-passage distributions (\ref{eq:lastw}) are normalized, therefore we can rewrite the no-return probabilities as $\Pi_{\sigma_0}(x,t)=1-\int_t^\infty dt'\, L_{\sigma_0}(x,t')$ as we did in section \ref{sec:lgtlimsurvival} and replace the integrand by its large $x$ limit. Note that a large $x$ limit implies a long $t$ limit due to the presence of the lower bound $t\geq t_m=x/(1-\mu)$ in (\ref{eq:piw}). Using the asymptotic expansion of the Bessel function $I_{0,1}(z)\sim \rme^{z}\,/\sqrt{2\pi z}$ for $z\rightarrow \infty$, we find
\begin{subequations}
   \begin{align}
   \Pi_{+}(x,t) &\sim 1-|\mu|\int_t^\infty dt' \,\frac{1}{\sqrt{2\pi\,h(t',x)}}\,\rme^{-t'+h(t',x)}\,  ,\\
   \Pi_{-}(x,t) &\sim 1-|\mu|\int_t^\infty dt'\,\frac{\sqrt{g(t',x)}}{\sqrt{2\pi\,f(t',x)\,h(t',x)}}\,\rme^{-t'+h(t',x)}\,  .
\end{align}
\label{eq:piltlx}
\end{subequations}
These integrals can be evaluated using the saddle-point method. The key point to note is that the argument of the exponential in the no-return probability (\ref{eq:piltlx}) is minimized for $t^*=x/|\mu|$ and is locally approximated by
\begin{align}
  -t+h(t,x) \sim  \frac{\mu^3}{2\,x}\left(t-t^*\right)^2\,,\quad t\rightarrow t^*\,,
\end{align} 
which means that the last-passage distribution is a Gaussian distribution centered around the mean value $t^*$ and with a standard deviation $w$ given by
\begin{align}
t^*=\frac{x}{|\mu|}\,,\quad w = \sqrt{\frac{x}{|\mu|^3}}\,.\label{eq:tsT}
\end{align} 
 Therefore, the no-return probability (\ref{eq:piltlx}) is simply the cumulative distribution of a Gaussian distribution with parameters (\ref{eq:tsT}):
\begin{align}
  \Pi_{\sigma_0}(x,t) &\sim 1 - \frac{1}{2} \,\text{erfc}\left(\frac{t-t^*}{\sqrt{2\,w}}\right)\,,\quad x\rightarrow \infty\,.
  \label{eq:noreturnerf}
\end{align}
The no-return probability (\ref{eq:noreturnerf}) can now be used to compute the cumulative distribution of the emptying time of the box by averaging it over uniformly distributed $\{x_i\}_N$ and plugging it into the expression (\ref{eq:Pempty}). These steps were recently taken in \cite{comtet2020last} where the authors studied the emptying time for a box of Brownian particles. As the steps to be taken here are identical, we quote their final result which takes the form of a Gumbel distribution
\begin{align}
  {\rm Pr.}\left(\tau\leq \frac{\ell}{|\mu|}+\sqrt{\frac{\ell}{2\, |\mu|^3}}\left(b(\zeta)+\frac{z}{b(\zeta)}\right)\bigg|\,\sigma_0\right) = \rme^{-\,\rme^{-z}}\,,
  \label{eq:emptyfi}
\end{align}
where
\begin{align}\label{def_b}
b(\zeta)=2\sqrt{{\cal W}\left(\frac{\zeta}{2\sqrt{\pi}}\right)} \quad, \quad \zeta=\rho\,\sqrt{\frac{\ell}{2\,|\mu|}}
\end{align}
and ${\cal W}(x)$ is called the Lambert function that satisfies ${\cal W}(x)\,\exp({\cal W}(x)) = x$. The distribution for the emptying time for a box of run-and-tumble particles (\ref{eq:emptyfi}) is therefore the same as the one for a box of Brownian particles. This is essentially because the emptying time will be dominated by the particles that start far away from the origin and such particles are well approximated by Brownian particles in the long-time limit when they will exit the box. This approximation might not hold for a different initial distribution. As pointed out in \cite{comtet2020last}, the Gumbel distribution in (\ref{eq:emptyfi}) appears as a consequence of taking the maximum of a large number of independent {\it but non-identically} distributed random variables (\ref{eq:Pempty}). This distribution is well-known in the field of extreme value statistics \cite{gumbel1958statistics} and we refer the reader to \cite{majumdar2020extreme} for a recent review on extreme value statistics.

\paragraph{Supercritical drift ($\mu<-1$).} For the case of a supercritical drift (see figure \ref{fig:survivalc} and \ref{fig:phase_diagram}), the survival probability $S_{\sigma_0}^*=1$ as the particle always moves in the same direction. Therefore the last-passage distribution (\ref{eq:lastpass}) simplifies to the expression of the first-passage distribution (\ref{eq:FPexprel}) found in section \ref{sec:sSs}. As expected, the first-passage and last-passage distributions coincide. We display our previous results (\ref{eq:Sstrongintro}) applied here to the last-passage distribution
\begin{subequations}
 \begin{align}
  L_+(x,t)  &= \left\{\begin{array}{ll}0\,, & t<t_m\,, \\
      \rme^{-t}\left(\delta\left(t-t_M\right)- \frac{1+\mu}{2}\,\sqrt{\frac{f(t,x)}{g(t,x)}}\,I_1\left[h(t,x)\right] +\frac{1-\mu}{2}\,I_0\left[h(t,x)\right]\right)\,, & t_m\leq t\leq t_M\,,\\
      0\, , & t>t_M\, .\end{array}\right. \\
     L_-(x,t)  &= \left\{\begin{array}{ll}0\,, & t<t_m\,,  \\
      \rme^{-t}\left( \delta\left(t-t_m\right)+\frac{1-\mu}{2}\,\sqrt{\frac{g(t,x)}{f(t,x)}}\,I_1\left[h(t,x)\right] -\frac{1+\mu}{2}\,I_0\left[h(t,x)\right]\right)\,, & t_m\leq t\leq t_M\,, \\
      0\, , &t>t_M\,  \;,\end{array}\right.
\end{align}
\label{eq:lasts}
\end{subequations}
where $t_m$, $t_M$, $f(t,x)$, $g(t,x)$ and $h(t,x)$ are given in (\ref{eq:tmfgh}). The analysis of the last-passage time is the same as the one done for the first-passage time in section \ref{sec:sSs}. The no-return probability is obtained as the cumulative density function of the last-passage time distributions (\ref{eq:lasts}). Performing similar steps as for the case of a subcritical drift (\ref{eq:piltlx}), we find that the large $x$ limit of the no-return probability (\ref{eq:lasts}) is given by
\begin{subequations}
   \begin{align}
   \Pi_{+}(x,t) &\sim 1-\int_t^\infty dt'\, \frac{1}{\sqrt{2\pi\,h(t',x)}}\left(\frac{1-\mu}{2}-\frac{1+\mu}{2}\sqrt{\frac{f(t',x)}{g(t',x)}}\right)\,\rme^{-t'+h(t',x)}\,  ,\\
   \Pi_{-}(x,t) &\sim 1-\int_t^\infty dt'\, \frac{1}{\sqrt{2\pi\,h(t',x)}}\left(\frac{1-\mu}{2}\sqrt{\frac{g(t',x)}{f(t',x)}}-\frac{1+\mu}{2}\right)\,\rme^{-t'+h(t',x)}\,   .
\end{align}
\label{eq:piltlxs}
\end{subequations}
These integrals can be evaluated using the saddle-point method and yields the same Gaussian result as for the case of a subcritical drift in (\ref{eq:noreturnerf}). The distribution for the emptying time is therefore the same as the one for the subcritical drift (\ref{eq:emptyfi}).

\subsection{Record statistics}
\begin{figure}[t]
    \centering
    \includegraphics[scale=0.5]{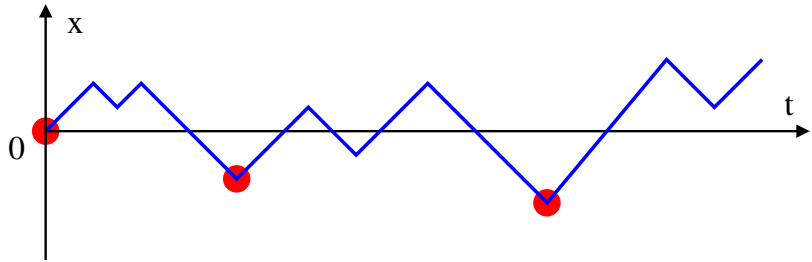}
    \caption{A typical run-and-tumble trajectory, starting at the origin, with its lower records marked in red.}
    \label{fig:records}
\end{figure}

Let us consider a single run-and-tumble particle in the presence of a drift $\mu$ and study the statistics of the number of records $R(t)$ as a function of time. We define the records as follows \cite{mori2020universal}. We consider a trajectory of the RTP up to time $t$ starting from $0$ and having $n$ tumblings. 
We mark the positions $\{0,\ell_1, \ell_1 + \ell_2 , \cdots, \ell_1 + \ell_2 + \cdots + \ell_n \}$ at the end of each tumbling of this trajectory. In this discrete sequence with $n+1$ entries, we say that an entry is a lower record if the position at the instant of the tumbling is lower than all the previous entries. Note that the number of tumblings $n$ is also a random variable for a given fixed $t$. Hence, the total number of records $R(t)$ is obtained by counting the number of records in every trajectory with $n$ tumblings and finally summing over all possible values of $n$ (see figure \ref{fig:records}).
A similar procedure has been used to define the number of records in continuous time random walks model (CTRW) \cite{sabhapandit2011record}. Note that, by symmetry, the lower records become upper records upon switching the sign of the drift $\mu$. In the absence of drift, the statistics of the number of lower records have been recently studied in \cite{mori2020universal}. Extending the approach devised in \cite{majumdar2008universal}, the authors in \cite{mori2020universal} have obtained the generating function of the average number of lower records $R(t)$ and found that the average number of records grows like
\begin{align}\langle R(t)\rangle \sim \frac{2\, \sqrt{t}}{\sqrt{\pi}} \,,\quad t\rightarrow \infty.
\label{eq:records0}
\end{align}
 In this section, we study how this result (\ref{eq:records0}) deviates from the square root growth in the presence of a drift. We restrict ourselves to the case of a subcritical drift $-1\!<\!\mu\!<\!1$ (see figure \ref{fig:survivalw} and \ref{fig:phase_diagram}) as a supercritical drift $\mu<-1$ would yield to the trivial result of records being broken at every tumbling. For simplicity, we will further assume that the particle starts in the state $\sigma_0=+1$.

\begin{figure}[t]
    \centering
    \includegraphics{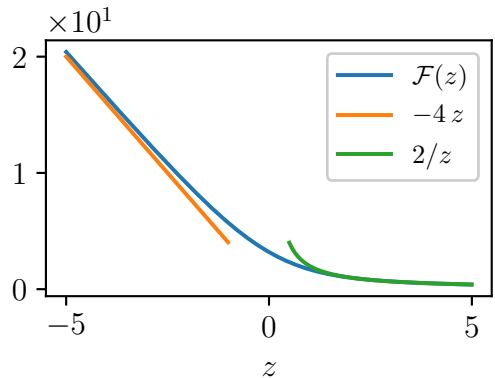}
    \caption{Plot of the scaling function $\mathcal{F}(z)$ (\ref{eq:scalingF}) that describes the scaling behavior of the average number of records $\langle R(t) \rangle$ in the scaling limit $t \to \infty$, $\mu \to 0$ keeping $\mu \sqrt{t}$ fixed. We have also indicated the asymptotic behaviors for $z \to - \infty$ and $z \to \infty$ as given in (\ref{F_rec}).}
    \label{fig:Mavg}
\end{figure}

To study the average number of records $\langle R(t)\rangle$, we rely again on the mapping to a discrete-time random walk developed in section \ref{sec:SA}. After $n$ tumbles, the average number of lower records $\langle R(t)\rangle_n$ done during a time $t$ can be written as
\begin{align}
  \langle R(t)\rangle_n = \int  d \vec{\ell}   \,P(\vec{\ell},n |\,t)\, \left(\sum_{j=0}^n \chi_j \right) \;,
  \label{eq:avgMt}
\end{align}
where $\chi_j$ is a binary variable which takes values $\chi_j=1$ if the position of the random walker at step $j$, i.e., $x_j = \ell_1 + \ell_2+\cdots+\ell_j $ is a record and $\chi_j=0$ otherwise. 
In (\ref{eq:avgMt}), $P(\vec{\ell},n |\,t)$ is the joint distribution of the run lengths and the number of tumblings $n$ within time $t$. We use the convention that the initial position is a record, as illustrated in figure \ref{fig:records}. Taking the Laplace transform of (\ref{eq:avgMt}) with respect to $t$ and summing over $n$ (following the same steps as in section \ref{sec:mapping}), we get
\begin{align}
  \langle\tilde R(s)\rangle =\sum_{n=0}^\infty \,  \langle\tilde R(s)\rangle_n =  \frac{1}{\gamma}\sum_{n=0}^\infty \,\left(\frac{\gamma}{\gamma+s}\right)^n\,\langle \mathcal{R}\rangle_n \label{eq:Mssum}\,,
\end{align}
where $\langle \mathcal{R}\rangle_n$ denotes the average number of records for a RW of $n$ steps and with jump distribution $\phi_s(\ell)$ (\ref{eq:stepdist0}). Therefore, on the right hand side, we recognize the generating function of the average number of records of the effective random walk $\sum_{n=0}^{\infty}\,\langle \mathcal{R}\rangle_n\, r^n $ evaluated at $r=\gamma/(\gamma+s)$. This generating function can be expressed in terms of the generating function of the survival probability $\bar{q}(x\!=\!0,\,r)$ presented in (\ref{eq:genfun}). The relation is given by \cite{majumdar2012record,godreche2017record}
\begin{align}
  \sum_{n=0}^{\infty}\,\langle \mathcal{R}\rangle_n \,r^n = \frac{1}{(1-r)^2\,\bar q(0,r)}\,.
\end{align}
Evaluating it at $r=\gamma/(\gamma+s)$ gives
\begin{align}
  \langle\tilde R(s)\rangle = \frac{1}{\gamma} \frac{1}{\left(1-\frac{\gamma}{\gamma+s}\right)^2\bar q\left(0,\frac{\gamma}{\gamma+s}\right)} \,.
\end{align}
Using explicitly the result for $\bar q(0,\gamma/(\gamma+s))$ given in (\ref{qbar0}), we get 
\begin{align}
  \langle\tilde R(s)\rangle = (v_0+\mu)\,\frac{(2 \tilde \gamma+s)(\lambda-\eta)}{\gamma\,s^2}\, ,
   \label{eq:Mavgs}
\end{align}
where we recall that $\tilde \gamma = \gamma/2$ and $\lambda$ and $\eta$ are given in (\ref{eq:lambda}) and (\ref{eq:eta}). For simplicity we set $v_0 = \tilde \gamma = 1$ and by performing the Laplace inversion we obtain the large $t$ behavior of the average number of lower records 
\begin{align}
  \langle R(t)\rangle \sim \left\{\begin{array}{lr}
    \frac{2\, (1+\mu)}{\mu}\, , & 0<\mu<1\, ,\\
    \frac{4\, \mu}{\mu-1}\, t + \frac{2\, (1+\mu^2)}{\mu\,(\mu-1)} \, , & -1<\mu<0\, .\end{array}\right.
    \label{eq:avgrecords}
\end{align}
For a positive subcritical drift $0\!<\!\mu\!<\!1$, the average number of lower records $\langle R(t)\rangle$ is constant (\ref{eq:avgrecords}). This is explained by the fact that, typically, the particle breaks a finite number of lower records in the beginning of the process while its position is still close to the negative axis, then has almost no chances to break newer records due to its position being drifted away from the negative axis. For a negative subcritical drift $-1\!<\!\mu\!<\!0$, $\langle R(t)\rangle$ has a linear growth (\ref{eq:avgrecords}) which is expected due to the drift pushing the particle towards breaking new lower records. Notice that the limit $\mu\rightarrow 0$ in the average number of lower records (\ref{eq:avgrecords}) does not match with the results for $\mu=0$ (\ref{eq:records0}). This indicates the existence of a scaling regime when $\mu \to 0$ and $t \to \infty$. Indeed comparing the result for $\mu=0$ in
(\ref{eq:records0}) and the result in (\ref{eq:avgrecords}) for $-1<\mu<0$, we anticipate a scaling form 
  \begin{align}\label{scaling_form_R}
    \langle R(t)\rangle \sim \sqrt{t}\, \mathcal{F}(z=\mu\,\sqrt{t})\,,
  \end{align} 
 where ${\mathcal F}(z)$ is a scaling function with asymptotic behaviors 
 \begin{eqnarray}\label{F_rec}
 {\cal F}(z) \sim
 \begin{cases}
 & -4 z \quad, \quad z \to -\infty \;,\\
 &\\
 & \dfrac{2}{\sqrt{\pi}} \quad, \quad \,\, z \to 0 \;, \\
 &\\
 & \dfrac{2}{z} \quad, \quad \;\,\;\; z \to + \infty \;. 
 \end{cases}
 \end{eqnarray}
These asymptotic behaviors ensure a smooth matching of (\ref{eq:records0}) and (\ref{eq:avgrecords}), with all three regimes
being part of the scaling form in (\ref{scaling_form_R}). We first insert this scaling form (\ref{scaling_form_R}) on the left hand 
side of (\ref{eq:Mavgs}) and evaluate the Laplace transform $\langle \tilde R(s)\rangle$. Then, on the right hand side of (\ref{eq:Mavgs})
we substitue the explicit forms of $\lambda$ and $\eta$ from (\ref{eq:lambda}) and (\ref{eq:eta}) and then take the scaling limit 
$s \to 0$, $\mu \to 0$ keeping the 
ratio $\mu/\sqrt{s}$ fixed (which corresponds to $\mu \to 0$, $t \to \infty$ keeping $\mu \sqrt{t}$ fixed). This gives an integral equation
for the scaling function ${\cal F}(z)$ which can be fortunately inverted explicitly leading to 
  \begin{align}
  \mathcal{F}(z) = 2\,\left(\sqrt{\frac{2}{\pi}}\rme^{-z^2/2} - z + \left(\frac{1}{z}+z\right)\text{erf}\left(\frac{z}{\sqrt{2}}\right)\right)\,.
  \label{eq:scalingF}
  \end{align}
 A plot of this scaling function ${\cal F}(z)$, together with its asymptotic behaviors, (\ref{F_rec}) is shown in figure \ref{fig:Mavg}. Interestingly,
 the same scaling function ${\cal F}(z)$, up to a multiplicative factor, coincides with the scaling function that describes
 the expected maximum of a Brownian motion in the presence of a drift $\mu$ over the time interval $[0,t]$ in the same scaling limit $\mu \to 0$, $t \to \infty$, keeping $z = \mu \sqrt{t}$
 fixed \cite{mounaix2018asymptotics}. This coincidence can be qualitatively understood by noting that the expected number of records $R_n$ of a random walk in the presence of a drift $\mu$ after $n$ steps is, for large $n$, proportional to the expected maximum of this random walk $M_n$, i.e. $\langle R_n \rangle \sim \alpha \, \langle M_n \rangle$ where $\alpha$ is independent of $n$ \cite{godreche2017record}. In the scaling limit $n \to \infty$, $\mu \to 0$ keeping $z = \mu \sqrt{n}$ fixed, it was shown \cite{mounaix2018asymptotics} that $\langle M_n \rangle$ is described by a scaling form similar to (\ref{scaling_form_R}), if $t$ is replaced by $n$, with the same scaling function ${\cal F}(z)$. The same scaling form thus also holds for $\langle R_n \rangle$. Applying this result to the random walk underlying the RTP (see figure \ref{fig:records}) explains qualitatively the scaling form obtained in (\ref{scaling_form_R}). 

  \section{Summary and conclusion}
  \label{sec:summary}
  In this paper, we first studied the survival probability of a run-and-tumble particle with an arbitrary velocity distribution using Spitzer's formula. We then focused on the two-state run-and-tumble particle in the presence of a drift. We distinguished the subcritical drift from the supercritical drift and obtained exact analytical results in both cases. In the former case, a comparison with the drifted Brownian motion was drawn and the main differences were highlighted. While the mean first-passage time coincides, the signature of activity of the run-and-tumble motion can be observed in the tail of the survival probability. In the latter case, we saw that a supercritical drift yields to a finite support for the first-passage time distribution, which has no diffusive equivalent. Additionally, it was shown that this distribution also coincides with the position distribution and the last-passage time distribution. The transition between subcritical and supercritical drift was discussed and several scaling regimes were found. Finally, we illustrated our results by applying them to the study of the emptying time of a box and the record statistics of a particle.
  
This work opens up several perspectives for further research. As suggested by our derivation of the survival probability using Spitzer's formula in section \ref{sec:SA}, we would like to generalize our results to higher dimensions. For instance, it would be interesting to extend the results on the convex hull of the run-and-tumble motion obtained in \cite{hartmann2020convex} for a run-and-tumble particle with an asymmetric velocity distribution. Another natural generalization is to replace the constant drift by a space dependent force \cite{singh2020run}, which can even be a random function in space as in the models of a particle moving in a random environment \cite{majumdar2002exact,dor2019ramifications}.

  \ack
  This work was partially supported by the Luxembourg National Research Fund (FNR) (App. ID 14548297).
  
\appendix
\section*{Appendix}

\section{Some properties of the telegraphic noise} \label{app:Poisson}

For pedagogical purposes, the telegraphic noise $\sigma(t)$ is briefly discussed here. During an infinitesimal time interval $dt$, the signal changes sign with probability $\tilde \gamma\, dt$ and remains constant with the complementary probability $1-\tilde \gamma\, dt$:
\begin{align}
  \sigma(t+dt) = \left\{\begin{array}{rl}\sigma(t)\, ,  & \text{with \; prob.~ }=1-\tilde \gamma\, dt\, , \\
  -\sigma(t)\, , &\text{with \; prob.~ } =\tilde \gamma\, dt\, . \end{array}\right. \label{eq:telegraphic_app}
\end{align}
The time $\tau$ between two consecutive switches is thus distributed according to an exponential distribution $p(\tau)=\tilde \gamma\,\rme^{-\tilde \gamma\,\tau}$. This can be seen from (\ref{eq:telegraphic_app}) by dividing $\tau$ into $n$ small intervals $dt=\tau/n$ during which the signal does not change sign and a last interval $d\tau$ during which the change of sign occurs. This yields
 \begin{align}
 p(\tau)\,d\tau= \lim_{n\rightarrow \infty} \left(1-\frac{\tilde \gamma \tau}{n}\right)^n\, \tilde \gamma\, d\tau = \tilde \gamma \, \rme^{-\tilde \gamma \tau}\, d\tau\, .\label{eq:tumblingdist}
 \end{align} 
More generally, in any time interval $[t_1,t_2]$ with $t_2>t_1$, it is easy to show that the distribution of the number of sign changes $m$ is given by a Poisson distribution
\begin{align} 
{\rm Pr.}(\text{``number of sign changes = $m$''}) = \frac{\tilde \gamma^m\,(t_2-t_1)^m}{m!}\, \rme^{-\tilde \gamma\,(t_2-t_1)}\,. \label{eq:poissondist}
 \end{align}
  In particular, the probability that $\sigma(t_1)$ has the same (respectively the opposite) sign as $\sigma(t_2)$ is given by the probabilities (\ref{eq:poissondist}) summed over even (respectively odd) values of $m$. The autocorrelation function $\langle\sigma(t_1)\sigma(t_2)\rangle$ is therefore given by:
 \begin{align}
   \langle \sigma(t_1)\sigma(t_2) \rangle &= \mathrm{Pr.}[\sigma(t_1)=\sigma(t_2)] + (-1)  \mathrm{Pr.}[\sigma(t_1)=-\sigma(t_2)]\, , \nonumber \\
   &= \sum_{m \text{ even}}\frac{\tilde \gamma^m(t_2-t_1)^m \rme^{-\tilde \gamma(t_2-t_1)}}{m!} - \sum_{m \text{ odd}}\frac{\tilde \gamma^m(t_2-t_1)^m \rme^{-(t_2-t_1)}}{m!}\, , \nonumber \\
   &= \rme^{-2\,\tilde \gamma\, (t_2-t_1)}\,. \label{eq:autocorell_app}
 \end{align}

\section{Derivation of the double Laplace transform of the survival probability of a $1d$ random walk}\label{app:spitzer}

We consider a discrete time random walker on the line, starting at the initial position $x \geq 0$ and jumping at each step
by a random length $\ell$ drawn from a normalized PDF $\phi_s(\ell)$, not necessarily symmetric. Let $q_n(x)$ denote the 
probability that the walker does not cross the origin up to step $n$. Given $\phi_s(\ell)$, can one calculate $q_n(x)$ for arbitrary
$x \geq 0$? In fact, this can be done by adapting a formula derived by Spitzer for the PDF of the maximum of a random walk
up to $n$ steps starting initially at the origin. This is the goal of this appendix. 

As a first step, we want to relate the survival probability $q_n(x)$ of the random walk with the jump distribution $\phi_s(\ell)$ with
the PDF of the maximum of a related random walk {\it starting at the origin}. Once we establish this relation, we can then directly 
use Spitzer's formula. Indeed, let us consider a random walk starting at the origin and performing jumps drawn from a normalized
PDF $\phi_s(-\ell)$. Let $y_k$ denote the position of the walker at step $k$ with $y_0=0$. Let $M_n$ denote the maximum up
to step $n$, i.e., $M_n = \max\{y_1, y_2, \cdots, y_ n\}$. Let $C(M,n) = {\rm Pr.}(M_n \leq M)$ be the cumulative distribution of the maximum up to step $n$.
Since the event that ``the maximum $M_n$ is smaller than $M$'' is equivalent to the one where all the positions up to step $n$ are smaller
than $M$, we can write  
\begin{eqnarray}\label{CMn}
C(M,n) = {\rm Pr.}[y_1 \leq M, y_2 \leq M, \cdots, y_n \leq M |y_0=0] \;.
\end{eqnarray}
Let us now make the change of variable $x_k = M-y_k$. Then we see that $x_k$ also denotes the position of a random walker at step $k$, 
starting initially at $x_0 = M-y_0 = M$ and with jumps distributed via $\phi_s(\ell)$. Hence from (\ref{CMn}) we obtain 
\begin{equation} \label{relation_CMn}
C(M,n) \Big \vert_{\rm with \; jump \; distribution \; \phi_s(-\ell)} = \theta(M)\,q_n(M) \Big \vert_{\rm with \; jump \; distribution \; \phi_s(\ell)}
\end{equation}
where we recall that $\theta(M)$ is the Heaviside theta function. 

Our next goal is to relate the Laplace transform of the PDF of the maximum with that of $q_n(M)$. The PDF of the maximum $M_n$ is simply 
$P(M,n) = \partial C(M,n)/\partial M$. Taking the Laplace transform with respect to $M$ and using integration by parts one gets
\begin{equation}\label{LT_CM}
\int_0^\infty dM\, \e^{-u\,M} \, P(M,n)  \Big \vert_{\phi_s(-\ell)}= u \, \int_0^\infty dM \, \e^{-u M}\,  q_n(M) \Big \vert_{\phi_s(\ell)}
\end{equation}
Spitzer derived an expression for the generating function of the quantity on the left hand side of (\ref{LT_CM}) for arbitrary jump distribution $\phi_s(-\ell)$. This formula reads \cite{spitzer1957wiener}
\begin{eqnarray}\label{Spitzer_app} 
&&\sum_{n=0}^\infty r^n \int_0^\infty dM\, \e^{-u\,M} \, P(M,n) \Big \vert_{\phi_s(-\ell)} \nonumber \\
&&= \frac{1}{1-r} \exp\left(\frac{1}{2\pi} \int_0^r d\tau \int_{-\infty}^\infty \frac{u}{k(k-{\rm i} u)} \frac{\hat \phi_s(k)-1}{(1-\tau)(1-\tau \hat \phi_s(k))}\,dk\right) \;,
\end{eqnarray}     
where 
\begin{eqnarray} \label{Fourier_phi}
\hat \phi_s(k) = \int_{-\infty}^\infty d\ell \, \phi_s(\ell)\, \e^{-{\rm i} k\, \ell}
\end{eqnarray}
is the Fourier transform of the jump distribution. In the integral over $k$ in (\ref{Spitzer_app}) one should interpret the integrand $1/k$ as $1/(k- {\rm i}\epsilon)$ where $\epsilon$ is 
a regulator and eventually take the $\epsilon \to 0$ limit after performing the integral over $k$. Indeed, this formula (\ref{Spitzer_app}) can be further simplified using a trick developed
in \cite{mounaix2018asymptotics} (see Appendix C) where it was used for a specific distribution (where the drift is just a constant). However, the same trick can used for arbitrary jump distribution $\phi_s(\ell)$ as outlined below.

We start with the right hand side of (\ref{Spitzer_app})  and denote it simply by RHS. We 
first perform the integral over $\tau$ explicitly. This gives
\begin{eqnarray} \label{rhs}
{\rm RHS} = \frac{1}{1-r} \exp\left( T_1 + T_2\right) 
\end{eqnarray}
where
\begin{eqnarray}
T_1 &=& \frac{u \ln(1-r)}{2 \pi} \int_{-\infty}^\infty \frac{dk}{(k-{\rm i}\epsilon)(k-{\rm i}u)} \label{T1}\\
T_2 &=& - \frac{u}{2\pi} \int_{-\infty}^\infty \frac{dk}{(k-{\rm i}\epsilon)(k-{\rm i}u)} \ln(1-r \hat\phi_s(k)) \label{T2} \;.
\end{eqnarray}
It turns out that, conveniently, $T_1 = 0$ (it simply follows by computing the residues at the two poles $k={\rm i} \epsilon$ and $k={\rm i} u$ which cancel each other exactly). Furthermore, $T_2$ can be simplified also by using $u/(k(u + {\rm i} k)) = 1/k - {\rm i}/(u+{\rm i}k)$. Using this expression of $T_2$ given in (\ref{T2}) we obtain an explicit expression for RHS in (\ref{rhs}). Finally, plugging this expression in (\ref{LT_CM}) we get our final simplified formula
\begin{equation}\label{Spitzer_app2} 
\sum_{n=0}^\infty r^n \int_0^\infty dM\, \e^{-u\,M} \, q_n(M) \Big \vert_{\phi_s(\ell)} = \frac{1}{u(1-r)} \exp\left( \frac{\Phi_s(0,r)-\Phi_s(u,r)}{2 \pi} \right) \;,
\end{equation}
where $\Phi_s(u,r)$ is given by
\begin{align}
  \Phi_s(u,r) = \int_{-\infty}^\infty \frac{dk}{u+\rmi k} \ln(1-r\,\hat\phi_s(k))\, \label{eq:Xi_app} \;.
\end{align}
The formula (\ref{Spitzer_app2}) together with (\ref{eq:Xi_app}) yields the result given in (\ref{eq:spitzer})-(\ref{FourierPhi_s}) in the text. 

Finally, using the result in (\ref{eq:Sqn}) we get
\begin{equation} \label{double_LT_S_app}
\int_0^\infty dx \, \int_0^\infty dt \, S(x,t) \e^{-st - x u } = \frac{\gamma + s}{\gamma \, u  \, s} \exp\left( \frac{\Phi_s(0,\frac{\gamma}{\gamma+s})-\Phi_s(u,\frac{\gamma}{\gamma+s})}{2 \pi} \right) - \frac{1}{\gamma \, u} 
\end{equation}
Using the expression for $\Phi_s(u,r)$ in (\ref{eq:xiua}) yields the expression in (\ref{eq:gformula}) in the text. 

\section{Survival probability conditioned on the sign of the initial velocity}\label{app:spitzerpm}

In this Appendix, we study the survival probabilities up to time $t$ for a general RTP with a velocity
distribution $W(v)$, conditioned to start from $x$ with a positive (respectively negative) velocity: we will denote 
this probability $S_\uparrow(x,t)$ (respectively $S_\downarrow(x,t)$). More formally, they can be written as
\begin{subequations} 
\begin{eqnarray}
S_{\uparrow}(x,t) &=& {\rm Pr.}\left[x(\tau) \geq 0 \,, \forall \tau \in [0,t] \, \Big | \, x(t=0) = x, \, \dot x(0) > 0 \right] \;, \label{eq:def_Sup} \\
S_{\downarrow}(x,t) &=& {\rm Pr.}\left[x(\tau) \geq 0 \,, \forall \tau \in [0,t] \, \Big | \, x(t=0) = x, \,\dot x(0) < 0 \right] \label{eq:def_Sdown}.
\end{eqnarray}
\end{subequations}
Of course, $S(x,t)$ can be obtained from $S_{\uparrow}(x,t)$ and $S_{\downarrow}(x,t)$ via the relation
\begin{eqnarray} \label{relationS}
S(x,t) = {\rm Pr.}(v>0) S_{\uparrow}(x,t) + {\rm Pr.}(v<0) S_{\downarrow}(x,t) \;,
\end{eqnarray} 
where ${\rm Pr.}(v>0) = \int_0^\infty dv \, W(v)$ and ${\rm Pr.}(v>0) = \int_{-\infty}^0 dv \, W(v)$.

As done for the full survival probability $S(x,t)$ in section \ref{sec:mapping}, we use the mapping between the RTP 
and a discrete-time random walk to relate the conditioned survival probabilities $S_{\uparrow}(x,t)$ and $S_{\downarrow}(x,t)$ to conditioned
survival probabilities in the random walk problem. Indeed, let us consider the discrete-time random walk with jump distribution 
$\phi_s(\ell)$ starting from the initial position $x$. We define $q^{+}_n(x)$ (respectively $q^{-}_n(x)$) as the survival probability up to step 
$n$ given that the first jump is positive (respectively negative). Following the same reasoning as explained in section \ref{sec:mapping} leading to 
the relation in (\ref{eq:Sqn}), we have here
\begin{subequations}
\begin{eqnarray}
\tilde S_{\uparrow}(x,s) = \int_{0}^{\infty} dt \, S_{\uparrow}(x,s) \e^{-st}  = \frac{1}{\gamma} \bar{q}^{+}\left(x , \frac{\gamma}{\gamma+s} \right) - \frac{1}{\gamma} \label{Sup_LT}\\
\tilde S_{\downarrow}(x,s)=\int_{0}^{\infty} dt\, S_{\uparrow}(x,s) \e^{-st}  = \frac{1}{\gamma} \bar{q}^{-}\left(x , \frac{\gamma}{\gamma+s} \right) - \frac{1}{\gamma} \label{Sdown_LT} \;.
\end{eqnarray} 
\end{subequations}
in terms of the generating functions $\bar q^{\pm}(x,r)$
\begin{equation} \label{qpm_GF}
\bar q^{\pm}(x,r) = \sum_{n=0}^\infty r^n q_n^{\pm}(x) \;.
\end{equation}

To compute these generating functions $\bar q^{\pm}(x,r)$, we notice that $q^\pm_{n}(x)$ satisfies the following backward equations 
\begin{subequations}
 \begin{align}
   q_{n+1}^+(x) = \frac{1}{\int_0^{\infty}\,d\ell\,\phi_s(\ell)} \int_0^\infty\,d\ell\,\phi_s(\ell)\,q_n(x+\ell)\,,\label{eq:qnpp}
 \end{align}
and
  \begin{align}
   q_{n+1}^-(x) = \frac{1}{\int_{-\infty}^{0}\,d\ell\,\phi_s(\ell)} \int_{-x}^0\,d\ell\,\phi_s(\ell)\,q_n(x+\ell)\,,\label{eq:pnpm}
 \end{align}
 \label{eq:qnp}
\end{subequations}
which are easily derived by considering what happens at the first step of the random walk. Note that in the equation for $q^-_{n+1}(x)$ in (\ref{eq:pnpm}) the integral over $\ell$ is bounded from below by $-x$ because the particle cannot cross zero at the first step and hence $x + \ell > 0$ (such a condition does not exist for $q^+_{n+1}(x)$ in (\ref{eq:pnpm}) since in this case the condition $x + \ell >0$ is automatically satisfied for all $\ell >0$). Note also that, from the definition of $\phi_s(\ell)$ in (\ref{eq:stepdist0}), one easily obtains
\begin{subequations}
\begin{eqnarray}
\int_0^\infty d\ell \, \phi_s(\ell) &=& \int_0^\infty dv\, W(v) = {\rm Pr.}(v>0) \,, \label{vpos}\\
\int_{-\infty}^0 d\ell \, \phi_s(\ell) &=& \int_{-\infty}^0 dv \, W(v)  = {\rm Pr.}(v<0) \;,  \label{vine}
\end{eqnarray}
\end{subequations}
as expected. 

To proceed, we notice that (\ref{eq:pnpm}) can be written, by performing the change of variable $\ell \to -\ell$, as
\begin{equation}\label{convolution}
q_{n+1}^{-}(x) = \frac{1}{{\rm Pr.}(v<0)}\int_0^x d\ell \, \phi_s(-\ell) q_{n}(x-\ell) \;,
\end{equation}
Interestingly, this equation (\ref{convolution}) has a convolution structure which we can exploit to solve it. First we introduce the generating function
\begin{eqnarray}\label{gf_qm}
\bar q^-(x,r) = \sum_{n=0}^\infty r^n q^-_n(x)  \;.
\end{eqnarray} 
The relation (\ref{convolution}) thus yields 
\begin{eqnarray} \label{gf_qm2}
\bar q^-(x,r) = 1 + \frac{r}{{\rm Pr.}(v<0)} \int_0^x d\ell \, \phi_s(-\ell) \, \bar q(x-\ell,r) 
\end{eqnarray}
where we have used that $q_{0}^{-}(x) = 1$. Let us denote $\tilde \phi^-_s(u)$ the Laplace transform of $\phi_s(-\ell)$
\begin{eqnarray}\label{LTF}
\tilde \phi^-_s(u) = \int_0^\infty \,d\ell  \, \e^{-u \ell} \phi_s(-\ell) \;.
\end{eqnarray}
By taking the Laplace transform of (\ref{gf_qm2}) with respect to $x$ one finds
\begin{eqnarray} \label{lt_gfqm2}
\hspace*{-2cm}\int_0^\infty dx \, \e^{-u x} \bar q^-(x,r) &=& \frac{1}{u} + \frac{r\,\tilde \phi^-_s(u)}{{\rm Pr.}(v<0)}  \int_0^\infty dx \, \e^{- u x} \bar q(x,r) \nonumber \\
\hspace*{-2cm}&=& \frac{1}{u} +  \frac{1}{{\rm Pr.}(v<0)}\frac{r\, \tilde \phi^-_s(u)}{u(1-r)} \exp\left(\frac{\Phi_s(0,r)-\Phi_s(u,r)}{2\pi} \right)
\end{eqnarray}
where in the second line we have used the result in (\ref{Spitzer_app2}) in terms of $\Phi_s(u,r)$ given in~(\ref{eq:Xi_app}).  Finally, using the relation (\ref{Sdown_LT}) together with (\ref{lt_gfqm2}) we obtain
\begin{equation}\label{doubleLT_Sdown}
\int_{0}^{\infty} dx\, \int_{0}^{\infty} dt\, S_{\downarrow}(x,t) \e^{-ux-st} =
 \frac{1}{{\rm Pr.}(v<0)}\frac{\tilde \phi^-_s(u)}{u\,s} \exp\left(\frac{\Phi_s(0,\frac{\gamma}{\gamma+s})-\Phi_s(u,\frac{\gamma}{\gamma+s})}{2\pi} \right)
\end{equation}
Finally, using the relation (\ref{relationS}) together with \ref{doubleLT_Sdown} and (\ref{double_LT_S_app}) we get
\begin{eqnarray}\label{doubleLT_Sup}
\hspace*{-2cm}&&\int_{0}^{\infty} dx\, \int_{0}^{\infty} dt\, S_{\uparrow}(x,t) \e^{-ux-st} = \nonumber \\
\hspace*{-2cm}&&=\frac{1}{{\rm Pr.}(v>0)} \left[\frac{1}{us} \left(\frac{\gamma+s}{\gamma} - \tilde \phi_s^{-}(u) \right) \exp\left(\frac{\Phi_s(0,\frac{\gamma}{\gamma+s})-\Phi_s(u,\frac{\gamma}{\gamma+s})}{2\pi} \right) - \frac{1}{\gamma u}\right] \;.
\end{eqnarray}

\section{Transition kernel}
\label{app:kernel}

We compute the transition kernel of a run-and-tumble particle in the presence of a drift. The approach used here is applicable to both supercritical and subcritical drift (see figure \ref{fig:survival}). The transition kernel $P(y,t,\sigma|x,\sigma(0)=\sigma_0)$ is the probability distribution that the particle is in a state $\sigma$ at position $y$ at a time $t$, given that it started at $x$ in a state $\sigma_0$. Because of translational invariance, we have that:
\begin{align}
  P(y,t,\sigma|x,\sigma_0) = P(y-x,t,\sigma|x=0,\sigma_0)\,, \label{eq:translinv}
\end{align}
 and we will use the shorthand notation $P(y,t,\sigma|\sigma_0)\equiv P(y,t,\sigma|x=0,\sigma_0)$. One can show that the transition kernel satisfies the forward master equations
\begin{subequations}
\begin{align}
    \partial_t P(y,t,+|\sigma_0) &= - (1+\mu)\partial_y P(y,t,+|\sigma_0)-  P(y,t,+|\sigma_0)+P(y,t,-|\sigma_0)\, , \\
      \partial_t P(y,t,-|\sigma_0) &= - (-1+\mu)\partial_y P(y,t,-|\sigma_0)+  P(y,t,-|\sigma_0)- P(y,t,+|\sigma_0) \, ,
\end{align}
\label{eq:propeq}
\end{subequations}
with the initial condition
\begin{align}
P(y,t=0,\sigma|\sigma_0)&=\delta_{\sigma_0,\sigma}\delta(y)\,,
\end{align} where we use a different notation to distinguish a Kronecker delta  $\delta_{\cdot,\cdot}$ from a Dirac delta $\delta(\cdot)$. To solve the system (\ref{eq:propeq}), it is convenient to go in a reference frame that moves along with the drift $\mu$ to cancel the effect of the drift. After a Galilean transform $y\rightarrow y- \mu\,t$, the system (\ref{eq:propeq}) reads
\begin{subequations}
\begin{align}
    \partial_t P(y,t,+|\sigma_0) &= - \partial_y P(y,t,+|\sigma_0)-  P(y,t,+|\sigma_0)+P(y,t,-|\sigma_0)\, , \\
      \partial_t P(y,t,-|\sigma_0) &= \partial_y P(y,t,-|\sigma_0)+  P(y,t,-|\sigma_0)- P(y,t,+|\sigma_0) \, ,
\end{align}
\end{subequations}
and the initial condition remains
\begin{align}
P_{\sigma_0}(y,t=0,\sigma|\sigma_0)&=\delta_{\sigma_0,\sigma}\,\delta(y)\label{eq:bcprop}\,.
\end{align}
 We solve this system in the Laplace domain. Noting that 
 \begin{subequations}
\begin{align}
  \mathcal{L}[\partial_t P(y,t,\sigma|\sigma_0)](s) &= \int_0^\infty dt\,  \partial_t P(y,t,\sigma|\sigma_0) \rme^{-st}\, ,\\
  &= s\int_0^\infty dt\,   P(y,t,\sigma|\sigma_0) \rme^{-st} - [P(y,t,\sigma|\sigma_0) \rme^{-st}]^{t=\infty}_{t=0}\, ,\\
  &= s\int_0^\infty dt\,   P(y,t,\sigma|\sigma_0) \rme^{-st} - \delta_{\sigma_0,\sigma}\,\delta(y)\, ,
\end{align}
 \end{subequations}
where we integrated by parts to go to the second line and used the boundary condition (\ref{eq:bcprop}) to go to the last line, we find that the system (\ref{eq:propeq}) writes in Laplace domain
\begin{subequations}
\begin{align}
   [\partial_y +1+s] \tilde P(y,s,+|\sigma_0) &= \tilde P(y,s,-|\sigma_0) + \delta_{\sigma_0,+}\delta(y)\, , \\
[-\partial_y +1+s] \tilde P(y,s,-|\sigma_0) &=\tilde P(y,s,+|\sigma_0) + \delta_{\sigma_0,-}\delta(y) \, .
\end{align}
\label{eq:pmes}
\end{subequations}
To solve the system (\ref{eq:pmes}), we first restrict $y$ to be either into $]-\infty,0[$ or $]0,\infty[$. On these intervals, the delta term is not present and the system reduces to
\begin{subequations}
\begin{align}
   [\partial_y +1+s] \tilde P(y,s,+|\sigma_0) &= \tilde P(y,s,-|\sigma_0)  , \\
[-\partial_y +1+s] \tilde P(y,s,-|\sigma_0) &=\tilde P(y,s,+|\sigma_0) \, .
\end{align}
\label{eq:Pwdelta}
\end{subequations}
 Then, we decouple the equations by applying the bracketed differential operator from the first equation to the second equation, and conversely:
\begin{subequations}
\begin{align}
   [\partial_y +1+s][-\partial_y +1+s] \tilde P(y,s,+|\sigma_0) &=\tilde P(y,s,+|\sigma_0) \, , \\
[-\partial_y +1+s][\partial_y +1+s]\tilde P(y,s,-|\sigma_0)  &=\tilde P(y,s,-|\sigma_0)  \, .
\end{align}
\label{eq:pmesd}
\end{subequations}
We notice that $\tilde P(y,s,+|\sigma_0)$ and $\tilde P(y,s,-|\sigma_0)$ satisfy the same telegraphic equation, whose general solution is 
\begin{align}
A\, \rme^{\lambda_0 \,y} + B\, \rme^{-\lambda_0\, y}\,,
\end{align}
 where $\lambda_0=\sqrt{s(s+2)}$ and $A$ and $B$ are integration constants. In total, there are $8$ integration constants to fix for $\tilde P(y,s,+|\sigma_0)$ and $\tilde P(y,s,-|\sigma_0)$ split into two sub-domains $]-\infty,0\,[$ and $]\,0,\infty[$. We use the fact that the solution must be bounded at $y\rightarrow \pm \infty$ to eliminate four of them:
 \begin{subequations}
 \begin{align}
   \tilde P(y,s,+|\sigma_0) &=\left\{\begin{array}{ll} A\, \rme^{-\lambda_0\, y}\, ,& y>0\, , \\ B\, \rme^{\lambda_0\, y}\, ,& y<0 \, ,\end{array}\right.  \\
 \tilde P(y,s,-|\sigma_0) &=\left\{\begin{array}{ll} C\, \rme^{-\lambda_0\, y}\, ,& y>0\, , \\ D\, \rme^{\lambda_0\, y}\, ,& y<0\, . \end{array}\right.\label{eq:Pmtoinj}
\end{align}
 \end{subequations}
Then, we inject (\ref{eq:Pmtoinj}) into (\ref{eq:Pwdelta}) to relate $C$ with $A$ and $D$ with $B$:
  \begin{subequations}
 \begin{align}
   \tilde P(y,s,+|\sigma_0) &=\left\{\begin{array}{ll} A\, \rme^{-\lambda_0\, y}\, ,& y>0\, , \\ B\, \rme^{\lambda_0\, y}\, ,& y<0\, , \end{array}\right.  \\
 \tilde P(y,s,-|\sigma_0) &=\left\{\begin{array}{ll} (-\lambda_0+1+s)A\, \rme^{-\lambda_0\, y}\, ,& y>0\, , \\ (\lambda_0+1+s)B\, \rme^{\lambda_0 \,y}\, ,& y<0 \, .\end{array}\right.
\end{align}
 \end{subequations}

 Finally, we impose a 'continuity condition' between the two sub-domains by integrating (\ref{eq:pmes}) around a small volume element centered on the origin. This 'continuity condition' writes
  \begin{subequations}
 \begin{align}
   P(y=0^+,s,+|\sigma_0) - P(y=0^-,s,+|\sigma_0) &= \delta_{\sigma_0,+}\,, \\
      P(y=0^+,s,-|\sigma_0) - P(y=0^-,s,-|\sigma_0) &=- \delta_{\sigma_0,-}\,.
 \end{align}
  \end{subequations}
Upon applying it, we find
\begin{subequations}
 \begin{align}
   \tilde P_{+}(y,s,+) &=  \tilde P_{-}(-y,s,-) = \frac{1+s+\text{sign}(y)\,\lambda_0}{2\, \lambda_0} \rme^{-\lambda_0 \,|y|}\, , \\
 \tilde P_{+}(y,s,-) &=\tilde P_{-}(-y,s,+)=\frac{1}{2\, \lambda_0} \rme^{-\lambda_0\, |y|}\, .
\end{align}
\label{eq:proplaps}
\end{subequations}
Finally, we carefully invert the Laplace transforms (\ref{eq:proplaps}). We find
\begin{align}\mathcal{L}^{-1}\left[\frac{1}{\lambda_0}\rme^{-\lambda_0 |y|}\right](t) &=  \mathcal{L}^{-1}\left[\frac{1}{\sqrt{s(2+s)}}\rme^{-|y|\sqrt{s(2+s)} }\right](t)\, ,\\
&=  \mathcal{L}^{-1}\left[\frac{1}{\sqrt{(s+1)^2-1}}\rme^{-|y|\sqrt{(s+1)^2-1} }\right](t)\, ,\\
&=  \rme^{-t}\mathcal{L}^{-1}\left[\frac{1}{\sqrt{s^2-1}}\rme^{-|y|\sqrt{s^2-1} }\right](t)\, ,\\
&= \left\{\begin{array}{ll} 0\,, & t<|y|\,,\\ \rme^{-t}\,I_0(\sqrt{t^2-y^2})\,, & t\geq |y|\,.
\end{array}\right.
\label{eq:lapexp0}
\end{align}
where we completed the square to go to the second line, used the Laplace transform property (\ref{eq:lapshift}) to go to the next one and finally used formula (36) from the integral tables \cite{bateman1954tables} to perform the last step. We also have that
\begin{align}\mathcal{L}^{-1}\left[\rme^{-\lambda_0\, |y|}\right](t) &= -\text{sign}(y)\,\partial_y \mathcal{L}^{-1}\left[\frac{1}{\lambda_0}\rme^{-\lambda_0\, |y|}\right](t) \, , \\
&= \left\{\begin{array}{ll} 0\,, & t<|y|\,, \\
\rme^{-t}\left(\text{sign}(y)\,\delta(t-|y|)+\frac{y}{\sqrt{t^2-y^2}}I_1(\sqrt{t^2-y^2})\right)\,, & t\geq |y|\,,
\end{array}\right.
\end{align}
where we took the derivative of (\ref{eq:lapexp0}) to go from the first line to the second one. The Dirac delta function appeared from the derivative of the Heavyside function $\Theta(t-|y|)$ that represents the constraint $t\geq |y|$ in (\ref{eq:lapexp0}). Furthermore, we have that
\begin{align}
\mathcal{L}^{-1}\left[\frac{s}{\lambda_0}\rme^{-\lambda_0 |y|}\right](t) &= \partial_t \mathcal{L}^{-1}\left[\frac{1}{\lambda_0}\rme^{-\lambda_0 |y|}\right](t) \\
 &= \left\{\begin{array}{ll} 0\,, & t<|y|\,, \\
\rme^{-t}\left(\delta(t-|y|)-I_0(\sqrt{t^2-y^2}) +\frac{t}{\sqrt{t^2-y^2}}\, I_1(\sqrt{t^2-y^2})\right)\,, & t\geq |y|\,.
\end{array}\right.
\end{align}
Combining these results, we find
\begin{subequations}
 \begin{align}
    P(y,t,+|+) &=   P(-y,t,-|-) = \left\{\begin{array}{ll}0\, , & t<|y|\,,\\\rme^{-t}\left(\delta(t-y)+\frac{\sqrt{t+y}}{\sqrt{t-y}}\,I_1(\sqrt{t^2-y^2})/2\right)\, ,& t\geq|y|\,, \\
   \end{array}\right. \\
P(y,t,-|+) &=P(-y,t,+|-)= \left\{\begin{array}{ll} 0\,, & t<|y|\,,\\ \rme^{-t}\, I_0(\sqrt{t^2-y^2})/2\,, & t\geq |y|\,.
\end{array}\right.
\end{align}
\end{subequations}
Finally, switching back to the original reference frame $y\rightarrow y+ \mu\,t$ gives
\begin{subequations}
 \begin{align}
    P(y,t,+|+) &= \left\{\begin{array}{ll} 0\, , & y<y_m\, \text{ or }\, y>y_M\,,\\\rme^{-t}\left(\delta(t-y+\mu t)+\frac{\sqrt{f(t,-y)}}{\sqrt{g(t,-y)}}\,I_1(h(t,-y))/2\right)\, , & y_m\leq y\leq y_M\, , \\
   \end{array}\right. \\
   P(y,t,-|+) &= \left\{\begin{array}{ll} 0\,, & y<y_m\, \text{ or }\, y>y_M\,,\\ \rme^{-t}\,I_0(h(t,-y))/2\,, & y_m\leq y\leq y_M\,
\end{array}\right. \\
  P(y,t,-|-) &= \left\{\begin{array}{ll}0\, , & y<y_m\, \text{ or }\, y>y_M\,,\\\rme^{-t}\left(\delta(t+y-\mu t)+\frac{\sqrt{g(t,-y)}}{\sqrt{f(t,-y)}}\,I_1(\sqrt{h(t,-y)})/2\right)\, ,& y_m\leq y\leq y_M\,, \\
   \end{array}\right. \\
P(y,t,+|-) &= \left\{\begin{array}{ll} 0\,, & y<y_m\, \text{ or }\, y>y_M\, ,\\ \rme^{-t}\,I_0(h(t,-y))/2\,, & y_m\leq y\leq y_M\, ,
\end{array} \right.
\end{align}
\label{eq:kernel}
\end{subequations}
where $y_m=(\mu-1)\,t$, $y_M=(\mu+1)\,t$ and $f(t,y)$, $g(t,y)$, $h(t,y)$ are given in (\ref{eq:tmfgh}).

\section{Laplace transform of the transition kernel}
\label{app:kernels}
We compute the Laplace transform of the transition kernel of a run-and-tumble particle in the presence of a subcritical drift. The approach used here is the same as the one in \ref{app:kernel} except that we do not perform the change of reference frame in order to obtain the Laplace transform of the kernel taking the drift into account. As in \ref{app:kernel}, we use the translation invariant notation $P(y,t,\sigma|x,\sigma_0)=P(y-x,t,\sigma|x=0,\sigma_0)$ (\ref{eq:translinv}). Following the same reasoning as in the beginning of \ref{app:kernel}, we find that the Fokker-Plank equations in Laplace domain are 
\begin{subequations}
\begin{align}
   [(\mu+1)\partial_y +1+s] \tilde P(y,s,+|\sigma_0) &= \tilde P(y,s,-|\sigma_0) + \delta_{\sigma_0,+}\delta(y)\, , \\
[(\mu-1)\partial_y +1+s] \tilde P(y,s,-|\sigma_0) &=\tilde P(y,s,+|\sigma_0) + \delta_{\sigma_0,-}\delta(y) \, .
\end{align}
\label{eq:pmesmu}
\end{subequations}
To solve them, we first restrict $y$ to be either into $]-\infty,0[$ or $]0,\infty[$. On these intervals, the delta term is not present:
\begin{subequations}
\begin{align}
   [(\mu+1)\partial_y +1+s]\, \tilde P(y,s,+|\sigma_0) &= \tilde P(y,s,-|\sigma_0)  , \\
[(\mu-1)\partial_y +1+s]\, \tilde P(y,s,-|\sigma_0) &=\tilde P(y,s,+|\sigma_0) \, .
\end{align}
\label{eq:Pwdeltamu}
\end{subequations}
 Then, we decouple the equations by applying the bracketed differential operator from the first equation to the second equation, and conversely:
\begin{subequations}
\begin{align}
   [(\mu+1)\partial_y +1+s][(\mu-1)\partial_y +1+s]\, \tilde P(y,s,+|\sigma_0) &=\tilde P(y,s,+|\sigma_0) \, , \\
[(\mu-1)\partial_y +1+s][(\mu+1)\partial_y +1+s]\,\tilde P(y,s,-|\sigma_0)  &=\tilde P(y,s,-|\sigma_0)  \, .
\end{align}
\label{eq:pmesdmu}
\end{subequations}
We notice that $\tilde P(y,s,+|\sigma_0)$ and $\tilde P(y,s,-|\sigma_0)$ satisfy the same telegraphic equation, whose general solution is 
\begin{align}
A \, \rme^{(\eta-\lambda)\,y} + B \, \rme^{(\eta+\lambda)\,y}\,,
\end{align}
 where $\lambda$ and $\eta$ are given in (\ref{eq:lambdaeta}) and $A$ and $B$ are integration constants. In total, there are $8$ integration constants to fix for $\tilde P(y,s,+|\sigma_0)$ and $\tilde P(y,s,-|\sigma_0)$ split into two sub-domains $]-\infty,0[$ and $]0,\infty[$. We use the fact that the solution must be bounded at $y\rightarrow \pm \infty$ to eliminate four of them:
  \begin{subequations}
 \begin{align}
   \tilde P(y,s,+|\sigma_0) &=\left\{\begin{array}{ll} A\, \rme^{(\eta-\lambda)\,y}\, ,& y>0\, , \\ B\, \rme^{(\eta+\lambda)\,y}\, ,& y<0 \, ,\end{array}\right.  \\
 \tilde P(y,s,-|\sigma_0) &=\left\{\begin{array}{ll} C\, \rme^{(\eta-\lambda)\,y}\, ,& y>0\, , \\ D\, \rme^{(\eta+\lambda)\,y}\, ,& y<0\, . \end{array}\right.\label{eq:pmap}
\end{align}
 \end{subequations}
Then, we inject (\ref{eq:pmap}) into (\ref{eq:Pwdeltamu}) to relate $C$ with $A$ and $D$ with $B$:
 \begin{subequations}
 \begin{align}
    \tilde P(y,s,+|\sigma_0) &=\left\{\begin{array}{ll} A\, \rme^{(\eta-\lambda)\,y}\, ,& y>0\, , \\ B\, \rme^{(\eta+\lambda)\,y}\, ,& y<0 \, ,\end{array}\right.  \\
 \tilde P(y,s,-|\sigma_0) &=\left\{\begin{array}{ll} \left[(\mu+1)(\eta-\lambda)+1+s\right]A\, \rme^{(\eta-\lambda)\,y}\, ,& y>0\, , \\ 
  \left[(\mu+1)(\eta+\lambda)+1+s\right] B\, \rme^{(\eta+\lambda)\,y}\, ,& y<0 \, ,\end{array}\right.  
\end{align}
 \end{subequations}
 Finally, we impose a 'continuity condition' between the two sub-domains by integrating (\ref{eq:pmesmu}) around a small volume element centered on the origin. This condition is
 \begin{subequations}
 \begin{align}
   P(y=0^+,s,+|\sigma_0) - P(y=0^-,s,+|\sigma_0) &= \frac{\delta_{\sigma_0,+}}{\mu+1}\, , \\
      P(y=0^+,s,-|\sigma_0) - P(y=0^-,s,-|\sigma_0) &= \frac{\delta_{\sigma_0,-}}{\mu-1}\, .
 \end{align}
 \end{subequations}
Upon applying it, we find
\begin{subequations}
 \begin{align}
   \tilde P(y,s,+|+) &= \frac{\eta/\mu+\text{sign}(y)\,\lambda}{2\,(1+\mu)\lambda} \rme^{\eta\, y -\lambda\, |y|} \, ,\\
   \tilde P(y,s,-|-) &= \frac{\eta/\mu-\text{sign}(y)\,\lambda}{2\,(1-\mu)\lambda} \rme^{\eta\, y -\lambda \,|y|}\, , \\
 \tilde P(y,s,-|+) &=\tilde P(y,s,+|-)=\frac{1}{2\,(1-\mu^2)\lambda} \rme^{\eta\, y-\lambda\, |y|}\, .
\end{align}
\end{subequations}

\section{Long-time limit of the survival probability at the transition $\mu=-v_0$}
\label{app:ltsurvivalz}

We take the long-time limit of survival probability when $\mu=-1$ (\ref{eq:ltSws}). As in section \ref{sec:lgtlimsurvival}, we know that the particle will eventually not survive so that we can rewrite (\ref{eq:ltSws}) as $S_{\sigma_0}(x,t)=\int_{t}^\infty dt'\, F_{\sigma_0}(x,t')$ and find 
\begin{subequations}
  \begin{align}
    S_+(x,t)&\sim \int_{t}^\infty dt' \rme^{-t'} I_0(\sqrt{x(2\, t'-x)})\, ,\\
     S_-(x,t)&\sim \int_{t}^{\infty} dt' \,\rme^{-t'}\, \frac{\sqrt{x}}{\sqrt{2t'-x}} \, I_1\left[\sqrt{x\,(2t'-x)}\right]\,.
\end{align}
\label{eq:ltzspma}
\end{subequations}
Using the asymptotic expansion of the Bessel function $I_{0,1}(x)\sim \rme^{x}\,/\sqrt{2\pi x}$ for $x\rightarrow \infty$ gives
\begin{subequations}
  \begin{align}
    S_+(x,t)&\sim \frac{ x^{-1/4}}{2^{3/4}\,\,\sqrt{\pi}}\int_{t}^\infty dt'\, \frac{\rme^{-t'+\sqrt{2\,x\,t'}}}{t'^{1/4}}\, ,\label{eq:Spltmapp}\\
     S_-(x,t)&\sim  \frac{ x^{1/4}}{2^{5/4}\,\sqrt{\pi}}\int_{t}^\infty dt'\, \frac{\rme^{-t'+\sqrt{2\,x\,t'}}}{t'^{3/4}}\,.
\end{align}
\end{subequations}
We now evaluate the integral in (\ref{eq:Spltmapp}) for large $t$. The second one can be done in a similar fashion. Let us first change variable $u=t'-t$:
\begin{align}
  \int_{0}^\infty du\, \frac{\rme^{-t-u+\sqrt{2\,x\,(u+t)}}}{(u+t)^{1/4}}\,.
\end{align}
Then we factor out $t$ in the denominator and in the argument of the exponential
\begin{align}
 I(t) =\frac{\rme^{-t}}{t^{1/4}} \int_{0}^\infty du\, \frac{\rme^{-u+\sqrt{2\, x\,t}(1+u/t)^{1/2}}}{(1+u/t)^{1/4}}\,,
\end{align}
and expand $(1+u/t)^{1/2}$ and $(1+u/t)^{1/4}$ for large $t$
\begin{align}
I(t) =\frac{\rme^{-t}}{t^{1/4}} \int_{0}^\infty du\,\left(1-\frac{4\,u}{t}+\ldots\right) \rme^{-u+\sqrt{2\, x\,t}\, (1+\frac{u}{2t}+\ldots)}\,,\\
 = \frac{\rme^{-t+\sqrt{2\, x\,t}}}{t^{1/4}} \int_{0}^\infty du\,\left(1-\frac{4\,u}{t}+\ldots\right) \rme^{-u+\, \frac{u\, \sqrt{x}}{\sqrt{2\,t}}+\ldots}\,,
\end{align}
where the dots correspond to higher order terms. Finally, we expand the exponential for large $t$
\begin{align}
I(t)= \frac{\rme^{-t+\sqrt{2\, x\,t}}}{t^{1/4}} \int_{0}^\infty du\,\left(1-\frac{4\,u}{t}+\ldots\right) \rme^{-u}\left(1+\frac{u\, \sqrt{x}}{\sqrt{2\,t}}+\ldots\right)\,,
\end{align}
and neglect the higher order term in the parenthesis to find
\begin{align}
I(t)\sim \frac{\rme^{-t+\sqrt{2\, x\,t}}}{t^{1/4}} \int_{0}^\infty du\, \rme^{-u} = \frac{\rme^{-t+\sqrt{2\, x\,t}}}{t^{1/4}} \,.
\end{align}
This yields the long time limit of the survival probability (\ref{eq:longtz}) displayed in the main text.
%

\section{Last-passage distribution with a finite observation time}
\label{app:lastfin}
For a supercritical drift, the finite observation time does not play a role as the the first-passage time is also the last-passage time. We therefore restrict ourselves to a subcritical drift $-1\!<\!\mu\!<\!1$ in this section. The formula for the last-passage distribution (\ref{eq:lastpass}) can be adapted to take a finite observation time $t_o$ into account:
  \begin{align}
  L_{\sigma_0}(x,t) = (\mu+1)\,P(y=0,t,+|x,\sigma_0)\, S_+(t_o-t) +  (1-\mu)\,P(y=0,t,-|x,\sigma_0)\, S_-(t_o-t)\, .\label{eq:lasto}
\end{align}
where $S_+(t)$ is the probability to survive on $[0,\infty[$ starting from the origin in the $\sigma=+1$ state and $S_-(t)$ is the probability to survive on $[-\infty,0[$ starting from the origin in the $\sigma=-1$ state. From the results on the survival probability in the presence of a subcritical drift (\ref{eq:Spmt}), we have
\begin{subequations}
\begin{align}
    S_+(t_o-t) &= 1-\sqrt{\frac{1-\mu}{1+\mu}} \int_0^{t_o-t}\, dt' \, \rme^{-t'}\,I_1(t'\sqrt{1-\mu^2})\,,  \label{eq:Spsubl}\\
        S_-(t_o-t) &= 1-\sqrt{\frac{1+\mu}{1-\mu}} \int_0^{t_o-t}\, dt' \, \rme^{-t'}\,I_1(t'\sqrt{1-\mu^2})\,.
         \end{align}\label{eq:Spmtl}
\end{subequations}
Upon defining
\begin{subequations}
\begin{align}
  R(y=0,t|x,\,\sigma_0) &= P(y=0,t,+|x,\,\sigma_0) + P(y=0,t,-|x,\,\sigma_0)\,,\\
   Q(y=0,tx,\,\sigma_0) &= P(y=0,t,+|x,\,\sigma_0) - P(y=0,t,-|x,\,\sigma_0)\,,
\end{align}
\label{eq:RQ}
\end{subequations}
we find that the relation (\ref{eq:lasto}) simplifies to
\begin{align}
   L_{\sigma_0}(x,t) = \mu\, &Q(y=0,t|x,\,\sigma_0) \\
   &+ R(y=0,t|x,\,\sigma_0) \left(1-\sqrt{1-\mu^2}\, \int_0^{t_o-t}\, dt' \, \rme^{-t'}\,I_1(t'\sqrt{1-\mu^2})\right)\,.\nonumber
\end{align}
Plugging the definitions of $R(y=0,t|x,\,\sigma_0)$ and $Q(y=0,t|x,\,\sigma_0)$ (\ref{eq:RQ}) and the expressions for the propagator derived in \ref{app:kernel}, we obtain the last-passage distribution with a finite observation time $t_o$. In particular, when $\mu\!=\!0$ and $x\!=\!0$, we find
\begin{align}
   L_{\sigma_0}(x=0,t) &= R(y=0,t|x,\,\sigma_0) \left(1-\, \int_0^{t_o-t}\, dt' \, \rme^{-t'}\,I_1(t')\right)\,,\\
   &= R(y=0,t|x,\,\sigma_0)\, e^{t-t_o} (I_0(t_o-t)+I_1(t_o-t))\,,\\
   &= e^{-t_o}\left(\delta(t)+\frac{I_1(t)+I_0(t)}{2}\right)\bigg(I_0(t_o-t)+I_1(t_o-t)\bigg)\,,\label{eq:last00}
\end{align}
where we used the expressions for the propagator derived in \ref{app:kernel} to go from the second to the third line. The result (\ref{eq:last00}) matches with the one obtained in \cite{singh2019generalised}.
\section{Laplace transform inversions}
\label{app:lapinv}
\subsection{$\mathcal{L}^{-1}[\rme^{-x (\lambda+\eta)}](t)$}
\label{app:lapinvFmw}

We start with the expression of $\lambda$ and $\eta$ given in (\ref{eq:lambda}) and (\ref{eq:eta}) and set $v_0 = \tilde \gamma =1$ which we recall here
\begin{subequations}
\begin{align}
\lambda &= \frac{\sqrt{s\, (2+s)+\mu^2}}{1-\mu^2}\,\label{eq:lambda_app} ,\\
\eta &= \frac{\mu\, (1+s)}{1-\mu^2}\,.\label{eq:eta_app}
\end{align}
  \label{eq:lambdaeta_app}
\end{subequations}
Denoting $y=x/(1-\mu^2)$, we find
\begin{align}
    \mathcal{L}^{-1}[\rme^{-x \eta}](u) &= \rme^{-u} \,\delta\left(u-y\mu\right) = \rme^{-y\mu}\, \delta\left(u-y\mu\right)\, ,
\end{align}
 and 
\begin{align}
     \mathcal{L}^{-1}[\rme^{-x \lambda}](u) &= \mathcal{L}^{-1}[\rme^{-y \sqrt{s(2+s)+\mu^2}}](u)\, ,\\
     &= \mathcal{L}^{-1}[\rme^{-y \sqrt{(s+1)^2+\mu^2-1}}](u)\, ,\\
       &= \rme^{-u}\,\mathcal{L}^{-1}[\rme^{-y \sqrt{s^2-(1-\mu^2)}}](u)\, ,\\
         &= \left\{\begin{array}{ll}0\,, & u< y\, ,\\
          \rme^{-u}\left(\delta(u-y)+\frac{ \sqrt{1-\mu^2}\,y}{\sqrt{u^2-y^2}}  I_1( \sqrt{1-\mu^2}\sqrt{u^2-y^2})\right)\,, & u\geq y \, .\end{array}\right.
         \label{eq:invexplambda}
\end{align}
where we completed the square to go to the second line, use the Laplace transform property
\begin{align} \mathcal{L}^{-1}[f(s+b)](t)=\rme^{-b\, t}\mathcal{L}^{-1}[f(s)](t)\, , \label{eq:lapshift}
\end{align}
 to go to the third line and used formula (35) from the integral table \cite{bateman1954tables} to go to the fourth line. We obtain $F_-(x,t)$ by taking the convolution of these two functions:
\begin{align}
   \mathcal{L}^{-1}[\rme^{-x (\lambda+\eta)}](t) &= \int du\, \mathcal{L}^{-1}[\rme^{-x \lambda}](t-u) \, \mathcal{L}^{-1}[\rme^{-x \eta}](u) \, ,\\
     &= \rme^{-y\, \mu}\mathcal{L}^{-1}[\rme^{-x \lambda}](t-y\,\mu)\, , \\
     &= \left\{\begin{array}{ll}0, & t<t_m \, ,\\
      \rme^{-t}\left(\delta\left(t-t_m\right)+\frac{x}{h(t,x)}  I_1\left[h(t,x)\right]\right), & t\geq t_m\, .\end{array}\right.
\end{align}
where $t_m$ and $h(t,x)$ are given in (\ref{eq:tmfgh}).

\subsection{$\mathcal{L}^{-1}[\rme^{-\eta x} \cosh (\lambda 
   x)](t)$}
   \label{app:lapinvFms1}
Developing $\cosh(x)=(\rme^x+\rme^{-x})/2$, we find:
 \begin{align}
    \mathcal{L}^{-1}[\rme^{- \eta x}\cosh(\lambda x)](t) =  \frac{1}{2}\left(\mathcal{L}^{-1}[\rme^{-x (\eta+\lambda)}](t) +  \mathcal{L}^{-1}[\rme^{-x (\eta-\lambda)}](t)\right)\, .
 \end{align} 
 The first term has been computed in \ref{app:lapinvFmw} and the second term term can be obtained from the first one by switching the sign of $x$ and $\mu$. This gives
 \begin{align}
     \mathcal{L}^{-1}[\rme^{- \eta x}\cosh(\lambda x)](t) &= \left\{\begin{array}{ll}0\,, & t<t_m \, ,\\
      \frac{\rme^{-t}}{2}\left(\delta\left(t-t_m\right)+\delta\left(t-t_M\right)+\frac{x}{h(t,x)}  I_1\left[h(t,x)\right]\right)\,, & t_m\leq t\leq t_M\, ,\\
      0\, , & t>t_M\, .\end{array}\right.
\end{align}
where $t_m$, $h(t,x)$ and $t_M$ are given in (\ref{eq:tm}), (\ref{eq:h}) and (\ref{eq:tM}) respectively. 

\subsection{$ \mathcal{L}^{-1}\left[\frac{1}{\lambda }\rme^{- \eta x}\sinh(\lambda x)\right](t)$}
\label{app:lapinvFms2}
Developing $\sinh(x)=(\rme^x-\rme^{-x})/2$, we find:
\begin{align}
  \mathcal{L}^{-1}\left[\frac{1}{\lambda }\rme^{- \eta x}\sinh(\lambda x)\right](t) = \frac{1}{2}\left(\mathcal{L}^{-1}\left[\frac{1}{\lambda}\rme^{-x (\eta-\lambda)}\right](t) -  \mathcal{L}^{-1}\left[\frac{1}{\lambda}\rme^{-x (\eta+\lambda)}\right](t)\right)\, .
\end{align}
Note that the second term term can be obtained from the first one by switching the sign of $x$ and $\mu$. The first term can be computed using the formula (36) from the integral tables \cite{bateman1954tables}). We find
 \begin{align}
     \frac{1}{(\mu^2-1)}\mathcal{L}^{-1}\left[\frac{1}{\lambda }\rme^{- \eta x}\sinh(\lambda x)\right](t) &= \left\{\begin{array}{ll}0\,, & t<t_m \, ,\\
      \rme^{-t} I_0\left[h(t,x)\right]/2\,, & t_m\leq t\leq t_M\, ,\\
      0\, , & t>t_M\, .\end{array}\right.
\end{align}

\subsection{$ \mathcal{L}^{-1}\left[\frac{s}{\lambda }\rme^{- \eta x}\sinh(\lambda x)\right](t)$}
\label{app:lapinvFms3}
We use the Laplace transform property
\begin{align}
  \mathcal{L}[\partial_t f(t)](s) = s\, \mathcal{L}[ f(t)](s)\, ,
\end{align}
to note that this inverse Laplace transform is the time derivative of the one derived in \ref{app:lapinvFms2}:
 \begin{align}
     &\frac{1}{(\mu^2-1)}\mathcal{L}^{-1}\left[\frac{s}{\lambda }\rme^{- \eta x}\sinh(\lambda x)\right](t)= \frac{1}{(\mu^2-1)}\partial_t\,  \mathcal{L}^{-1}\left[\frac{1}{\lambda }\rme^{- \eta x}\sinh(\lambda x)\right](t)\\
     &= \left\{\begin{array}{ll}0\,, & t<t_m \, ,\\
      \frac{\rme^{-t}}{2}\left( \delta\left(t-t_m\right) + \delta\left(t-t_M\right) +\frac{t(1-\mu^2)-x\mu}{h(t,x)}\,I_1\left[h(t,x)\right] -I_0\left[h(t,x)\right]\right)\,, &t_m\leq t\leq t_M\, ,\\
      0\, , & t>t_M\, .\end{array}\right.
\end{align}

\section{Equivalence with the results from Cinque and Orsingher}
\label{app:equiv}

\begin{figure}[ht]
    \centering
    \begin{subfigure}[t]{0.4\textwidth}
        \centering
        \includegraphics[width=\textwidth]{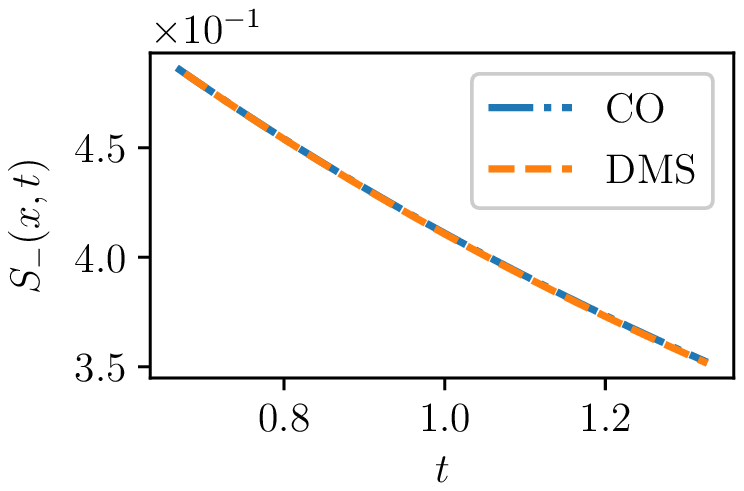}
        \caption{$\mu=-0.5$}
    \end{subfigure}%
    ~ 
    \begin{subfigure}[t]{0.4\textwidth}
        \centering
        \includegraphics[width=\textwidth]{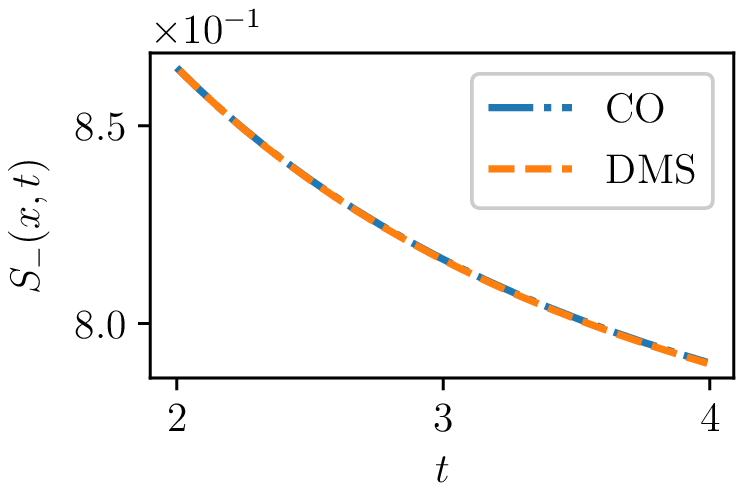}
        \caption{$\mu=0.5$}
    \end{subfigure}\\
       \centering
    \begin{subfigure}[t]{0.4\textwidth}
        \centering
        \includegraphics[width=\textwidth]{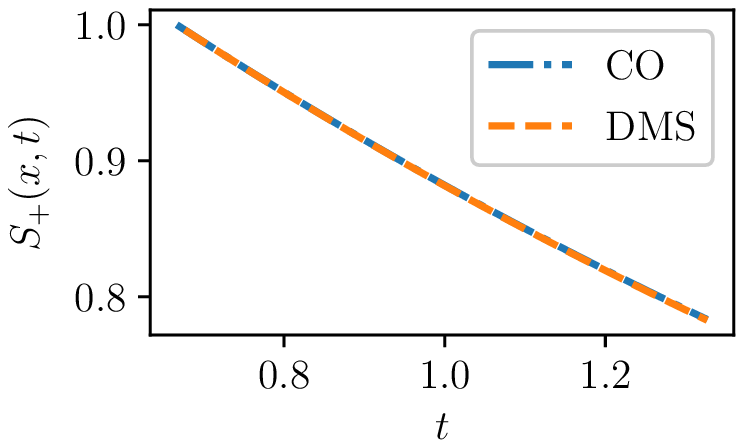}
        \caption{$\mu=-0.5$}
    \end{subfigure}%
    ~ 
    \begin{subfigure}[t]{0.4\textwidth}
        \centering
        \includegraphics[width=\textwidth]{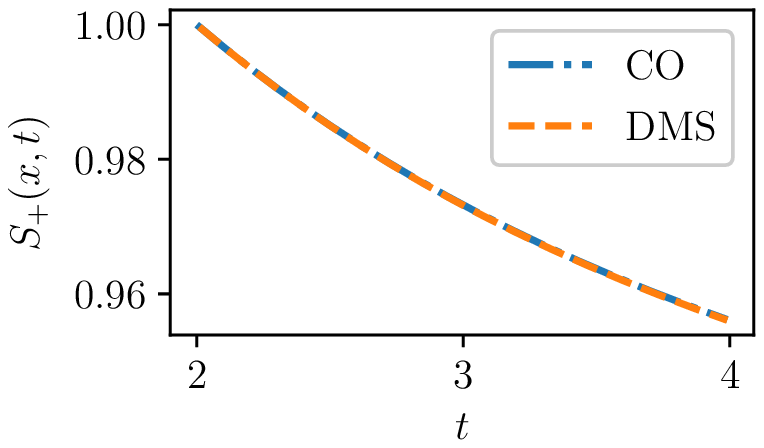}
        \caption{$\mu=0.5$}
    \end{subfigure}
    \caption{Numerical comparison between the results obtained by Cinque and Orsingher (CO) and our results (DMS) for both initial states $\sigma_0=\pm 1$ and different values of subcritical drift~$\mu$.}
    \label{fig:comp}
\end{figure}

We consider a PRW process $y(s)$ with $s \in [0,t]$. The process starts at the initial position $y(0)=0$ with initial velocity $\dot y(0)$. Cinque and Orsingher obtained the following results \cite{cinque2020exact} for the cumulative distribution of the maximum in the subcritical regime ($-1<\mu<1$) 
\begin{align}
  &{\rm Pr.}\left[\text{max}_{0\leq s\leq t}\;y(s) \leq M | \; \dot y(0)=c_1\right] = \\
  &\rme^{-\lambda t} \sum_{r=1}^\infty I_r\left(\frac{2\lambda}{c_1+c_2}\sqrt{(c_1t-M)(c_2 t+M)}\right)\left[\left(\sqrt{\frac{c_2 t+M}{c_1 t-M}}\right)^r-\left(\frac{c_2}{c_1}\sqrt{\frac{c_1 t - M}{c_2 t +M}}\right)^r\right]\, 
\end{align}
for $0\leq M \leq c_1 t$, where $I_r(z)$ is the modified Bessel function of index $r$ and 
\begin{align}
  &{\rm Pr.}\left[\text{max}_{0\leq s\leq t}\;y(s) \leq M | \; \dot y(0)=-c_2\right] = \\
  &\rme^{-\lambda t}\bigg[ \sum_{r=0}^\infty I_r\left(\frac{2\lambda}{c_1+c_2}\sqrt{(c_1t-M)(c_2 t+M)}\right)\left(\sqrt{\frac{c_2 t+M}{c_1 t-M}}\right)^r\\
  & -\frac{c_1}{c_2}\sum_{r=2}^\infty I_r\left(\frac{2\lambda}{c_1+c_2}\sqrt{(c_1t-M)(c_2 t+M)}\right)\left(\frac{c_2}{c_1}\sqrt{\frac{c_1 t-M}{c_2 t+M}}\right)^r\bigg]\, .
\end{align}
for $0\leq M\leq c_1 t$. Converting to our notations [and using the relation (\ref{CDF_2})],
\begin{subequations}
\begin{align}
  c_1&=1-\mu\, ,\\
  c_2&=1+\mu\, ,\\
  M&=x\, , \\
  \lambda&=1\, , \\
  {\rm Pr.}\left[\text{max}_{0\leq s\leq t}\;y(s) \leq M | \; \dot y(0)=c_1\right]&= S_-(x,t)\, , \label{I6a}\\
  {\rm Pr.}\left[\text{max}_{0\leq s\leq t}\;y(s) \leq M | \; \dot y(0)=-c_2\right] &= S_+(x,t)\, \label{I6b} ,
\end{align}
\end{subequations}
it becomes 
\begin{subequations}
\begin{align}
  S^{\text{CO}}_-(x,t)  &=\rme^{- t} \sum_{r=1}^\infty I_r\left(h(t,x)\right)\left[\left(\sqrt{\frac{g(t,x)}{f(t,x)}}\right)^r-\left(\frac{1+\mu}{1-\mu}\sqrt{\frac{f(t,x)}{g(t,x)}}\right)^r\right]\, ,\label{eq:scom}
\end{align}
where the superscript 'CO' refers to Cinque and Orsingher, and 
\begin{align}
  S^{\text{CO}}_+(x,t) &=\rme^{- t}\bigg[ \sum_{r=0}^\infty I_r\left(h(t,x)\right)\left(\sqrt{\frac{g(t,x)}{f(t,x)}}\right)^r \nonumber \\&
   -\frac{1-\mu}{1+\mu}\sum_{r=2}^\infty I_r\left(h(t,x)\right)\left(\frac{1+\mu}{1-\mu}\sqrt{\frac{f(t,x)}{g(t,x)}}\right)^r\bigg]\, ,
\end{align}
\label{eq:CO}
\end{subequations}
where $f(t,y)$, $g(t,y)$ and $h(t,y)$ are given in (\ref{eq:tmfgh}). We recall that in (\ref{I6a}) and (\ref{I6b}) the expression for $S_-(x,t)$ and $S_+(x,t)$ are given in (\ref{eq:Spmt}). 
A numerical comparison of the survival probabilities (\ref{eq:CO}) derived by Cinque and Orsingher 
with the one derived in this work (\ref{eq:Spmt}) shows an excellent agreement (see figure \ref{fig:comp}).

It is actually possible to show analytically that the two formulae coincide, i.e. $S^{\text{CO}}_-(x,t)=S_-(x,t)$ and $S^{\text{CO}}_+(x,t)=S_+(x,t)$. Below we derive in detail the first equality -- actually it is more convenient to show that $\partial_t S^{\text{CO}}_-(x,t)=\partial_t S_-(x,t)$. Let us write the survival probability (\ref{eq:scom}) as
\begin{align}
  S^{\text{CO}}_-(x,t) = S^{\text{CO}}_{-,1}(x,t) - S^{\text{CO}}_{-,2}(x,t)\,,
\end{align}
where
\begin{align}
S^{\text{CO
}}_{-,1}(x,t) &= \rme^{- t} \sum_{r=1}^\infty I_r\left(h(t,x)\right)\left(\sqrt{\frac{g(t,x)}{f(t,x)}}\right)^r\,,\label{eq:scom1}\\
S^{\text{CO}}_{-,2}(x,t) &= \rme^{- t} \sum_{r=1}^\infty I_r\left(h(t,x)\right)\left(\frac{1+\mu}{1-\mu}\sqrt{\frac{f(t,x)}{g(t,x)}}\right)^r\,.
\end{align}
where $f(t,y)$, $g(t,y)$ and $h(t,y)$ are given in (\ref{eq:tmfgh}). We start by converting the series in (\ref{eq:scom1}) into an integral by using the formula 5.8.3.1 p.~694 from the table of integrals \cite{prudnikov1986integral}:
\begin{align}
  \sum_{k=0}^\infty t^k\, I_{k+\nu}(z)=z^{-\nu}\,\rme^{t\,z/2}\int_{0}^z \tau^\nu \, \rme^{-\frac{t\, \tau^2}{2\, z}}\, I_{\nu-1}(\tau)\, d\tau\,,
\end{align}
specified to $z=h(t,x)$, $t=\sqrt{g(t,x)/f(t,x)}$ and $\nu=1$. It gives
\begin{align}
  S^{\text{CO}}_{-,1}(x,t)  = \frac{\rme^{-t+\frac{g(t,x)}{2}}}{f(t,x)} \int_0^{h(t,x)}d\tau\, \tau\, \rme^{-\frac{\tau^2}{2 f(t,x)}} I_0(\tau)\,.\label{eq:scom1i}
\end{align}
Performing integration by parts in (\ref{eq:scom1i}) yields
\begin{align}
  S^{\text{CO}}_{-,1}(x,t)  = \rme^{-t+\frac{g(t,x)}{2}} \left[1-I_0(h(t,x))\, \rme^{-\frac{h(t,x)^2}{2f(t,x)}}+\int_0^{h(t,x)}d\tau\, \rme^{-\frac{\tau^2}{2 f(t,x)}} I_1(\tau)\right]\,.\label{eq:sco1}
\end{align}
where we used the fact that $\partial_\tau I_1(\tau)=I_0(\tau)$. In a similar way, we find for $S^{\text{CO}}_{-,2}(x,t)$ that
\begin{align}
  S^{\text{CO}}_{-,2}(x,t)  = \rme^{-t+\frac{f(t,x)}{2}\frac{1+\mu}{1-\mu}} \left[1-I_0(h(t,x))\, \rme^{-\frac{h(t,x)^2}{2g(t,x)}\frac{1+\mu}{1-\mu}}+\int_0^{h(t,x)}d\tau\, \rme^{-\frac{\tau^2}{2 g(t,x)}\frac{1+\mu}{1-\mu}} I_1(\tau)\right]\,.\label{eq:sco2}
\end{align}
Taking the difference of the two results (\ref{eq:sco1}) and (\ref{eq:sco2}), and noting that $h(t,x)^2=f(t,x)\,g(t,x)$ we obtain
\begin{align}
\begin{split}   S^{\text{CO}}_{-}(x,t) =
  & \rme^{-t+\frac{g(t,x)}{2}}\left[1+\int_0^{h(t,x)}d\tau\, \rme^{-\frac{\tau^2}{2\, f(t,x)}}\,I_1(\tau) \right]\\
  &-\rme^{-t+\frac{f(t,x)}{2}\frac{1+\mu}{1-\mu}}\left[1 + \int_0^{h(t,x)}d\tau\, \rme^{-\frac{\tau^2}{2 g(t,x)}\frac{1+\mu}{1-\mu}}I_1(\tau)\right]\,.
  \end{split}\label{eq:SCOint}
\end{align}
As a check, we evaluate (\ref{eq:SCOint}) at $t=t_m$. Noting that
\begin{align}
  t_m &= \frac{x}{1-\mu}\,, \\
  g(t_m,x)&= 2\,t_m\,,\\
  f(t_m,x) &= 0\,, \\
  h(t_m,x) &= 0\,,
\end{align}
we find 
\begin{align}
  S^{\text{CO}}_{-}(x,t_m) = 1 - \rme^{-t_m}\, ,
\end{align}
which matches with our results. We will now take the time derivative of (\ref{eq:SCOint}). Noting that
\begin{align}
  -t+\frac{g(t,x)}{2} &= \frac{x}{2} + \frac{\mu-1}{2}\, t\, , \\
  -t+\frac{\mu+1}{\mu-1}\frac{f(t,x)}{2}&=\frac{\mu+1}{2(\mu-1)}x +\frac{\mu-1}{2}\, t\, ,\\
  \partial_t f(t,x) &= 1-\mu\, , \\
  \partial_t g(t,x) &= \mu+1 \, ,
\end{align}
we find
\begin{align}
\begin{split}
  \partial_t S^{\text{CO}}_{-}(x,t) =\, &\frac{\mu-1}{2} \rme^{-t+\frac{g(t,x)}{2}}\left[1+\int_0^{h(t,x)} d\tau \, \rme^{-\frac{\tau^2}{2\,f(t,x)}} I_1(\tau)\right] \\
  & +\frac{\mu-1}{2}\, \rme^{-t+\frac{f(t,x)}{2}\frac{1+\mu}{1-\mu}}\left[1+\int_0^{h(t,x)} d\tau \, \rme^{-\frac{\tau^2}{2\,g(t,x)}\frac{1+\mu}{1-\mu}}I_1(\tau)\right]\\
    & -\frac{\mu-1}{2} \rme^{-t+\frac{g(t,x)}{2}} \int_0^{h(t,x)} d\tau\, \, \frac{\tau^2}{ f(t,x)^2}\rme^{-\frac{\tau^2}{2\,f(t,x)}}I_1(\tau)\\
  & - \frac{\mu-1}{2}\, \rme^{-t+\frac{f(t,x)}{2}\frac{1+\mu}{1-\mu}} \int_0^{h(t,x)} d\tau\, \frac{(1+\mu)^2}{(1-\mu)^2}\,  \frac{\tau^2}{g(t,x)^2}  \rme^{-\frac{\tau^2}{2\,g(t,x)}\frac{1+\mu}{1-\mu}}I_1(\tau)\, .
  \end{split}
\end{align}
Note that the terms coming from taking the derivative in the upper-bound of the integrals cancel out. Combining the first line with the third one, and the second one with the fourth one, and noting that
\begin{align}
\begin{split}
  1 +& \int_0^{h(t,x)}d\tau \left(1-\frac{\tau^2}{f(t,x)^2}\right)\rme^{-\frac{\tau^2}{2\, f(t,x)}} I_1(\tau) = \\
  & \left(I_0(h(t,x)) 
  + \frac{h(t,x)}{f(t,x)}I_1(h(t,x))\right) \rme^{-\frac{h(t,x)^2}{2\, f(t,x)}}\, ,
  \end{split}
\end{align} 
along with 
\begin{align}
\begin{split}
    1 + &\int_0^{h(t,x)}d\tau  \left(1-\frac{(1+\mu)^2}{(1-\mu)^2}\frac{\tau^2}{2\,g(t,x)^2}\right)\rme^{-\frac{1+\mu}{1-\mu}\frac{\tau^2}{2\, g(t,x)}} I_1(\tau) =\\ 
    &\, \left(I_0(h(t,x))
  + \frac{1+\mu}{1-\mu}\frac{h(t,x)}{g(t,x)}I_1(h(t,x))\right) \rme^{-\frac{h(t,x)^2}{2\, g(t,x)}}\, ,
  \end{split}
\end{align}
we find
\begin{align}
   \partial_t S^{\text{CO}}_{-}(x,t) &= \frac{\rme^{-t}}{2}\left[ (\mu-1)\sqrt{\frac{g(t,x)}{f(t,x)}} +(\mu+1)\sqrt{\frac{f(t,x)}{g(t,x)}} \right]I_1(h(t,x))\, , \\
   &= \frac{\rme^{-t}}{2} \frac{1}{h(t,x)}\left[ (\mu-1)g(t,x) +(\mu+1)f(t,x) \right]I_1(h(t,x))\, ,\\
   &= -\rme^{-t}\frac{x}{h(t,x)} I_1(h(t,x))\, , \\
   &= \partial_t  S_{-}(x,t)\, ,
\end{align}
which is the desired result. The second equality $S^{\text{CO}}_+(x,t)=S_+(x,t)$ can be proved along the same lines.

\section*{References}
\bibliography{reference.bib}{}
\bibliographystyle{iopart-num}
\end{document}